\documentclass[article]{jss}
\usepackage[utf8]{inputenc}

\providecommand{\tightlist}{%
  \setlength{\itemsep}{0pt}\setlength{\parskip}{0pt}}

\author{
Jared D. Huling\\The Ohio State University \And Menggang Yu\\University of Wisconsin-Madison
}
\title{Subgroup Identification Using the \pkg{personalized} Package}

\Plainauthor{Jared D. Huling, Menggang Yu}
\Plaintitle{Subgroup Identification Using the personalized Package}
\Shorttitle{\pkg{personalized}: A Package for Subgroup Identification}

\Abstract{
A plethora of disparate statistical methods have been proposed for
subgroup identification to help tailor treatment decisions for patients.
However a majority of them do not have corresponding \proglang{R}
packages and the few that do pertain to particular statistical methods
or provide little means of evaluating whether meaningful subgroups have
been found. Recently, the work of \cite{chen2017ageneral} unified many
of these subgroup identification methods into one general, consistent
framework. The goal of the \pkg{personalized} package is to provide a
corresponding unified software framework for subgroup identification
analyses that provides not only estimation of subgroups, but evaluation
of treatment effects within estimated subgroups. The \pkg{personalized}
package allows for a variety of subgroup identification methods for many
types of outcomes commonly encountered in medical settings. The package
is built to incorporate the entire subgroup identification analysis
pipeline including propensity score diagnostics, subgroup estimation,
analysis of the treatment effects within subgroups, and evaluation of
identified subgroups. In this framework, different methods can be
accessed with little change in the analysis code. Similarly, new methods
can easily be incorporated into the package. Besides familiar
statistical models, the package also allows flexible machine learning
tools to be leveraged in subgroup identification. Further estimation
improvements can be obtained via efficiency augmentation.
}

\Plainkeywords{subgroup identification, heterogeneity of treatment effect, interaction modeling, inverse weighting, individualized treatment rules, precision medicine}

\Address{
    Jared D. Huling\\
  The Ohio State University\\
  1958 Neil Ave.\\
  Columbus, Ohio 43210\\
  United States of America\\
  E-mail: \email{huling.7@osu.edu}\\
  URL: \url{http://jaredhuling.org}\\~\\
      Menggang Yu\\
  University of Wisconsin-Madison\\
  600 Highland Ave.\\
  Madison, WI 53792\\
  United States of America\\
  E-mail: \email{meyu@biostat.wisc.edu}\\
  
  }

\usepackage{amsmath,amssymb,amsfonts,bm,bbm,amsthm} \usepackage{longtable} \usepackage{booktabs} \usepackage{bbm, bm}                 

\begin{document}

\hypertarget{introduction}{%
\section{Introduction}\label{introduction}}

Many studies of medical interventions, especially clinical trials, often
focus on population average treatment effects. However, it is widely
recognized that the effects of treatments can have substantial
differences across a population. With the increasing interest to improve
the efficacy and effectiveness of health care, there has been a
significant effort in the statistics community to develop methodology
for optimal allocation of treatments to patients. Optimal treatment
allocation can be thought of as a subgroup identification task, where
subgroups are determined based on the heterogeneity of treatment effect.
Heterogeneity of treatment effect can be characterized by the
interaction of the treatment with patient characteristics. Thus, the
goal in subgroup identification is to characterize and estimate these
interactions in order to construct an optimal mapping from patient
characteristics to a treatment assignment. This mapping is called an
individualized treatment rule (ITR). An optimal ITR is one that, when
enacted on a population, results in the largest expected patient
outcome, assuming without loss of generality that larger outcomes are
preferred. The overall patient outcome is impacted by both the main
effects of patient characteristics and the treatment-covariate
interactions and thus many approaches, such as
\citet{qian2011performance}, focus on modeling this full relationship to
estimate ITRs. In their work, \citet{qian2011performance} show 
robustness properties to model misspecification under certain
conditions. Many recent works have instead focused on methods which do
not require correct specification of the entire relationship between
patient characteristics, treatment, and outcome, but only the parts
relevant to the optimal ITR and are thus often more robust to modeling
choices. Regardless of the general modeling approach, a vast majority of
methods for ITR estimation do not have corresponding \proglang{R}
packages and those that do often pertain to particular statistical
methods for optimal ITR estimation
\citep{SubgrpID, SIDES, quint, EffectTreat, DynTxRegime, FindIt}. In
addition, there has been much focus on estimation of subgroups based on
patient characteristics, yet not enough emphasis on evaluation of the
treatment effects within the resulting estimated subgroups, which is an
equally important but challenging aspect of any subgroup analysis.

\citet{chen2017ageneral} revealed that a wide range of existing
statistical methods for optimal ITR estimation fall under the umbrella
of a unified estimation framework. This unified framework focuses on the
estimation of treatment scores, which rank patients based on their
individualized treatment effect. The scoring system encompasses optimal
ITR estimation in the sense that a threshold for the treatment score can
be used as a treatment assignment mechanism. The \pkg{personalized}
package aims to be a versatile tool in the \proglang{R} statistical
language for optimal ITR estimation and treatment scoring corresponding
to the framework of \citet{chen2017ageneral}. Further, two valid
approaches for estimation of treatment effects within the estimated
subgroups are provided and can be used straightforwardly with any of the
available methods for estimation of treatment scores, enabling
validation of fitted subgroup identification models. The
\pkg{personalized} package offers an entire subgroup analysis workflow
with intuitive and easy-to-use structure. Thus, a wide range of subgroup
identification methods can be accessed with little change in the
analysis workflow of the user. Furthermore the subgroup identification
framework allows the practitioner to conduct a subgroup identification
analysis using familiar statistical modeling concepts. The package is designed to accommodate a wide range
of subgroup identification and treatment decision-making analyses. 

The features of the \pkg{personalized} package include:

\begin{enumerate}
\def\labelenumi{\arabic{enumi}.}
\tightlist
\item
  A wide range of loss function-based subgroup identification methods
\item
  Modeling options for continuous, binary, count, and time-to-event
  outcomes
\item
  Accommodation of observational studies via either propensity
  score-based analysis or matching
\item
  Handling of both binary and multiple treatment scenarios
\item
  Efficiency improvements through loss augmentation
\item
  Evaluation of estimated subgroups with correction for overfitting.
\item
  Options for utilizing custom loss functions
\end{enumerate}

The package is available on the Comprehensive \proglang{R} Archive
Network (CRAN) at \url{https://cran.r-project.org/package=personalized}
in addition to a development version available as a GitHub repository at
\url{https://github.com/jaredhuling/personalized}.

In Section \ref{subgroup-identification-framework} we provide background
on the methodological underpinnings of \pkg{personalized}, followed by a
detailed description of the package itself in Section
\ref{the-personalized-package}. In Section \ref{numerical-comparisons}
we evaluate the various methods offered in the \pkg{personalized}
package in a numerical comparison with several methods from other
packages. Finally, in Section
\ref{analysis-of-national-supported-work-study} we demonstrate the use
of the package on a subgroup identification analysis of the National
Supported Work Study \citep{lalonde1986evaluating} and conclude with
some discussion in Section \ref{discussion}.

\hypertarget{subgroup-identification-framework}{%
\section{Subgroup identification
framework}\label{subgroup-identification-framework}}

\hypertarget{modeling-setup-and-notation}{%
\subsection{Modeling setup and
notation}\label{modeling-setup-and-notation}}

\hypertarget{individualized-treatment-effects-benefit-scores-and-individualized-treatment-rules}{%
\subsubsection{Individualized treatment effects, benefit scores, and
individualized treatment
rules}\label{individualized-treatment-effects-benefit-scores-and-individualized-treatment-rules}}

In this section we provide a formal overview of the subgroup
identification framework of \citet{chen2017ageneral}. We first consider
binary treatments and then provide extensions to multi-category
treatments in a later section. Let the treatment assignment be denoted
as \(T \in \mathcal{T} = \{-1,1\}\), where \(T = 1\) indicates that a
patient received the treatment, and \(T = -1\) indicates a patient
received the control. We also observe the patient outcome \(Y\), where
larger values are assumed to be preferable without loss of generality.
We further observe a length \(p\) vector of patient covariate
information \(\mathbf X\in \mathcal{X}\). Note that the first element of
\(\mathbf X\) is an intercept term. The covariates may modify the effect
of \(T\) on \(Y\), resulting in treatment effect heterogeneity. Relating
the above to observable quantities, we observe data from \(n\) patients
\(\{(Y_i, T_i, \mathbf x_i), i = 1, \dots, n \}\) consisting of \(n\)
independent, identically distributed copies of \((Y, T, \mathbf X)\). In
identifying subgroups of patients who may benefit from \(T\)
differently, we are often interested in estimating the contrast function
\begin{eqnarray}
\Delta(\mathbf X) \equiv E(Y|T=1, \mathbf X) - E(Y|T=-1, \mathbf X). \label{definition of contrast function}
\end{eqnarray} Note that \eqref{definition of contrast function}
involves a difference of means. There can be cases where a ratio,
\begin{eqnarray}
\Gamma(\mathbf{X}) \equiv E(Y|T=1, \mathbf X)/E(Y|T=-1, \mathbf X), \label{definition of gamma function}
\end{eqnarray} may instead be a more interpretable or relevant estimand,
especially for positive \(Y\). Treatment effect heterogeneity is clearly
reflected only through either \(\Delta(\mathbf{X})\) or
\(\Gamma(\mathbf{X})\). To see how these quantities relate to the full
regression model, note that a completely unspecified regression model
\(E(Y| T, \mathbf X)\) can be expressed in terms of main covariate
effects and interactions of the covariates and treatment status:
\begin{align*}
E(Y|T, \mathbf{X}) = {} & I(T=1)\cdot E(Y|T=1, \mathbf{X}) + I(T=-1)\cdot E(Y|T=-1, \mathbf{X}) \\
\equiv {} & g(\mathbf{X}) + T\cdot \Delta(\mathbf{X})/2,
\end{align*} where \(I(\cdot)\) is an indicator function and
\(g(\mathbf{X}) \equiv \frac{1}{2}[E(Y|T=1,\mathbf{X}) + E(Y|T=-1, \mathbf{X})]\)
represents the covariate main effects. Regardless of the form of
\(E(Y| T, \mathbf X)\), the only components that guide which patients
benefit from a treatment are the treatment-covariate interactions,
\(\Delta(\mathbf X)\). A similar re-expression of \(E(Y|T, \mathbf{X})\)
can be shown in terms of \(\Gamma(\mathbf X)\) for positive \(Y\).

Not all subgroup identification methods target \(\Delta(\mathbf{X})\) or
\(\Gamma(\mathbf{X})\) directly, but may rather target some useful
transformation of them. To formalize this notion, we define a
\emph{benefit score} to be any mapping \(f(\mathbf X)\) that possesses
the following two properties: i) it reflects the degree to which
individual patients ``benefit'' from a treatment, i.e.~is monotone in
the treatment effect \(\Delta(\mathbf X)\), \(\Gamma(\mathbf X)\), or
otherwise; ii) it has a meaningful, known cutpoint value \(c\) such that
for a given level of covariates \(\mathbf x\), \(f(\mathbf x) > c\)
implies that the treatment is more effective than control (e.g.
\(\Delta(\mathbf X) > 0\)) and \(f(\mathbf x) \leq c\) implies control
is more effective than treatment (e.g. \(\Delta(\mathbf X) \leq 0\)).
Clearly \(\Delta(\mathbf X)\) is itself a benefit score as it reflects
how much a patient is expected to benefit from a treatment in terms of
his or her outcome. For a patient with \(\mathbf X= \mathbf x\),
\(\Delta(\mathbf x) > 0\) indicates that the treatment is ``better'' in
terms of the expected outcome whereas \(\Delta(\mathbf x) < 0\)
indicates that the control is better. Hence, estimation of
\(\Delta(\mathbf X)\) or its sign allows recommending different
subgroups of patients to different treatments in an optimal manner. By
definition, \(\Delta(\mathbf X)\) can also be used to rank patients by
the magnitude of treatment effect. \(\Gamma(\mathbf{X})\) is clearly
also a benefit score where \(\Gamma(\mathbf x) > 1\) indicates that the
treatment is better in terms of the expected outcome and
\(\Gamma(\mathbf x) \leq 1\) indicates the reverse. It is easily seen
that the use of either \(\Delta(\mathbf x)\) or \(\Gamma(\mathbf x)\)
should lead to similar subgroups.

In Section 2.2 we will introduce benefit score estimators
\(\hat{f}(\mathbf X)\) which can either identify the patients for whom a
treatment is better than a control or rank patients by the degree of
``benefit'' a treatment has. These estimators are not always estimators
of \(\Delta(\mathbf X)\) directly, but often monotone transformations of
\(\hat{f}(\mathbf X)\) will yield estimators of \(\Delta(\mathbf X)\),
i.e. \(\widehat{\Delta}(\mathbf X) = h(\hat{f}(\mathbf X))\) for some
monotone \(h(\cdot)\).

Another quantity of interest is an ITR, which is a map from patient
characteristics to treatment decisions
\(d(\mathbf X): \mathcal{X} \mapsto \mathcal{T}\). An optimal ITR
maximizes the value function \(V(d) = E^d(Y) = \int Y \mathrm{d}P^d\),
where \(P^d\) is the distribution of \((\mathbf X, T, Y)\) given
\(T = d(\mathbf X)\). Essentially, optimal ITRs make treatment decisions
for patients in a manner such that the average outcomes across the
population are maximized. Both \(\Delta(\mathbf X)\) and
\(\Gamma(\mathbf X)\) can be used to construct optimal ITRs. In
particular, \(\text{sign}\{\Delta(\mathbf X)\}\) and
\(\text{sign}\{\Gamma(\mathbf X) - 1\}\) are optimal ITRs.

\hypertarget{assumptions-for-causal-interpretations}{%
\subsubsection{Assumptions for causal
interpretations}\label{assumptions-for-causal-interpretations}}

To deal with non-randomized treatment assignment in observational
studies, as in \citet{chen2017ageneral}, we adopt the notation from the
potential outcome framework of \citet{rubin2005causal}. Let \(Y^{(1)}\)
and \(Y^{(-1)}\) be the potential outcomes if the patient receives
\(T = 1\) and \(T = -1\), respectively. In reality only one of the
potential outcomes can be observed for each individual. Formally, this
statement can be enforced by the relation
\(Y = I(T = 1)Y^{(1)} + I(T = -1)Y^{(-1)}\), where \(I(\cdot)\) is the
indicator function, under the stable unit treatment value assumption
(SUTVA)\citep{rubin2005causal}. In essence, SUTVA requires the potential
outcome of a unit when exposed to a treatment will be the same no matter
what mechanism is used to assign the treatment. We also assume ``strong
ignorability'' \citep{rosenbaum83, rubin2005causal}, that is,
\(T \perp\!\!\perp (Y^{(1)}, Y^{(-1)}) \,|\, \mathbf X\). Violations of
SUTVA can occur when there are spillover effects from treated units to
other units, however in this paper we always assume SUTVA holds. We
assume that the treatment assignment mechanism is either known, as is
the case in randomized controlled trials, or is unknown and can be
estimated, as is the case when there are no unmeasured confounders. In
other words, \(\pi(\mathbf X) = Pr(T = 1 | \mathbf X)\) is either a
known function or can be consistently estimated via regression modeling.
Further, a ``positivity'' assumption that all patients have a chance of
receiving the treatment, i.e. \(0 < \pi(\mathbf x) < 1\) for all
\(\mathbf x\in \mathcal{X}\), is required. Under these assumptions,
\(\Delta(\mathbf X) = E(Y^{(1)}|\mathbf X) - E(Y^{(-1)}|\mathbf X)\) and
\(\Gamma(\mathbf X) = E(Y^{(1)}|\mathbf X) / E(Y^{(-1)}|\mathbf X)\) are
treatment effects conditional on patient characteristics. Note that in
the potential outcome notation, the value function is
\(V(d) = E[Y^{(d)}]\). Many matching strategies
\citep{imbens_rubin_2015} can also be used instead of direct modeling of
\(\pi(\mathbf X) = Pr(T = 1 | \mathbf X)\). However, note that under
matching, the targeted estimand can depend on the matching mechanism.
For example, if matching is based on the treated subjects, then the
estimand is the treatment effect on the treated conditional on patient
characteristics, i.e.
\(\Delta_1(\mathbf X) = E(Y^{(1)}|\mathbf X, T=1) - E(Y^{(-1)}|\mathbf X, T=1)\)
or
\(\Gamma_1(\mathbf X) = E(Y^{(1)}|\mathbf X, T=1) / E(Y^{(-1)}|\mathbf X, T=1)\).

\hypertarget{benefit-score-estimators-and-their-properties}{%
\subsection{Benefit score estimators and their
properties}\label{benefit-score-estimators-and-their-properties}}

\hypertarget{subgroup-identification-and-benefit-score-estimation-via-loss-functions}{%
\subsubsection{Subgroup identification and benefit score estimation via
loss
functions}\label{subgroup-identification-and-benefit-score-estimation-via-loss-functions}}

The framework of \citet{chen2017ageneral} covers two classes of benefit
score estimators. The two methods, called the weighting method and the
Advantage-learning (A-learning) method, are both quite general
approaches for estimating \(\Delta(\mathbf X)\) or \(\Gamma(\mathbf X)\)
(or transformations of \(\Delta(\mathbf X)\) or \(\Gamma(\mathbf X)\))
via loss functions. Both the weighting and the A-learning methods do not
require specification of the full outcome regression model and focus on
direct estimation of \(\Delta(\mathbf X)\), \(\Gamma(\mathbf X)\), or
transformations thereof. As we will explore in later sections, outcome
regression models can, however, be incorporated into both the weighting
and A-learning methods in order to improve efficiency. A major benefit
of both the weighting and A-learning methods is that even when full
outcome regression models are utilized, misspecification of the full
outcome regression model does not impact the validity of the resulting
estimators.

Consider a convex loss function \(\mathbf M(y, v)\) used for the purpose
of estimating benefit scores. A useful example is the squared error
loss, \(\mathbf M(y, v) = (y-v)^2\). In their original work,
\citet{chen2017ageneral} require \(\mathbf M(y, v)\) to meet the
following conditions i)
\(\mathbf M_v(y, v) = \partial \mathbf M(y, v)/\partial v\) is
increasing in \(v\) for every fixed \(y\) and ii) \(\mathbf M_v(y, 0)\)
is monotone in \(y\). These requirements are sufficient for Fisher
consistent subgroup identification, however, they are not necessary.
Conditions i) and ii) can be relaxed to incorporate a wider class of
losses such as the hinge loss \(\mathbf M(y, v) = y\max(1 - v, 0)\). In
Section 2.4, we point out that the conditions specified by
\citet{chen2017ageneral} on \(\mathbf M\) for the multi-category
treatment setting can also be relaxed.

\hypertarget{weighting-method}{%
\subsubsection{Weighting Method}\label{weighting-method}}

The first estimation method is called the weighting method. Given a
sample of \(n\) patients, the weighting method estimates
\(\Delta(\mathbf X)\) or \(\Gamma(\mathbf X)\) (or transformations
thereof) by minimizing the following objective function with respect to
\(f(\mathbf X)\): \begin{equation} \label{eqn:weighting}
L_W(f) = \frac{1}{n}\sum_{i = 1}^n\frac{\mathbf M(Y_i, T_i\times f(\mathbf x_i))}{ {T_i\pi(\mathbf x_i)+(1-T_i)/2} },
\end{equation} where \(W\) indicates the weighting method and
\(\pi(\mathbf x) = Pr(T = 1|\mathbf X= \mathbf x)\) is the propensity
score function. The weighting estimator is then
\(\hat{f}_W = \mbox{argmin}_f L_W(f)\). The corresponding population
level weighting estimator is the minimizer of
\begin{equation} \label{eqn:weighting_truth}
\ell_W(f, \mathbf x) = E\left[\frac{\mathbf M(Y, T\times f(\mathbf x))}{ {T\pi(\mathbf x)+(1-T)/2} }\vert \mathbf X= \mathbf x\right],
\end{equation} with respect to \(f\), where \(W\) again indicates the
weighting method. The weighting method is valid without specification of
the full outcome regression model, as the inverse weights result in the
interactions \(T\times f(\mathbf X)\) being uncorrelated with the main
effects \(g(\mathbf X)\). The estimated benefit score under the
weighting method, \(\hat{f}_W\), can be used to estimate
\(\Delta(\mathbf X)\) under many different loss functions. See Table
\ref{tab:losses_estimands} for examples.

\hypertarget{a-learning-method}{%
\subsubsection{A-learning method}\label{a-learning-method}}

The A-learning estimator involves minimizing\\
\begin{equation} \label{eqn:alearning}
L_A(f) = \frac{1}{n}\sum_{i = 1}^n \mathbf M(Y_i, {\{(T_i+1)/2 -\pi(\mathbf x_i)\} } {\times f(\mathbf x_i))},
\end{equation} where \(A\) indicates the A-learning method and
\((T_i+1)/2 = I(T_i =1)\). The A-learning estimator is then
\(\hat{f}_A = \mbox{argmin}_f L_A(f)\). The A-learning method works
without specification of the full regression model, because the centered
interaction \({\{(T+1)/2 -\pi(\mathbf X)\} } \times f(\mathbf X)\) is
uncorrelated with, and in fact orthogonal to, the main effects
\(g(\mathbf X)\). This property follows from the fact that
\(E[(T+1)/2 -\pi(\mathbf X)|\mathbf X]\) is zero. The corresponding
population level A-learning estimator is the minimizer of
\begin{equation} \label{eqn:alearning_truth}
\ell_A(f, \mathbf x) = E\left[\mathbf M(Y, {\{(T+1)/2 -\pi(\mathbf x)\} } {\times f(\mathbf x))} \vert \mathbf X= \mathbf x\right]
\end{equation} with respect to \(f\), where again \(A\) indicates the
A-learning method and \((T+1)/2 = I(T = 1)\).

\hypertarget{benefit-score-properties-and-estimands}{%
\subsubsection{Benefit score properties and
estimands}\label{benefit-score-properties-and-estimands}}

Although \(\hat{f}_W(\cdot)\) and \(\hat{f}_A(\cdot)\) are not always
themselves estimates of \(\Delta(\cdot)\), the zero point for
\(\hat{f}_W(\cdot)\) and \(\hat{f}_A(\cdot)\) is \emph{always}
meaningful and can be used as a threshold for determining subgroups. For
example, assuming that larger outcomes are preferred, all patients with
covariates \(\mathbf x\) such that \({f}_W(\mathbf x)>0\) should have
better outcomes under the treatment than under the control on average.
More formally, denote the population estimators to be
\[f_{W0}(\mathbf x) = \mbox{argmin}_{f}E\{\ell_W(f, \mathbf x)\} \mbox{ and } f_{A0}(\mathbf x) = \mbox{argmin}_{f}E\{\ell_A(f, \mathbf x)\}.\]
Under both the weighting and A-learning methods and conditions i) and
ii) of \(M\) from above, for patients with a negative benefit score
score (\({f}_{A0}(\mathbf x) < 0\) or \({f}_{W0}(\mathbf x) < 0\)), we
have
\(E\{ \mathbf U(Y^{(1)}) | \mathbf X= \mathbf x\} > E\{ \mathbf U(Y^{(-1)}) | \mathbf X= \mathbf x\}\)
where \(\mathbf U(y) = \partial \mathbf M(y, v)/\partial v |_{v=0}\) and
for those with a positive benefit score, we have
\(E\{ \mathbf U(Y^{(1)}) | \mathbf X= \mathbf x\} < E\{ \mathbf U(Y^{(-1)}) | \mathbf X= \mathbf x\}\).
Thus, \(d_{W0}(\mathbf x) = \text{sign}(f_{W0}(\mathbf x))\) and
\(d_{A0}(\mathbf x) = \text{sign}(f_{A0}(\mathbf x))\) are optimal
decision rules for mapping patient characteristics \(\mathbf X\) to
treatment decision \(T\). Note that treatment assignment desicions here,
while based on the overall cutoff value of 0, are determined by the
individual treatment effects. It was shown in \citet{chen2017ageneral}
that the estimates resulting from both the weighting and A-learning
methods result in Fisher-consistent treatment decision rules under a
wide class of types of outcomes and \(\mathbf M\) functions. Hence, the
estimated benefit scores can be used to optimally assign patients to
treatment groups. For non-differentiable losses such as the hinge loss,
similar arguments can be made.

Furthermore, the benefit scores themselves can reflect the magnitude of
the individual treatment effect and thus can be used for ranking
patients by how effective the treatment is. For example if we use
\(\mathbf M(y, v) = (y - v) ^ 2\), then
\[2{f}_{W0}(\mathbf x) = E(Y^{(1)} | \mathbf X= \mathbf x) - E(Y^{(-1)} | \mathbf X= \mathbf x)=\Delta(\mathbf X)\hphantom{.}\]
and
\[\hphantom{2}{f}_{A0}(\mathbf x) = E(Y^{(1)} | \mathbf X= \mathbf x) - E(Y^{(-1)} | \mathbf X= \mathbf x)=\Delta(\mathbf X).\]
Other choices of \(\mathbf M(y, v)\) lead to different interpretations.
See Table \ref{tab:losses_estimands} for more examples of the
relationship between \({f}_{W0}\), \({f}_{A0}\) and
\(\Delta(\mathbf X)\) or \(\Gamma(\mathbf X)\). As pointed out in
\citet{chen2017ageneral}, similar to using surrogate loss functions in
classification setting \citep{BartlettJASA2006}, the final form of the
solution, \({f}_{W0}\) or \({f}_{A0}\), depends on the choice of the
loss functions. However, not all choices of \(\mathbf{M}(y, v)\) lead to
such direct interpretation. For example, the hinge loss
\(\mathbf{M}(y, v) = y\max(0, 1 - v)\) does not seem to have direct link
with \(\Delta(\mathbf X)\), though the zero point of both \({f}_{W0}\)
and \({f}_{A0}\) under the hinge loss is still meaningful. The
\pkg{personalized} package offers estimation under more losses than are
listed in Table \ref{tab:losses_estimands}, however the additional
losses lead to less interpretable estimates. A listing of all losses
implemented in the \pkg{personalized} package is available in Table
\ref{tab:losses}.

In scenarios where there are limited resources to allocate treatments,
it may be of interest to find a smaller subgroup of patients to
recommend the treatment than the subgroup resulting from patients with
\({f}_{W0} > 0\). Since \({f}_{W0}\) and \({f}_{A0}\) rank patients by
magnitude of treatment effect under most losses, and thus using
\({f}_{W0} > c\) with \(c>0\) yields a smaller subgroup with larger
treatment effect than \({f}_{W0} > 0\). Thus, given limited resources
for treatment allocation, using \({f}_{W0} > c\) can be useful to find a
small subgroup of patients for whom the treatment is highly beneficial.

\begin{table*}[h]
\centering
\begin{tabular}{@{}cccc@{}}\toprule
  $\mathbf{M}(y, v)$ & Estimand & Weighting &  A-learning \\ \midrule
 $(y - v) ^ 2$ & $\Delta(\mathbf{X})$ &  $2f_{W0}(\mathbf{X})$ & $f_{A0}(\mathbf{X})$ \\
\\[-0.5ex]   $-[yv - \exp(v)]$ & $\Delta(\mathbf{X})$ &  $\exp\{f_{W0}(\mathbf{X})\}$ & $\exp\{(1-\pi(\mathbf{X}))f_{A0}(\mathbf{X})\}$ \\
  & & $- \exp\{-f_{W0}(\mathbf{X})\}$ & $- \exp\{-\pi(\mathbf{X})f_{A0}(\mathbf{X})\}$\\
\\[-0.5ex] $-[yv - \log(1 + \exp\{-v\})]$ & $\Delta(\mathbf{X})$ & $\frac{\exp\{f_{W0}(\mathbf{X})\} - 1}{\exp\{f_{W0}(\mathbf{X})\} + 1}$ & $\frac{(\exp\{f_{A0}(\mathbf{X})\} - 1) }{(\exp\{\pi(\mathbf{X})f_{A0}(\mathbf{X})\}+1)}$  \\
 & & & $\times \frac{1}{1 + \exp\{(1-\pi(\mathbf{X}))f_{A0}(\mathbf{X})\}}$ \\
 \\[-0.5ex]
 $ y\log(1 + \exp\{-v\})$ & $\Gamma(\mathbf{X})$ &  $\exp\{f_{W0}(\mathbf{X})\} $ & $\frac{1+\exp\{(1-\pi(\mathbf{X}))f_{A0}(\mathbf{X}) \}}{1+\exp\{-\pi(\mathbf{X})f_{A0}(\mathbf{X}) \}}$ \\
\\[-0.5ex]  $-\left\{ \int_0^\tau\left(  v - \log[E\{ e^vI(X \geq u) \}]  \right) \right.$ &  $\Gamma_M^*(\mathbf{X})^\dagger$  & $\exp\{-f_{W0}(\mathbf{X})\}$ & $\exp\{-f_{A0}(\mathbf{X})\}$ \\
 $\left. \vphantom{\int_0^\tau} \times\mathrm{d} N(u) \right\}$ & & \\
\bottomrule \\[-1.25ex]
\multicolumn{3}{l}{$\dagger$ censoring rates are assumed to be equal within treatment arms }
\end{tabular}
\caption{The last loss above is for survival outcomes with $y = (X, \delta) = \{ \widetilde{X} \wedge C, I(\widetilde{X} \leq t) \}$, $\widetilde{X}$ is the survival time, $C$ is the censoring time, $N(t) = I(\widetilde{X} \leq t)\delta$, and $\tau$ is a fixed point such that $P(X \geq \tau) > 0$. The term $\Gamma_M^*(\mathbf{X})$ for $M \in \{W, A\}$ above is defined as $\frac{E[\Lambda_M^*(Y^\dagger)|T=1, \mathbf{X}]}{E[\Lambda_M^*(Y^\dagger)|T=-1, \mathbf{X}]}$, where $\Lambda_W^*(t)$ is a monotone increasing function described in the Appendix of \citet{tian2014asimple} and $\Lambda_A^*(t)$ is quite quite similar to $\Lambda_W^*(t)$. $\Gamma_W^*(\mathbf{X})$ corresponds to the estimand of the weighting method and $\Gamma_A^*(\mathbf{X})$ corresponds to the estimand of the A-learning method. Under randomization into treatment and control groups with equal probability, the forms above for $\Gamma(\mathbf{X})$ or $\Delta(\mathbf{X})$ for the A-learning method simplify dramatically. For example, under equal randomization and $\mathbf{M}(y,v) = y\log(1 + \exp\{-v\})$, $\Gamma(\mathbf{X}) = \exp\{f_{A0}(\mathbf{X})/2\}$.}
\label{tab:losses_estimands}
\end{table*}

\hypertarget{loss-function-choices-and-relationship-with-other-methods}{%
\subsubsection{Loss function choices and relationship with other
methods}\label{loss-function-choices-and-relationship-with-other-methods}}

Although many of the loss functions in Table \ref{tab:losses_estimands}
are related to negative log-likelihoods from specific models and
distributions, in general there is no distributional requirement of
outcomes for specific choices of losses, except for the loss
corresponding to the Cox proportional hazards model. Thus, if the
outcome of interest is a count outcome, it is valid to use losses other
than \(\mathbf M(y, v) = -[yv - \exp(v)]\), such as the squared loss,
hinge loss, or others. Similarly, it is valid to use losses other than
the logistic loss, \(\mathbf M(y, v) = -[yv - \log(1 + \exp\{-v\})]\).

Each combination of the A-learning or weighting methods with a valid
loss function results in a different estimator, allowing for a high
degree of versatility in estimation. For example,
\(\mathbf M(y, v) = y\log\{1 + \exp(-v)\}\) corresponds to the method
developed in \citet{xu2015regularized},
\(\mathbf M(y, v) = y\max(1 - v, 0)\) corresponds to the Outcome
Weighted Learning (OWL) method of \citet{zhao2012estimating}, under a
randomized clinical trial setting with treatments assigned with equal
probability, both the A-learning and weighting methods with
\(\mathbf M(y, v) = (y-v)^2\) reduce to the modified covariate method of
\citet{tian2014asimple} for continuous responses, the A-learning method
with \(\mathbf M(y, v) = (y-v)^2\) corresponds to the approach of
\citet{lu2013variable} and \citet{ciarleglio2015treatment}, among
others. Using the A-learning method with the squared error loss and loss
augmentation (described below in Section
\ref{efficiency-improvement-via-loss-function-augmentation}) is
equivalent, after accounting for any variable selection penalties, to
the estimation method utilized in \citet{zhao2017selective},
\citet{shi2016robust}, \citet{shi2018high}, and the method behind the
de-sparsified estimator of \citet{jeng2018high}.

\hypertarget{modeling-choices-for-the-benefit-score}{%
\subsubsection{Modeling choices for the benefit
score}\label{modeling-choices-for-the-benefit-score}}

Modeling choices must be made for the form of \({f}_W\) and \({f}_A\).
One can use a simple form of \(f\) such as a linear combination of the
covariates, i.e. \(f(\mathbf X) = \mathbf X^\top\boldsymbol \beta\).
Hence \(\hat{f}(X) = \mathbf X^\top\hat{\boldsymbol \beta}\). Such a
choice leads to interpretable models. For most loss functions, if the
effect \(\beta_j\) of variable \(j\) is positive, then increased values
of variable \(j\) lead to an increase in treatment benefit and negative
effects lead to decreased benefit. Beyond linear forms, regression
trees, smoothing splines, or other nonparametric and flexible approaches
may be used for \({f}_W\) or \({f}_A\).

\hypertarget{loss-function-example-and-implementation-details}{%
\subsection{Loss function example and implementation
details}\label{loss-function-example-and-implementation-details}}

One of the key benefits of the framework of \citet{chen2017ageneral} is
the relative ease of implementation. Many combinations of method
(weighting or A-learning) and loss function can be computed by existing
regression software. Either \eqref{eqn:weighting} or
\eqref{eqn:alearning} can be minimized for a given loss by providing
existing software a modified covariate \(\widetilde{\boldsymbol X}\),
provided the existing software can accept observation weights.

To see how this is accomplished, consider the familiar example of the
squared error loss function \(\mathbf M(y, v) = (y - v)^2\). Under this
loss and the assumption that
\(\Delta(\mathbf X) = \mathbf X^\top\boldsymbol \beta\), we can minimize
(\ref{eqn:weighting}) using existing software. First, denote the
\(n \times p\) design matrix of patient covariate information to be
\(\boldsymbol X\). Denote a modified design matrix
\(\widetilde{\boldsymbol X}\) to be
\(\mbox{diag}(\boldsymbol T) \boldsymbol X\), where
\(\boldsymbol T= (T_1, \dots, T_n)^\top\) and
\(\mbox{diag}(\boldsymbol T)\) is the diagonal matrix with diagonal
elements as the elements of the vector \(\boldsymbol T\). Then the
minimizer of (\ref{eqn:weighting}) is simply the weighted least squares
estimator
\[(\widetilde{\boldsymbol X}^\top\mbox{diag}(\boldsymbol W)\widetilde{\boldsymbol X})^{-1}\widetilde{\boldsymbol X}^\top\mbox{diag}(\boldsymbol W)\boldsymbol Y.\]
where \(\boldsymbol W\) is a vector of weights with the \(i\)th element
as \(1 / ({T_i\pi(\mathbf x_i)+(1-T_i)/2})\) and
\(\boldsymbol Y= (Y_i, \dots, Y_n)^\top\). If
\(\widetilde{\boldsymbol X}\) is high dimensional and variable selection
is desired, \(\widetilde{\boldsymbol X}\) and \(\boldsymbol Y\) along
with a vector of observation weights can be supplied to existing
software, such as the \pkg{glmnet} \proglang{R} package \citep{glmnet}.
More details on how this is handled in \pkg{personalized} package are
provided in Section \ref{the-personalized-package}.

More generally, existing software can be used to minimize
(\ref{eqn:weighting}) and (\ref{eqn:alearning}) by appropriate
construction of weights and modified design matrices. The modified
design matrix for the weighting method is defined as in the example
above, and the modified design matrix for the A-learning method is
defined as
\(\widetilde{\boldsymbol X} = \mbox{diag}(({\boldsymbol T} + 1) / 2 - \boldsymbol \pi(\boldsymbol X))\boldsymbol X\)
where \(\boldsymbol \pi(\boldsymbol X)\) is a vector with \(i\)th
element equal to \(\pi(\mathbf x_i)\).

\hypertarget{extension-to-multi-category-treatments}{%
\subsection{Extension to multi-category
treatments}\label{extension-to-multi-category-treatments}}

Often, more than two treatments are available for patients and the
researcher may wish to understand which of all treatment options are the
best for which patients. Extending the above methodology to
multi-category treatment results in added complications, and in
particular there is no straightforward extension of the A-learning
method for multiple treatment settings. In the supplementary materials
of \citet{chen2017ageneral}, the weighting method was extended to
estimate a benefit score corresponding to each level of a treatment
subject to a sum-to-zero constraint for identifiability. In particular,
we are interested in estimating (the sign of) \begin{eqnarray}
\Delta_{kl}(\mathbf x) \equiv 
 E(Y | T = k, \mathbf X= \mathbf x) - E(Y | T = l, \mathbf X= \mathbf x).  \label{definition of Delta_kl}
\end{eqnarray} If \(\Delta_{kl}(\mathbf x) > 0\), then treatment \(k\)
is preferable to treatment \(l\) for a patient with
\(\mathbf X= \mathbf x\). For each patient, evaluation of all pairwise
comparisons of the \(\Delta_{kl}(\mathbf x)\) indicates which treatment
leads to the largest expected outcome. The weighting estimators of the
benefit scores are the minimizers of the following loss function:
\begin{equation} \label{eqn:weighting_mult}
L_W(f_1, \dots, f_{K}) = \frac{1}{n}\sum_{i = 1}^n\frac{\mathbf M(Y_i,  \sum_{k = 1}^{K}I(T_{i} = k)\times f_k(\mathbf x_i) ) }{ { Pr(T = T_i | \mathbf X= \mathbf x_i)} }
\end{equation} subject to \(\sum_{k = 1}^{K}f_k(\mathbf x_i) = 0\).
Clearly when \(K = 2\), this loss function is equivalent to
(\ref{eqn:weighting}).

Estimation of the benefit scores in this model is still challenging
without added modeling assumptions, as enforcing
\(\sum_{k = 1}^{K}f_k(\mathbf x_i) = 0\) may not always be feasible
using existing estimation routines. However, if each
\(\Delta_{kl}(\mathbf X)\) has a linear form, i.e.
\(\Delta_{kl}(\mathbf X) = \mathbf X^\top\boldsymbol \beta_k\) where
\(l\) represents a reference treatment group, estimation can then easily
be fit into the same computational framework as for the simpler two
treatment case by constructing an appropriate design matrix. Thus, for
multiple treatments the \pkg{personalized} package is restricted to
linear estimators of the benefit scores. For instructive purposes,
consider a scenario with three treatment options, \(A\), \(B\), and
\(C\). Let
\(\boldsymbol X= ({\boldsymbol X}_A^\top, {\boldsymbol X}_B^\top, {\boldsymbol X}_C^\top )^\top\)
be the design matrix for all patients, where each
\({\boldsymbol X}_k^\top\) is the sub-design matrix of patients who
received treatment \(k\). Under
\(\Delta_{kl}(\mathbf X) = \mathbf X^\top\boldsymbol \beta_k\) with
\(l\) as the reference treatment, we can construct a new design matrix
which can then be provided to existing estimation routines in order to
minimize (\ref{eqn:weighting_mult}). With treatment \(C\) as the
reference treatment, the design matrix is constructed as \[
\widetilde{{\boldsymbol X}} = \mbox{diag}(\boldsymbol J)\begin{pmatrix}
{\boldsymbol X}_A & \boldsymbol 0 \\
\boldsymbol 0 & {\boldsymbol X}_B \\
{\boldsymbol X}_C & {\boldsymbol X}_C
\end{pmatrix},
\] where the \(i\)th element of \(\boldsymbol J\) is
\(2I(T_i \neq C) - 1\) and the weight vector \(\boldsymbol W\) is
constructed with the \(i\)th element set to
\(1 / Pr(T = T_i | \boldsymbol X= \boldsymbol x_i)\). Furthermore denote
\(\widetilde{\boldsymbol \beta} = (\boldsymbol \beta_A^\top, \boldsymbol \beta_B^\top)^\top\).
Hence
\(\widetilde{{\boldsymbol X}}^\top\widetilde{\boldsymbol \beta} = {\boldsymbol X}_A^\top\boldsymbol \beta_A + {\boldsymbol X}_B^\top\boldsymbol \beta_B - {\boldsymbol X}_C^\top(\boldsymbol \beta_A + \boldsymbol \beta_B)\),
and thus the sum-to-zero constraints on the benefit scores hold by
construction.

The conditions specified for the loss functions in the supplementary
material of \citet{chen2017ageneral} are too restrictive. In general, we
find that all Fisher-consistent margin based classification loss
functions
\citep{BartlettJASA2006, JMLR:Tewari2007, zou2008, JMLR:Tewari2007} can
be adopted for the weighting method in the multi-category treatment
setting. The sum to zero constraint may be elegantly solved by mimicking
the angle-based reformulation proposed by \cite{ZhangLiuBiometrika2014}
in the classification setting. However this approach has not yet been
adopted in the \pkg{personalized} package.

\hypertarget{efficiency-improvement-via-loss-function-augmentation}{%
\subsection{Efficiency improvement via loss function
augmentation}\label{efficiency-improvement-via-loss-function-augmentation}}

As mentioned in previous sections, the weighting and A-learning
approaches do not require the specification of a full outcome regression
model to consistently recover optimal subgroups. However, gains in
efficiency can be made through augmentation of the loss function. Loss
augmentation involves positing and fitting a full outcome regression
model and using the predictions from this model to construct a modified
loss function. This approach has a few benefits: first, if the full
outcome regression model is indeed correctly specified, then the
resulting estimator will be efficient and second, if the outcome
regression model is incorrectly specified, it does not impact the Fisher
consistency of the estimated subgroups. Further, in practice even when
the outcome regression model is incorrect, efficiency gains are often
realized.

The basic approach to augmentation is the following:

\begin{enumerate}
\def\labelenumi{\arabic{enumi}.}
\tightlist
\item
  Fit a regression model for the conditional mean
  \(E[Y|T, \mathbf X] = g(\mathbf X) + T\Delta(\mathbf X)\) and create
  predictions \(\widehat{E}[Y|T, \mathbf X]\). When the conditional mean
  is linked to predictors via a link function, fit a regression model
  for \(h(E[Y|T, \mathbf X]) = g(\mathbf X) + T\Delta(\mathbf X)\),
  where \(h(\cdot)\) is a known link function, and generate predictions
  on the scale of the linear predictor,
  \(\text{pred}(\mathbf X, T) = h(\widehat{E}[Y|T, \mathbf X])\).
\item
  Create the augmentation function by integrating predictions over both
  treatment options:
  \(a(\mathbf X_i) = \sum_{T \in \mathcal{T}}a_T\text{pred}(\mathbf X_i, T)\),
  where \(a_T\) are weights. In practice, a simple average with
  \(a_T = 1/2\) works well.
\item
  Construct an augmented loss function
  \(\widetilde{\mathbf{M}}(y, v) = {\mathbf{M}}(y, v) + {\mathbf{g}}(a(\mathbf X), v)\),
  where \({\mathbf{g}}(y, v)\) meets the same conditions required of
  \({\mathbf{M}}(y, v)\).
\end{enumerate}

The augmented loss function \(\widetilde{\mathbf{M}}(y, v)\) does not
change the optimality of the resulting decision rules, and thus allows
for potential reductions in variance. The most efficient augmentation
function under the class of augmentation functions of the multiplicative
form \(v\mathbf{g}(\mathbf X)\) was derived in the Supplementary
Material of \citet{chen2017ageneral}. However, in the \pkg{personalized}
package, we consider a more limited class of augmentation functions
\({\mathbf{g}}^*(\mathbf X) = {\mathbf{g}}(a(\mathbf X))\) that allows
for simpler implementation using the functionality for offsets provided
in existing software.

\hypertarget{validating-estimated-subgroups-via-subgroup-conditional-treatment-effects}{%
\subsection{Validating estimated subgroups via subgroup-conditional
treatment
effects}\label{validating-estimated-subgroups-via-subgroup-conditional-treatment-effects}}

A subgroup analysis under the framework of \citet{chen2017ageneral}
involves estimating a set of benefit scores \(\hat{f}(\mathbf x_i)\)
based on \(\{(Y_i, T_i, \mathbf x_i), i = 1, \dots, n \}\) and then
using the benefit scores to construct subgroups, i.e.~a subgroup of
patients for whom the treatment is ``recommended'' via
\(\hat{f}(\mathbf x_i) > 0\) and a subgroup of patients for whom the
treatment is not recommended, \(\hat{f}(\mathbf x_i) \leq 0\). Upon
using the data for this purpose, some natural questions to ask are
``what is the effect of the treatment among those with
\(\hat{f}(\mathbf x_i) > 0\)?'', ``is the effect of the treatment among
those with \(\hat{f}(\mathbf x_i) > 0\) meaningfully different from
zero?'', and ``what would be the improvement in outcomes over the
population of interest if all patients followed the recommendations of
\(\hat{f}(\mathbf x_i)\)?''. Such questions are nontrivial to answer
using the same \(n\) samples used to construct the subgroups. The
subgroups themselves are conditional on the observed outcomes, thus
simply evaluating treatment effects within subgroups will yield biased
and overly-optimistic estimates of the subgroup-conditional treatment
effects. See \citet{athey2016recursive} and \citet{qiu2018estimation}
for more discussion.

Denote the decision rule under benefit scores \(\hat{f}\) as
\(d(\mathbf X) = \text{sign}(\hat{f}(\mathbf X))\). Then the overall
benefit of the treatment assignment rule \(d\) over the population is
\begin{align*}
\delta(d) = {} & E(Y^{(d(\mathbf X))}|d(\mathbf X) = T) - E(Y^{(-d(\mathbf X))}|d(\mathbf X) = T).
\end{align*}

Further define the following subgroup-conditional treatment effects as
\begin{align*}
\delta_1(d) = {} &  E(Y^{(1)}|d(\mathbf X)=1) - E(Y^{(-1)}|d(\mathbf X)=1)  \\
= {} & E(Y|T = 1, d(\mathbf X)=1) - E(Y|T = -1, d(\mathbf X)=1)
\end{align*} and \begin{align*}
\delta_{-1}(d) = {} &  E(Y^{(-1)}|d(\mathbf X)=-1) - E(Y^{(1)}|d(\mathbf X)=-1) \\
= {} & E(Y|T = -1, d(\mathbf X)=-1) - E(Y|T = 1, d(\mathbf X)=-1).
\end{align*} The quantities \(\delta_1(d)\) and \(\delta_{-1}(d)\) may
also be of interest. Directly calculating empirical versions,
\(\hat{\delta}(d)\); \(\hat{\delta}_{1}(d)\); and
\(\hat{\delta}_{-1}(d)\) as defined below, of the above quantities using
the following will yield biased estimates. The biased, empirical
estimates are
\[\hat{\delta}_{t}(d) = \frac{\sum_{i}I\{d(\mathbf x_i) =  T_i = t\}Y_i}{\sum_i I\{ d(\mathbf x_i) =  T_i = t\}} - \frac{\sum_{i} I\{d(\mathbf x_i)= T_i = -t\}Y_i}{\sum_{i} I\{d(\mathbf x_i)= T_i = -t\}}\]
for \(t \in \{1, -1\}\) and
\[\hat{\delta}(d) = \frac{\sum_{i} I\{d(\mathbf x_i) = T_i\}Y_i}{\sum_i I\{d(\mathbf x_i) = T_i\}} - \frac{\sum_{i}I\{d(\mathbf x_i) \neq T_i\} Y_i}{\sum_i I\{d(\mathbf x_i) \neq T_i\} }.\]
Thus, alternative approaches are needed to estimate \(\delta(d)\),
\(\delta_1(d)\), and \(\delta_{-1}(d)\). Another potentially interesting
statistic to measure benefit of subgroup recommendations is the
C-for-benefit statistic of \citet{van2018proposed}, however this is not
used in the \pkg{personalized} package.

\hypertarget{bootstrap-bias-correction}{%
\subsubsection{Bootstrap bias
correction}\label{bootstrap-bias-correction}}

The first approach used in the \pkg{personalized} package to estimate
\(\delta(d)\), \(\delta_1(d)\), and \(\delta_{-1}(d)\) is the bootstrap
bias correction approach of \citet{Harrell1996}. The bootstrap bias
correction approach seeks to estimate the bias in the estimates of the
subgroup treatment effects that arise from using the same data to
estimate these effects as was used to construct the subgroups and then
corrects for this bias. This bootstrap bias correction method was
introduced in \citet{Harrell1996} and later used in \citet{Foster2011}
for the purpose of evaluating subgroup effectiveness. For any statistic
\(S\), let \(S_{full}(\boldsymbol X)\) be the statistic estimated with
the full training data \(\{(Y_i, T_i, \mathbf x_i), i = 1, \dots, n \}\)
and evaluated on data \(\boldsymbol X\) and \(S_{b}(\boldsymbol X)\) be
the statistics estimated using a bootstrap sample \(\boldsymbol X_b\)
(samples from \(\{(Y_i, T_i, \mathbf x_i), i = 1, \dots, n \}\)) and
evaluated on \(\boldsymbol X\). The general outline of the bootstrap
bias correction method is as follows:

\begin{itemize}
\item
  Construct \(B\) bootstrap samples of size \(n\) with replacement. For
  the \(b\)th bootstrap sample calculate the statistic
  \(S_{b}(\boldsymbol X)\) and \(S_b(\boldsymbol X_b)\)
\item
  The bootstrap estimate of the amount of bias with regards to the
  statistic \(S\) is \[
  \text{bias}_S(\boldsymbol X) = \frac{1}{B}\sum_{b = 1}^B \left[S_b(\boldsymbol X_b) - S_b(\boldsymbol X)\right]
  \]
\item
  The bias-corrected estimate of the statistic \(S\) is then calculated
  as \[S_{full}(\boldsymbol X) - \text{bias}_S(\boldsymbol X)\]
\end{itemize}

The term \(S_b(\boldsymbol X_b)\) involves evaluating statistic \(S\) on
the same data as was used to construct the underlying estimator. The
term \(S_b(\boldsymbol X)\) involves evaluating the \(S\) on the
original dataset, which acts like an external dataset. Thus,
\(S_b(\boldsymbol X_b) - S_b(\boldsymbol X)\) mimics the bias that
arises from evaluating a statistic on the same dataset that was used to
construct the statistic/estimator.

The \pkg{personalized} package uses the bootstrap bias correction
procedure to estimate \(\delta(d)\), \(\delta_1(d)\), and
\(\delta_{-1}(d)\).

\hypertarget{repeated-trainingtesting-splitting}{%
\subsubsection{Repeated training/testing
splitting}\label{repeated-trainingtesting-splitting}}

The training and testing splitting approach we outline in this section
is similar to the sample-splitting scheme of \citet{qiu2018estimation},
however their approach is based on a \(K\)-fold type procedure, whereas
ours is based on repeated splitting of the data. The \pkg{personalized}
package has functionality for using the training/testing splitting
procedure to estimate \(\delta(d)\), \(\delta_1(d)\), and
\(\delta_{-1}(d)\). The procedure involves repeatedly randomly
partitioning the data into a training portion and a testing portion.
Each replication partitions \(\tau\times 100\%\) of the data into the
training data and \((1-\tau)\times 100\%\) into the testing data. Using
similar notation as for the bootstrap bias correction approach, define
for the \(b\)th replication \(S_{train,b}(\boldsymbol X)\) to be
statistic \(S\) constructed using the \(b\)th training sample and
evaluated on data \(\boldsymbol X\) and \(\boldsymbol X_{test,b}\) to be
the covariates from the \(b\)th test data. The repeated training/testing
splitting is as follows:

\begin{itemize}
\item
  Construct \(B\) random partitions of the data using training fraction
  \(\tau\) and for each \(b\) calculate
  \(S_{train,b}(\boldsymbol X_{test,b})\)
\item
  The training/testing splitting estimate of statistic \(S\) is then
  \(\frac{1}{B}\sum_{b = 1}^BS_{train,b}(\boldsymbol X_{test,b})\)
\end{itemize}

\citet{Foster2011} explored a variety of approaches for estimating
quantities similar to subgroup-conditional treatment effects and found
bootstrap bias correction approaches to be the least biased and lower
variability than cross-validation based approaches. \citet{Foster2011}
found that cross validation approaches tend to underestimate effects of
interest. Thus, we advocate the use of the bootstrap bias correction
approach, however the training and testing splitting approach is
appropriate as well and tends to give more conservative estimates of the
subgroup-conditional treatment effects.

\hypertarget{the-personalized-package}{%
\section{The personalized package}\label{the-personalized-package}}

In this section we provide detailed information about the
\pkg{personalized} package and how it is utilized in subgroup
identification analyses. We begin by providing an outline of the
workflow of the \pkg{personalized} package. The remainder of this
section roughly follows the order of this workflow and explains each
function involved in each step. Along the way, key arguments of each of
these functions are described in detail with usage examples intermixed.
Finally, this section concludes with a demonstration of an entire
subgroup identification analysis on simulated dataset with a
multi-category treatment.

\hypertarget{workflow-of-subgroup-identification-analysis}{%
\subsection{Workflow of subgroup identification
analysis}\label{workflow-of-subgroup-identification-analysis}}

Regardless of the specific modeling choices, the workflow of subgroup
identification analyses in the \pkg{personalized} package has the
following four steps:

\begin{enumerate}
\def\labelenumi{\arabic{enumi}.}
\tightlist
\item
  Construct propensity score function and check propensity score
  diagnostics (\code{check.overlap()})
\item
  Fit a subgroup identification model using \code{fit.subgroup()}
\item
  Estimate the resulting treatment effects among estimated subgroups
  using \code{validate.subgroup()}
\item
  Visualize and examine model (\code{plot()}), subgroup treatment
  effects (\code{print()}), and characteristics of the subgroups
  (\code{summarize.subgroups()})
\end{enumerate}

We will create a simulated dataset where we know the underlying
data-generating mechanism. We will use this dataset throughout this
paper for illustration. In this simulation, the treatment assignment
depends on covariates and hence we must model the propensity score
\(\pi(\boldsymbol x) = Pr(T = 1 | \boldsymbol X= \boldsymbol x)\). We
also assume that larger values of the outcome are better. We generate
1000 samples with 50 covariates and consider continuous, binary, and
time-to-event outcomes. The covariates in \(\boldsymbol X\) in this
dataset are generated from a Normal distribution and are uncorrelated,
the propensity function \(\pi(\boldsymbol x)\) depends only on the
\(21\)st and \(41\)st covariates, the optimal treatment rule depends on
covariates 3, 11, 1, and 12 including linear terms an an interaction
between covariates 1 and 12, and the main effects in the outcome
regression model depend on covariates 1, 11, 12, 13, and 15 and include
both nonlinear and linear terms.

\begin{CodeChunk}

\begin{CodeInput}
R> library("personalized")
R> 
R> set.seed(123)
R> n.obs <- 1000 
R> n.vars <- 50
R> x <- matrix(rnorm(n.obs * n.vars, sd = 3), n.obs, n.vars)
R> 
R> # simulate non-randomized treatment assignment
R> xbetat <- 0.5 + 0.25 * x[,21] - 0.25 * x[,41]
R> trt.prob <- exp(xbetat) / (1 + exp(xbetat))
R> trt <- rbinom(n.obs, 1, prob = trt.prob)
R> 
R> # simulate Delta(x)
R> delta <- (0.5 + x[,2] - 0.5 * x[,3] - 1 * x[,11] + 1 * x[,1] * x[,12] )
R> 
R> # simulate main effects g(X)
R> xbeta <- x[,1] + x[,11] - 2 * x[,12] ** 2 + x[,13] + 0.5 * x[,15] ** 2
R> 
R> # add both main effects and interactions
R> xbeta <- xbeta + delta * (2 * trt - 1)
R> 
R> 
R> trt <- ifelse(trt == 1, "Trt", "Ctrl")
R> 
R> # simulate continuous outcomes
R> y <- xbeta + rnorm(n.obs)
R> 
R> # simulate binary outcomes
R> y.binary <- 1 * (xbeta + rnorm(n.obs, sd = 2) > 0 )
R> 
R> # create time-to-event outcomes
R> surv.time <- exp(-20 - xbeta + rnorm(n.obs, sd = 1))
R> cens.time <- exp(rnorm(n.obs, sd = 3))
R> y.time.to.event <- pmin(surv.time, cens.time)
R> status <- 1 * (surv.time <= cens.time)
\end{CodeInput}
\end{CodeChunk}

Note that for the continuous outcomes \code{y} in the code above,
\code{delta} aligns with \eqref{definition of contrast function},
however \code{delta} does not exactly correspond to
\eqref{definition of contrast function} for the binary outcomes
\code{y.binary} and the time-to-event outcomes \code{y.time.to.event}
above. Still, \code{delta} in all types of outcomes above drives
heterogeneity of treatment effect.

\hypertarget{the-check.overlap-function}{%
	\subsection[The check.overlap() function]{The \texorpdfstring{\code{check.overlap()}}{}
		function}\label{the-check.overlap-function}}

\hypertarget{observational-studies}{%
\subsubsection{Observational studies}\label{observational-studies}}

To deal with non-randomized treatment assignment in subgroup
identification analysis for observational studies, we usually construct
a model for the propensity score, which is the probability of treatment
assignment conditional on observed baseline characteristics
\citep{imbens_rubin_2015}. Our package also allows matched analysis for
which this step or the \code{check.overlap()} function is not needed. In
the \pkg{personalized} package, we need to wrap the propensity score
model in a function which inputs covariate values and the treatment
statuses and outputs propensity scores between 0 and 1. It is crucial
for the \pkg{personalized} package to utilize the propensity score model
as a function instead of simply a vector of probabilities. Later in this
paper when we seek to evaluate the subgroup treatment effects, we must
use either bootstrap resampling or repeated training and test splitting
of our data. In both of these approaches we need to re-fit our subgroup
identification model, including the propensity score model and hence it
would be invalid to assume the propensity scores remain constant for
each bootstrap iteration. A simple example of how one construct their
propensity score function is as follows:

\begin{CodeChunk}

\begin{CodeInput}
R> propensity.func <- function(x, trt)
R+ {
R+     # save data in a data.frame
R+     data.fr <- data.frame(trt = trt, x)
R+     
R+     # fit propensity score model using binomial GLM
R+     propensity.model <- glm(trt ~ ., family = binomial(), data = data.fr)
R+     
R+     # create estimated probabilities
R+     pi.x <- predict(propensity.model, type = "response")
R+     return(pi.x)
R+ }
R> 
R> propensity.func(x, trt)[101:105]
\end{CodeInput}

\begin{CodeOutput}
      101       102       103       104       105 
0.2251357 0.2786683 0.9021204 0.4400091 0.8250830 
\end{CodeOutput}

\begin{CodeInput}
R> trt[101:105]
\end{CodeInput}

\begin{CodeOutput}
[1] "Ctrl" "Ctrl" "Trt"  "Trt"  "Trt" 
\end{CodeOutput}
\end{CodeChunk}

The above function uses a binomial generalized linear model (GLM) as the
propensity score model and then uses the \code{predict()} function to
return the estimated propensity scores.

To assess the positivity assumption, propensity scores should be checked
to ensure sufficient overlap between treatment groups. This is a
requirement for valid use of propensity scores. The \pkg{personalized}
package offers a visual aid for checking overlap via the
\texttt{check.overlap()} function, which plots densities or histograms
of the propensity scores for each of the treatment groups. The following
code generates Figure \ref{fig:plot_overlap}:

\begin{CodeChunk}

\begin{CodeInput}
R> check.overlap(x, trt, propensity.func)
\end{CodeInput}
\begin{figure}

{\centering \includegraphics{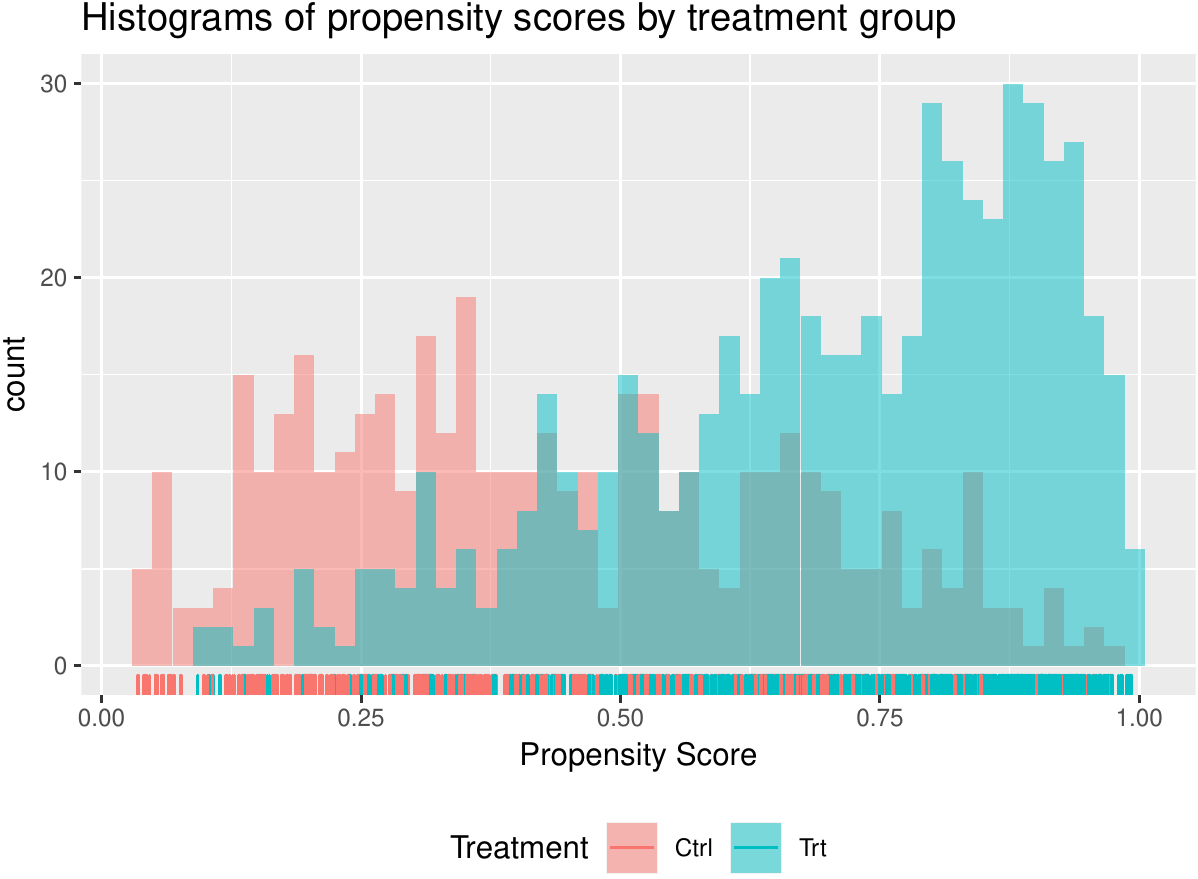} 

}

\caption[Histograms illustrating overlap of propensity scores]{Histograms illustrating overlap of propensity scores.}\label{fig:plot_overlap}
\end{figure}
\end{CodeChunk}

To help assess whether there is sufficient overlap of the propensity
score distributions between treated and untreated, the user should check
whether the regions near 0 or 1 where there is either an area where
there is a positive density of propensity scores for the treatment group
but not the control group or for the control gorup and not the treatment
group. The overlap in Figure \ref{fig:plot_overlap} has reasonable
overlap, however there is a slight region near 0 with a positive density
of propensity scores for the control group but no density of propensity
scores for the treatment group. One may consider corrective measures to
mitigate this. In the presence of insufficient overlap, techniques such
as those proposed in \citet{crump2009dealing} may be utilized. Further
discussion on identification of support overlap issues and approaches
for mitigating these issues can be found in \citet{caliendo2008some} and
\citet{garrido2014methods}.

\hypertarget{randomized-controlled-trials}{%
\subsubsection{Randomized controlled
trials}\label{randomized-controlled-trials}}

If the data to be analyzed come from a randomized controlled trial, it
is still valid to construct a propensity score model as above, but is
not necessary. If the modeler knows that patients were randomized to the
treatment group with probability 0.5, for example, the propensity
function can simply be constructed as the following:

\begin{CodeChunk}

\begin{CodeInput}
R> constant.propensity.func <- function(x, trt) 0.5
\end{CodeInput}
\end{CodeChunk}

\hypertarget{the-fit.subgroup-function}{%
	\subsection[The fit.subgroup() function]{The \texorpdfstring{\code{fit.subgroup()}}{} function}\label{the-fit.subgroup-function}}

The \code{fit.subgroup()} function is the main workhorse of the
\pkg{personalized} package. It provides fitting capabilities for a wide
range of subgroup identification models for different types of outcomes.
We will first show a basic usage of the \code{fit.subgroup()} function
and then provide detailed information about its arguments and more
involved examples.

A basic usage of \code{fit.subgroup()} for continuous outcomes based on
the A-learning method is as follows:

\begin{CodeChunk}

\begin{CodeInput}
R> subgrp.model <- fit.subgroup(x = x, y = y, trt = trt,
R+     propensity.func = propensity.func,
R+     method = "a_learning",
R+     loss   = "sq_loss_lasso",
R+     nfolds = 10) # option for cv.glmnet
R> 
R> summary(subgrp.model)
\end{CodeInput}

\begin{CodeOutput}
family:    gaussian 
loss:      sq_loss_lasso 
method:    a_learning 
cutpoint:  0 
propensity 
function:  propensity.func 

benefit score: f(x), 
Trt recom = Trt*I(f(x)>c)+Ctrl*I(f(x)<=c) where c is 'cutpoint'

Average Outcomes:
                Recommended Ctrl   Recommended Trt
Received Ctrl  -7.8102 (n = 171) -18.589 (n = 239)
Received Trt  -18.9831 (n = 258) -7.5232 (n = 332)

Treatment effects conditional on subgroups:
Est of E[Y|T=Ctrl,Recom=Ctrl]-E[Y|T=/=Ctrl,Recom=Ctrl] 
                                     11.1729 (n = 429) 
    Est of E[Y|T=Trt,Recom=Trt]-E[Y|T=/=Trt,Recom=Trt] 
                                     11.0658 (n = 571) 

NOTE: The above average outcomes are biased estimates of
      the expected outcomes conditional on subgroups. 
      Use 'validate.subgroup()' to obtain unbiased estimates.

---------------------------------------------------

Benefit score quantiles (f(X) for Trt vs Ctrl): 
     0%     25%     50%     75%    100% 
-18.144  -2.746   1.020   4.298  16.332 

---------------------------------------------------

Summary of individual treatment effects: 
E[Y|T=Trt, X] - E[Y|T=Ctrl, X]

    Min.  1st Qu.   Median     Mean  3rd Qu.     Max. 
-18.1440  -2.7457   1.0201   0.8175   4.2980  16.3316 

---------------------------------------------------

4 out of 50 interactions selected in total by the lasso (cross validation criterion).

The first estimate is the treatment main effect, which is always selected. 
Any other variables selected represent treatment-covariate interactions.

            Trt     V2      V3    V11    V13
Estimate 0.7957 1.2542 -0.5189 -0.884 0.5292
\end{CodeOutput}
\end{CodeChunk}

The above code fits a model with linear interaction terms with a squared
error loss as \(M(\cdot, \cdot)\) with a lasso penalty for variable
selection. The squared error loss is used due to the outcome \code{y}
being continuous. The A-learning method is used for demonstrative
purposes and in this situation, the weighting method could have been
used as well. The \code{nfolds} argument is passed to the underlying
fitting function \code{cv.glmnet()} from the \pkg{glmnet} package. The
output provides some basic summary statistics of the resulting
subgroups, benefit scores, and estimated conditional treatment effects
for each sample and shows the estimated interaction coefficients.

\hypertarget{explanation-of-major-function-arguments}{%
\subsubsection{Explanation of major function
arguments}\label{explanation-of-major-function-arguments}}

\hypertarget{section}{%
\paragraph{\texorpdfstring{\code{x}}{}}\label{section}}

The argument \code{x} is for the design matrix. Each column of \code{x}
corresponds to a variable to be used in the model for
\(\Delta(\mathbf X)\) and each row of \code{x} corresponds to an
observation. Every variable in \code{x} will be used for the subgroup
identification model (however some variables may be removed if a
variable selection procedure is specified for \code{loss}).

\hypertarget{section-1}{%
\paragraph{\texorpdfstring{\code{y}}{}}\label{section-1}}

The argument \code{y} is for the response vector. Each element in
\code{y} is a patient observation. In the case of time-to-event outcomes
\code{y} should be specified as a \code{Surv} object of the
\pkg{survival} package \citep{survival}. For example the user should
specify \code{y = Surv(time, status)}, where \code{time} is the observed
time and \code{status} is an indicator that the observed time is the
survival time.

\hypertarget{section-2}{%
\paragraph{\texorpdfstring{\code{trt}}{}}\label{section-2}}

The argument \code{trt} corresponds to the vector of observed treatment
statuses. The vector \code{trt} can either be a character vector
specifying the levels of the treatments (e.g. \code{"Trt"} vs
\code{"Ctrl"}), a factor vector, or and integer vector (e.g.~for binary
treatment, 1 or 0 in the \(i\)th position indicates patient \(i\)
received the treatment or control). For character vectors or integer
vectors it is assumed that the first level alphabetically or
numerically, respectively, is the reference treatment in the sense that
the estimated benefit score will represent the benefit of the second
treatment level with respect to the reference level. For example, if
\code{trt} is a character vector with two treatment options \code{"Trt"}
and \code{"Ctrl"}, the estimated benefit score reflects the benefit of
\code{"Trt"} versus \code{"Ctrl"} in the sense that postive estimated
benefit scores indicate \code{"Trt"} is preferable to \code{"Ctrl"}. For
a factor vector, the first level of the factor will be the reference
treatment. Without specifying otherwise, for \code{trt} vectors with
more than 2 treatment levels, the reference treatment will be chosen in
the same way. However, the user may specify which level of the treatment
should be the reference via the \code{reference.trt} argument. For
example, if \code{trt} has the levels \code{c("TrtA", "TrtB", "Ctrl")},
setting \code{reference.trt = "Ctrl"} will ensure that \code{"Ctrl"} is
the reference level.

\hypertarget{section-3}{%
\paragraph{\texorpdfstring{\code{propensity.func}}{}}\label{section-3}}

The argument \code{propensity.func} corresponds to a function which
returns a propensity score. While it seems cumbersome to have to specify
a function instead of a vector of probabilities, it is crucial for later
validation for the propensity scores to be re-estimated using the
resampled or sampled data (this will be explained further in the section
below for the \code{validate.subgroup()} function). The user should
specify a function which inputs two arguments: \code{trt} and \code{x},
where \code{trt} corresponds to the \code{trt} argument for the
\code{fit.subgroup()} function and \code{x} corresponds to the \code{x}
argument for the \code{fit.subgroup()} function. The function supplied
to the \code{propensity.func} argument should contain code that uses
\code{x} and \code{trt} to fit a propensity score model and then return
an estimated propensity score for each observation in \code{x}. If there
are many covariates, the modeler may wish to use variable selection
techniques in constructing the propensity score model. In the following
code we construct the wrapper function for the propensity score model,
which is a logistic regression model with the lasso penalty where the
tuning parameter is selected by 10-fold cross validation using the
\code{cv.glmnet()} function of the \pkg{glmnet} package \citep{glmnet}:

\begin{CodeChunk}

\begin{CodeInput}
R> # create function for fitting propensity score model
R> propensity.func.lasso <- function(x, trt)
R+ {
R+     # fit propensity score model 
R+     # with binomial model and lasso penalty.
R+     # tuning parameter selected by 10-fold CV
R+     propens.model <- cv.glmnet(y = trt, x = x, 
R+         family = "binomial")
R+     pi.x <- predict(propens.model, s = "lambda.min",
R+         newx = x, type = "response")[,1]
R+     pi.x
R+ }
\end{CodeInput}
\end{CodeChunk}

For randomized controlled trials with equal probability of assignment to
treatment and control, the user can simply define \code{propensity.func}
as:

\begin{CodeChunk}

\begin{CodeInput}
R> propensity.func.const <- function(x, trt) 0.5
\end{CodeInput}
\end{CodeChunk}

which always returns the constant \(1/2\).

For cases with multi-category treatments, the user must specify a
propensity function that returns \(Pr(T = T_i | \mathbf X= \mathbf x)\)
for patient \(i\). In other words, it should return the probability of
receiving the treatment that was actually received for each patient. For
example:

\begin{CodeChunk}

\begin{CodeInput}
R> propensity.func.multinom <- function(x, trt)
R+ {
R+     require(nnet)
R+     df <- data.frame(trt = trt, x)
R+     mfit <- nnet::multinom(trt ~ . -trt, data = df)
R+     # predict returns a matrix of probabilities:
R+     # one column for each treatment level
R+     propens <- predict(mfit, type = "probs")
R+ 
R+     if (is.factor(trt))
R+     {
R+         values <- levels(trt)[trt]
R+     } else 
R+     {
R+         values <- trt
R+     }
R+     # return the probability corresponding to the
R+     # treatment that was observed
R+     probs <- propens[cbind(1:nrow(propens), 
R+         match(values, colnames(propens)))]
R+     probs
R+ }
\end{CodeInput}
\end{CodeChunk}

Optionally the user can specify the function to return a matrix of
treatment probabilities, however, the columns \emph{must} be ordered by
the levels of \code{trt}. An example of this is the following:

\begin{CodeChunk}

\begin{CodeInput}
R> propensity.func.multinom <- function(x, trt)
R+ {
R+     require(nnet)
R+     df <- data.frame(trt = trt, x)
R+     mfit <- multinom(trt ~ . -trt, data = df)
R+     # predict returns a matrix of probabilities:
R+     # one column for each treatment level
R+     propens <- predict(mfit, type = "probs")
R+ 
R+     if (is.factor(trt))
R+     {
R+         levels <- levels(trt)
R+     } else 
R+     {
R+         levels <- sort(unique(trt))
R+     }
R+     # return the probability corresponding to the
R+     # treatment that was observed
R+     probs <- propens[,match(levels, colnames(propens))]
R+     probs
R+ }
\end{CodeInput}
\end{CodeChunk}

For more information on the construction of propensity scores for
multi-category treatments, see \citet{mccaffrey2013tutorial}.

\hypertarget{section-4}{%
\paragraph{\texorpdfstring{\code{loss}}{}}\label{section-4}}

The \code{loss} argument specifies the combination of \(M\) function
(i.e.~loss function) and underlying model \(f(\boldsymbol X)\). The name
of each possible value for \texttt{loss} has two parts: the first part
corresponds to the \(M\) function and the second part corresponds to
\(f(\boldsymbol X)\) and whether variable selection via the lasso is
used. The available \(M\) functions are listed in Table
\ref{tab:losses}.

\newcommand{\ra}[1]{\renewcommand{\arraystretch}{#1}}

\begin{table*}[h]
\centering
\begin{tabular}{@{}cccl@{}}\toprule
 \code{loss} prefix &  Outcomes  &  $M(y, v)$ \\ \midrule
 \code{`sq_loss'} & C/B/CT & $(y - v) ^ 2$  \\
 \code{`logistic_loss'} & B & $-[yv - \log(1 + \exp\{-v\})]$  \\
 \code{`owl_logistic'}$^\dagger$ & C/B/CT & $ y\log(1 + \exp\{-v\})$ \\
 \code{`owl_hinge'}$^\dagger$ & C/B/CT & $y\max(0, 1 - v)$ \\
 \code{`owl_logistic_flip'} & C/B/CT & $\vert y\vert \log(1 + \exp\{-\mbox{sign}(y)v\})$ \\
 \code{`owl_hinge_flip'} & C/B/CT & $\vert y\vert\max(0, 1 - \mbox{sign}(y)v)$ \\
 \code{`poisson_loss'}$^\dagger$ & CT & $-[yv - \exp(v)]$ \\
 \code{`cox_loss'} & TTE & $-\left\{ \int_0^\tau\left(  v - \log[E\{ e^vI(X \geq u) \}]  \right)\mathrm{d} N(u) \right\}$ \\
 && where $y = (X, \delta) = \{ \widetilde{X} \wedge C, I(\widetilde{X} \leq t) \}$, $\widetilde{X}$ is the \\
 && survival time, $C$ is the censoring time, $N(t) = I(\widetilde{X} \leq t)\delta$, \\
 && and $\tau$ is a fixed point such that $P(X \geq \tau) > 0$. \\
\bottomrule \\[-1.25ex]
\multicolumn{3}{l}{$\dagger$ the outcomes need to be non-negative. }
\end{tabular}
\caption{Listed above are the forms of the loss function $M(y, v)$ available in the \pkg{personalized} package. In the outcomes column, ``C'' indicates a loss is available for continuous outcomes, ``B'' for binary outcomes, ``CT'' for count outcomes, and ``TTE'' for time-to-event outcomes. Note that positive continuous outcomes may be used for \code{`poisson\_loss'} as well. In general there are fewer restrictions in theory about the types of outcomes used for the above losses, however the imposed restrictions in this package are due to implementation limitations.}
\label{tab:losses}
\end{table*}

All \code{loss} options that have \code{`lasso'} in their suffix use
\(f(\boldsymbol X) = \boldsymbol X^\top\boldsymbol \beta\) and have the
penalty term \(\sum_{j = 1}^p|\beta_j|\) added to the overall objective
function \(L_W(f)\) or \(L_A(f)\). Adding the penalty term makes the
benefit score estimate
\(\hat{f}(\boldsymbol X) = \boldsymbol X^\top\hat{\boldsymbol \beta}\)
sparse in the sense that some elements of \(\hat{\boldsymbol \beta}\)
will be exactly zero, allowing a simpler form of the benefit score. An
example is \code{`sq_loss_lasso'}, which corresponds to using
\(M(y, v) = (y - v) ^ 2\), a linear form of \(f\), i.e.
\(f(\boldsymbol X) = \boldsymbol X^\top\beta\), and an additional
penalty term \(\sum_{j = 1}^p|\beta_j|\) added to the loss function for
variable selection. All options containing \code{`lasso'} in the name
use the \code{cv.glmnet()} function of the \pkg{glmnet} package
\citep{glmnet} for the underlying model fitting and variable selection.
A \(K\)-fold cross validation is used to select the penalty tuning
parameter. Please see the documentation of \code{cv.glmnet()} for
information about other arguments which can be passed to it.

Any options for \code{loss} which end with \code{`lasso_gam'} have a
two-stage model. Variables are selected using a linear or GLM in the
first stage and then the selected variables are used in a generalized
additive model in the second stage. Univariate nonparametric smoother
terms are used in the second stage for all continuous variables. Binary
variables are used as linear terms in the model. All \code{loss} options
containing \code{`gam'} in the name use the \code{gam()} function of the
\proglang{R} package \pkg{mgcv} \citep{mgcv}. Please see the
documentation of \code{gam()} for information about other arguments
which can be passed to it.

All options that end in \code{`gbm'} use gradient-boosted decision trees
models for \(f(\boldsymbol X)\). Such machine learning models can
provide more flexible forms of estimation by essentially using a sum of
decision trees models. However, these ``black box'' models can be more
challenging to interpret. The \code{gbm}-based models are fit using the
\pkg{gbm} \citep{gbm} \proglang{R} package. Please see the documentation
for the \code{gbm()} function of the \pkg{gbm} package for more details
on the possible arguments. Tuning the values of the hyperparameters
\code{shrinkage}, \code{n.trees}, and \code{interaction.depth} is
crucial for a successful gradient-boosting model. These arguments can be
passed to the \code{fit.subgroup()} function. By default, when
\code{gbm}-based models are used, a plot of the cross validation error
versus the number of trees is displayed. If this plot has values which
are still decreasing significantly by the maximum value of the number of
trees, then it is recommended to either increase the number of trees
(\code{n.trees}), the maximum tree depth (\code{interaction.depth}), or
the step size of the algorithm (\code{shrinkage}).

The loss \code{`owl_hinge'} options are based on the hinge loss function
and thus correspond to support vector machine type of optimization
procedures. Optimization of hinge-based losses is done via the
\pkg{kernlab} package \citep{kernlab}. As such, the underlying model for
\(f(\boldsymbol X)\) depends on the kernel chosen by the user. A linear
kernel will yield a linear decision rule \(f(\boldsymbol X)\), whereas a
nonlinear kernel such as the Gaussian radial basis function kernel will
yield a more flexible, nonlinear decision rule. Available kernels are
listed in the \pkg{kernlab} package and can be displayed by running
\code{?kernels}.

The outcome-weighted learning based losses with \code{`flip'} in their
name allow for non-positive outcomes. In these cases they may offer
substantial finite sample efficiency gains compared with using the
original outcome-weighted learning losses and shifting the response such
that it is positive.

\hypertarget{section-5}{%
\paragraph{\texorpdfstring{\code{method}}{}}\label{section-5}}

The \code{method} argument is used to specify whether the weighting or
A-learning method is used. Specify \code{`weighting'} for the weighting
method that uses \(L_W(f)\) and specify \code{`a_learning'} for the
A-learning method that uses \(L_A(f)\).

\hypertarget{section-6}{%
\paragraph{\texorpdfstring{\code{match.id}}{}}\label{section-6}}

This argument allows the user to specify that the analysis dataset are
based on matched groups of cases and controls. If used, it should be
either a character, factor, or integer vector with length equal to the
number of observations in \code{x} indicating which patients are in
which matched groups. Defaults to \code{NULL} and assumes the samples
are not from a matched cohort. Matched case-control groups can be
created using any method such as propensity score matching, optimal
matching, etc \citep{imbens_rubin_2015}. If each case is matched with a
control or multiple controls, this would indicate which case-control
pairs or groups go together. If \code{match.id} is supplied, then it is
unnecessary to specify a function via the \code{propensity.func}
argument. A quick usage example: if the first patient is a case and the
second and third are controls matched to it, and the fouth patient is a
case and the fifth through seventh patients are matched with it, then
the user should specify \code{match.id = c(1,1,1,2,2,2,2)} or
\code{match.id = c(rep("Grp1", 3),rep("Grp2", 4)) }.

\hypertarget{section-7}{%
\paragraph{\texorpdfstring{\code{augment.func}}{}}\label{section-7}}

The \code{augment.func} argument is used to allow the user to specify an
efficiency augmentation function. The basic idea of efficiency
augmentation is to construct a model for the main effects of the outcome
model and shift the outcome based on these main effects. The resulting
estimator based on the shifted outcome can be more efficient than using
the outcome itself.

For the same reason that the \code{propensity.func} must be specified as
a function, the user should specify a wrapper function for
\code{augment.func} which inputs the covariate information \code{x} and
the outcome \code{y} and outputs a prediction for the outcome for each
observation in \code{x}. The predictions should be returned on the link
scale, in other words on the scale of the linear predictors. The
augmentation function may be from a nonlinear or nonparametric model,
however the predictions should still be returned on the link scale. An
example of an augmentation function that uses linear regression with a
lasso penalty for this model is as follows:

\begin{CodeChunk}

\begin{CodeInput}
R> augment.func.simple <- function(x, y)
R+ {
R+     cvmod <- cv.glmnet(y = y, x = x, nfolds = 10)
R+     predictions <- predict(cvmod, newx = x, s = "lambda.min")
R+     predictions
R+ }
\end{CodeInput}
\end{CodeChunk}

A more involved example that models the full conditional outcome
\(E[Y|T, \mathbf X]\) and integrates over the treatment levels is:

\begin{CodeChunk}

\begin{CodeInput}
R> augment.func <- function(x, y, trt) 
R+ {
R+     data <- data.frame(x, y, trt = ifelse(trt == "Trt", 1, -1))
R+     xm <- model.matrix(y~trt*x-1, data = data)
R+ 
R+     cvmod <- cv.glmnet(y = y, x = xm)
R+     ## get predictions when trt = 1
R+     data$trt <- 1
R+     xm1 <- model.matrix(y~trt*x-1, data = data)
R+     preds_1  <- predict(cvmod, xm1, s = "lambda.min")
R+ 
R+     ## get predictions when trt = -1
R+     data$trt <- -1
R+     xm2 <- model.matrix(y~trt*x-1, data = data)
R+     preds_n1  <- predict(cvmod, xm2, s = "lambda.min")
R+ 
R+     ## return predictions averaged over trt
R+     return(0.5 * (preds_1 + preds_n1))
R+ }
\end{CodeInput}
\end{CodeChunk}

For binary outcomes, one must define the augmentation function such that
it returns predictions on the link scale as follows:

\begin{CodeChunk}

\begin{CodeInput}
R> augment.func.bin <- function(x, y)
R+ {
R+     cvmod <- cv.glmnet(y = y, x = x, family = "binomial")
R+     predict(cvmod, newx = x, s = "lambda.min", type = "link")
R+ }
\end{CodeInput}
\end{CodeChunk}

Then the defined augmentation function can be used in
\code{fit.subgroup()} by passing the function to the argument
\code{augment.func}. A usage example using the above-defined
augmentation function is the following:

\begin{CodeChunk}

\begin{CodeInput}
R> subgrp.model.eff <- fit.subgroup(x = x, y = y, trt = trt,
R+     propensity.func = propensity.func,
R+     loss   = "sq_loss_lasso",
R+     augment.func = augment.func,
R+     nfolds = 10) # option for cv.glmnet
R> 
R> summary(subgrp.model.eff)
\end{CodeInput}

\begin{CodeOutput}
family:    gaussian 
loss:      sq_loss_lasso 
method:    weighting 
cutpoint:  0 
augmentation 
function: augment.func 
propensity 
function:  propensity.func 

benefit score: f(x), 
Trt recom = Trt*I(f(x)>c)+Ctrl*I(f(x)<=c) where c is 'cutpoint'

Average Outcomes:
                Recommended Ctrl    Recommended Trt
Received Ctrl  -8.6343 (n = 178) -18.0961 (n = 232)
Received Trt  -19.7439 (n = 254)  -7.0398 (n = 336)

Treatment effects conditional on subgroups:
Est of E[Y|T=Ctrl,Recom=Ctrl]-E[Y|T=/=Ctrl,Recom=Ctrl] 
                                     11.1095 (n = 432) 
    Est of E[Y|T=Trt,Recom=Trt]-E[Y|T=/=Trt,Recom=Trt] 
                                     11.0563 (n = 568) 

NOTE: The above average outcomes are biased estimates of
      the expected outcomes conditional on subgroups. 
      Use 'validate.subgroup()' to obtain unbiased estimates.

---------------------------------------------------

Benefit score quantiles (f(X) for Trt vs Ctrl): 
     0%     25%     50%     75%    100% 
-12.750  -1.885   0.832   3.248  10.467 

---------------------------------------------------

Summary of individual treatment effects: 
E[Y|T=Trt, X] - E[Y|T=Ctrl, X]

   Min. 1st Qu.  Median    Mean 3rd Qu.    Max. 
-25.500  -3.770   1.664   1.363   6.496  20.934 

---------------------------------------------------

13 out of 50 interactions selected in total by the lasso (cross validation criterion).

The first estimate is the treatment main effect, which is always selected. 
Any other variables selected represent treatment-covariate interactions.

            Trt     V1     V2      V3     V8      V9     V11    V13    V17
Estimate 0.7119 0.2294 0.6364 -0.3792 -0.016 -0.0467 -0.7971 0.4577 0.0512
             V27    V36     V42     V43    V50
Estimate -0.1035 0.0489 -0.1526 -0.1059 0.0434
\end{CodeOutput}
\end{CodeChunk}

From the Online Supplementary Material of \citet{chen2017ageneral}, the
optimal efficiency augmentation function may depend on the treatment
statuses. Hence the user is allowed to specify \code{augment.func} as
additionally a function of \code{trt}, i.e.
\code{augment.func <- function(x, y, trt)}.

\hypertarget{section-8}{%
\paragraph{\texorpdfstring{\code{fit.custom.loss}}{}}\label{section-8}}

The \code{fit.custom.loss} argument allows the user to provide a
function which \textit{minimizes} a custom loss function for use in the
\code{fit.subgroup()} function. The loss function, \(\mathbf M(y, v)\),
to be minimized must meet the criteria outlined in Section
\ref{benefit-score-estimators-and-their-properties}. The user must
provide \code{fit.custom.loss} a function which minimizes a sample
weighted version of the loss function and returns a list with the
solution of the minimization in addition to a function which takes
covariates as an argument and returns predictions of the benefit score
\(\hat{f}(\mathbf x)\) under the estimator resulting from the
minimization of the custom loss.

If provided, this function should take the modified design matrix as an
argument and the responses and optimize a custom weighted loss function.
The provided function \textit{must} be a function with the following
arguments:

\begin{itemize}
\tightlist
\item
  \code{x} --- design matrix
\item
  \code{y} --- vector of responses
\item
  \code{weights} --- vector for observations weights. The underlying
  loss function MUST have samples weighted according to this vector. See
  the example below.
\item
  \code{...} --- additional arguments passed via \code{...}. This can be
  used so that users can specify more arguments to the underlying
  fitting function via \code{fit.subgroup()} if so desired.
\end{itemize}

The provided function \textit{must} return a list with the following
elements:

\begin{itemize}
\tightlist
\item
  \code{predict} --- A function that inputs a design matrix and a `type'
  argument for the type of predictions and outputs a vector of
  predictions on the scale of the linear predictor. Note that the matrix
  provided to \code{fit.custom.loss} has a column appended to the first
  column of \code{x} corresponding to the treatment main effect. Thus,
  the prediction function should deal with this, e.g.
  \code{predict(model, cbind(1, x))}
\item
  \code{model} --- A fitted model object returned by the underlying
  fitting function. This can be an arbitrary \proglang{R} object
\item
  \code{coefficients} --- If the underlying fitting function yields a
  vector of coefficient estimates, they should be provided here.
\end{itemize}

The provided function can also optionally take the following arguments
which may be optionally used in the custom fitting routine:

\begin{itemize}
\tightlist
\item
  \code{match.id} --- Vector of case/control cluster identifiers. This
  is useful if cross validation is used in the underlying fitting
  function in which case it is advisable to sample whole clusters
  randomly instead of individual observations.
\item
  \code{offset} --- If efficiency augmentation is used, the predictions
  from the outcome model from \code{augment.func} will be provided via
  the \code{offset} argument, which can be used as an offset in the
  underlying fitting function as a means of incorporating the efficiency
  augmentation model's predictions.
\item
  \code{trt} --- Vector of treatment statuses
\item
  \code{family} --- Family of outcome
\end{itemize}

An example of \code{fit.custom.loss} is a minimization of the
exponential loss \(\mathbf M(y,v) = y\exp(-v)\) for positive outcomes:

\begin{CodeChunk}

\begin{CodeInput}
R> fit.expo.loss <- function(x, y, weights, ...) {
R+     ## define loss
R+     expo.loss <- function(beta, x, y, weights) {
R+         sum(weights * y * exp(-drop(x %*% beta)))
R+     }
R+ 
R+     ## use optim() to minimize loss function
R+     opt <- optim(rep(0, NCOL(x)), fn = expo.loss, 
R+         x = x, y = y, weights = weights)
R+ 
R+     coefs <- opt$par
R+ 
R+     ## define prediction function which 
R+     ## inputs a design matrix 
R+     ## and returns benefit scores
R+     pred <- function(x, type = "response") {
R+         cbind(1, x) %*% coefs
R+     }
R+ 
R+     # return list of required components
R+     list(predict = pred, model = opt, coefficients = coefs)
R+ }
\end{CodeInput}
\end{CodeChunk}

\hypertarget{section-9}{%
\paragraph{\texorpdfstring{\code{larger.outcome.better}}{}}\label{section-9}}

The argument \code{larger.outcome.better} is a boolean variable
indicating whether larger values of the outcome are better or preferred.
If \code{larger.outcome.better = TRUE}, then \code{fit.subgroup()} will
seek to estimate subgroups in a way that maximizes the population
average outcome and if \code{larger.outcome.better = FALSE},
\code{fit.subgroup()} will seek to minimize the population average
outcome.

\hypertarget{section-10}{%
\paragraph{\texorpdfstring{\code{reference.trt}}{}}\label{section-10}}

As mentioned in the \code{trt} section, the user may specify which level
of the treatment should be the reference via the \code{reference.trt}
argument. For example, if \code{trt} has the levels
\code{c("TrtA", "TrtB", "Ctrl")}, setting \code{reference.trt = "Ctrl"}
will ensure that \code{"Ctrl"} is the reference level. This argument is
not used for multi-category treatment fitting with OWL-type losses, as
the underlying multinomial outcome-weighted model is parameterized such
that there is not a reference treament group. This parameterization is
described in \citet{friedman2010regularization}.

\hypertarget{section-11}{%
\paragraph{\texorpdfstring{\code{cutpoint}}{}}\label{section-11}}

The cutpoint is the numeric value of the benefit score
\(f(\boldsymbol X)\) above which patients will be recommended the
treatment. In other words for outcomes where larger values are better
and a cutpoint with value \(c\) if \(f(\boldsymbol x) > c\) for a
patient with covariate values \(\boldsymbol X= \boldsymbol x\), then
they will be recommended to have the treatment instead of recommended
the control. If lower values are better for the outcome, \(c\) will be
the value below which patients will be recommended the treatment (i.e.~a
patient will be recommended the treatment if \(f(x) < c\)). By default
the cutpoint value is 0. Users may wish to increase this value if there
are limited resources for treatment allocation. The cutpoint argument is
available for multi-category treatments and is still a single value
applied to each comparison with the reference treatment.

The user can also set \code{cutpoint = "median"}, which will use the
median value of the benefit scores as the cutpoint. Similarly, the user
can set specific quantile values via \code{"quantx"} where \code{"x"} is
a number between 0 and 100 representing the quantile value; e.g.
\code{cutpoint = "quant75"} will use the 75th percent upper quantile of
the benefit scores as the cutpoint value.

\hypertarget{section-12}{%
\paragraph{\texorpdfstring{\code{retcall}}{}}\label{section-12}}

The argument \code{retcall} is a boolean variable which indicates
whether to return the arguments passed to \code{fit.subgroup()}. It must
be set to \code{TRUE} if the user wishes to later validate the fitted
model object from \code{fit.subgroup()} using the
\code{validate.subgroup()} function. This is necessary because when
\code{retcall = TRUE}, the design matrix \code{x}, response \code{y},
and treatment vector \code{trt} must be re-sampled in either the
bootstrap procedure or training and testing resampling procedure of
\code{validate.subgroup()}. The only time when \code{retcall} should be
set to \code{FALSE} is when the design matrix is too big to be stored in
the fitted model object.

\hypertarget{section-13}{%
\paragraph{\texorpdfstring{\code{...}}{}}\label{section-13}}

The argument \code{...} is used to pass arguments to the underlying
modeling functions. For example, if the lasso is specified in the
\code{loss} argument, \code{...} is used to pass arguments to the
\code{cv.glmnet()} function from the \pkg{glmnet} package. If \code{gam}
is present in the name for the \code{loss} argument, the underlying
model is fit using the \code{gam()} function of \pkg{mgcv}, so arguments
to \code{gam()} can be passed using \code{...}. The only tricky part for
\code{gam()} is that it also has an argument titled \code{method} and
hence instead, to change the \code{method} argument of \code{gam()}, the
user can pass values using \code{method.gam} which will then be passed
as the argument for \code{method} in the \code{gam()} function. For all
\code{loss} options with `hinge', this will be passed to both
\code{weighted.ksvm()} from the \pkg{personalized} package and
\code{ipop} from the \pkg{kernlab} package.

\hypertarget{continuous-outcomes}{%
\subsubsection{Continuous outcomes}\label{continuous-outcomes}}

The \code{loss} argument options that are available for continuous
outcomes are:

\begin{itemize}
\item
  \code{`sq_loss_lasso'}
\item
  \code{`owl_logistic_loss_lasso'}
\item
  \code{`owl_hinge_loss'}
\item
  \code{`owl_logistic_flip_loss_lasso'}
\item
  \code{`owl_hinge_flip_loss'}
\item
  \code{`sq_loss_gam'}
\item
  \code{`owl_logistic_loss_gam'}
\item
  \code{`owl_logistic_flip_loss_gam'}
\item
  \code{`sq_loss_lasso_gam'}
\item
  \code{`owl_logistic_loss_lasso_gam'}
\item
  \code{`owl_logistic_flip_loss_lasso_gam'}
\item
  \code{`sq_loss_gbm'}
\end{itemize}

Note that the \code{`owl_logistic_loss_lasso'},
\code{`owl_logistic_loss_gam'}, and

\code{`owl_logistic_loss_lasso_gam'} require the outcome to be positive
whereas the corresponding options with \code{`_flip_'} in them have no
such requirement. Similarly, \code{`owl_hinge_loss'} requires the
outcome to be positive whereas \code{`owl_hinge_flip_loss'} does not.

Flexible gradient-boosted decision trees models can also be used. A
typical usage of such models for continuous outcomes is as follows:

\begin{CodeChunk}

\begin{CodeInput}
R> subgrp.model.gbm <- fit.subgroup(x = x, y = y, trt = trt,
R>     propensity.func = propensity.func.lasso,
R>     loss   = "sq_loss_gbm",
R>     shrinkage = 0.025, # options for gbm
R>     n.trees = 1000,
R>     interaction.depth = 2,
R>     cv.folds = 5) 
\end{CodeInput}
\end{CodeChunk}

\hypertarget{binary-outcomes}{%
\subsubsection{Binary outcomes}\label{binary-outcomes}}

All loss options for continuous outcomes can also be used for binary
outcomes. Additionally, the \code{loss} argument options that are
exclusively available for binary outcomes are:

\begin{itemize}
\item
  \code{`logistic_loss_lasso'}
\item
  \code{`logistic_loss_lasso_gam'}
\item
  \code{`logistic_loss_gam'}
\item
  \code{`logistic_loss_gbm'}
\end{itemize}

\begin{CodeChunk}

\begin{CodeInput}
R> subgrp.bin <- fit.subgroup(x = x, y = y.binary, trt = trt,
R+     propensity.func = propensity.func.lasso,
R+     loss   = "logistic_loss_lasso",
R+     nfolds = 10) # option for cv.glmnet
\end{CodeInput}
\end{CodeChunk}

When gradient-boosted decision trees are used for \(f(X)\) by the
package \pkg{gbm}, care must be taken to choose the hyperparameters
effectively. Specifically, \code{shrinkage} (similar to the step-size in
gradient descent), \code{n.trees} (the number of trees to fit), and
\code{interaction.depth} (the maximum depth of each tree) should be
tuned according to the data at hand. By default for gradient-boosting
models, \code{fit.subgroup()} plots the cross validation error versus
the number of trees to enable the users to assess their choice of tuning
parameters.

\hypertarget{count-outcomes}{%
\subsubsection{Count outcomes}\label{count-outcomes}}

All loss options for continuous outcomes can also be used for count
outcomes. Additionally, the \code{loss} argument options that are
exclusively available for count outcomes are:

\begin{itemize}
\item
  \code{`poisson_loss_lasso'}
\item
  \code{`poisson_loss_lasso_gam'}
\item
  \code{`poisson_loss_gam'}
\item
  \code{`poisson_loss_gbm'}
\end{itemize}

\hypertarget{time-to-event-outcomes}{%
\subsubsection{Time-to-event outcomes}\label{time-to-event-outcomes}}

The \code{loss} argument options that are available for continuous
outcomes are:

\begin{itemize}
\item
  \code{`cox_loss_lasso'}
\item
  \code{`cox_loss_gbm'}
\end{itemize}

For subgroup identification models for time-to-event outcomes, the user
should provide \code{fit.subgroup()} with a \code{Surv} object of the
\pkg{survival} package for \code{y}. This can be done as follows:

\begin{CodeChunk}

\begin{CodeInput}
R> library("survival")
R> set.seed(123)
R> subgrp.cox <- fit.subgroup(x = x, y = Surv(y.time.to.event, status),
R+     trt = trt, propensity.func = propensity.func.lasso,
R+     loss   = "cox_loss_lasso",
R+     nfolds = 10)      # option for cv.glmnet
\end{CodeInput}
\end{CodeChunk}

The subgroup treatment effects are estimated using the restricted mean
survival time statistic
\citep{irwin1949standard,zhao1997consistent,zhao1999efficient,chen2001causal}
and can be displayed with \code{summary.subgroup_fitted()} or
\code{print.subgroup_fitted()} as follows:

\begin{CodeChunk}

\begin{CodeInput}
R> summary(subgrp.cox)
\end{CodeInput}

\begin{CodeOutput}
family:    cox 
loss:      cox_loss_lasso 
method:    weighting 
cutpoint:  0 
propensity 
function:  propensity.func 

benefit score: f(x), 
Trt recom = Trt*I(f(x)>c)+Ctrl*I(f(x)<=c) where c is 'cutpoint'

Average Outcomes:
                Recommended Ctrl    Recommended Trt
Received Ctrl 275.7499 (n = 255)  11.9909 (n = 155)
Received Trt  367.1772 (n = 369) 162.6475 (n = 221)

Treatment effects conditional on subgroups:
Est of E[Y|T=Ctrl,Recom=Ctrl]-E[Y|T=/=Ctrl,Recom=Ctrl] 
                                    -91.4273 (n = 624) 
    Est of E[Y|T=Trt,Recom=Trt]-E[Y|T=/=Trt,Recom=Trt] 
                                    150.6566 (n = 376) 

NOTE: The above average outcomes are biased estimates of
      the expected outcomes conditional on subgroups. 
      Use 'validate.subgroup()' to obtain unbiased estimates.

---------------------------------------------------

Benefit score quantiles (f(X) for Trt vs Ctrl): 
      0%      25%      50%      75%     100% 
-0.51188 -0.16824 -0.05754  0.07037  0.66802 

---------------------------------------------------

Summary of individual treatment effects: 
E[Y|T=Trt, X] / E[Y|T=Ctrl, X]

Note: for survival outcomes, the above ratio is 
E[g(Y)|T=Trt, X] / E[g(Y)|T=Ctrl, X], 
where g() is a monotone increasing function of Y, 
the survival time

   Min. 1st Qu.  Median    Mean 3rd Qu.    Max. 
 0.5127  0.9320  1.0592  1.0686  1.1832  1.6684 

---------------------------------------------------

8 out of 49 interactions selected in total by the lasso (cross validation criterion).

The first estimate is the treatment main effect, which is always selected. 
Any other variables selected represent treatment-covariate interactions.

            Trt     V1     V2      V3     V11     V12    V13   V17     V47
Estimate 0.0461 0.0065 0.0473 -0.0101 -0.0207 -0.0014 0.0065 2e-04 -0.0024
            V50
Estimate 0.0176
\end{CodeOutput}
\end{CodeChunk}

\hypertarget{the-summarize.subgroups-function-for-summarizing-subgroups}{%
	\subsection[The summarize.subgroups() function for summarizing
	subgroups]{The \texorpdfstring{\code{summarize.subgroups()}}{} function for summarizing
		subgroups}\label{the-summarize.subgroups-function-for-summarizing-subgroups}}

The \code{summarize.subgroups()} function provides a quick way of
comparing the covariate values between the subgroups recommended the
treatment and the control respectively. P-values for the differences of
covariate values between subgroups are computed and adjusted for
multiple comparisons using the approach of \citet{hommel1988stagewise}.
For continuous variables the p-values come from t-test and for discrete
variables the p-values come from a chi-squared test. The p-values are
computed and used as a means to filter out covariates without meaningful
differences between subgroups, however they are not displayed as they do
not represent valid statistical inferences due to their post-hoc nature.

\begin{CodeChunk}

\begin{CodeInput}
R> comp <- summarize.subgroups(subgrp.cox)
\end{CodeInput}
\end{CodeChunk}

The user can optionally print only the covariates which have
``significant'' differences between subgroups with a multiple
comparisons-adjusted p-value below a given threshold like the following:

\begin{CodeChunk}

\begin{CodeInput}
R> print(comp, p.value = 0.01)
\end{CodeInput}

\begin{CodeOutput}
    Avg (recom Ctrl) Avg (recom Trt) Ctrl - Trt SE (recom Ctrl)
V1            0.4021         -0.5387     0.9408         0.11842
V2            1.7412         -2.5508     4.2920         0.09051
V11          -0.5269          1.0827    -1.6095         0.11687
V50           0.8054         -1.0275     1.8329         0.12070
    SE (recom Trt)
V1          0.1503
V2          0.1085
V11         0.1500
V50         0.1564
\end{CodeOutput}
\end{CodeChunk}

\hypertarget{the-validate.subgroup-function-for-evaluating-identified-subgroups}{%
	\subsection[The validate.subgroup() function for evaluating identified
	subgroups]{The \texorpdfstring{\code{validate.subgroup()}}{} function for evaluating identified
		subgroups}\label{the-validate.subgroup-function-for-evaluating-identified-subgroups}}

It is crucial to evaluate the findings by assessing the improvement in
outcomes with the estimated subgroups. Ideally, the treatment should
have a positive impact on the outcome within the subgroup of patients
who are recommended to the treatment and the control should have a
positive impact on the outcome within the subgroup of patients who were
not recommended the treatment.

In general it is quite challenging to obtain valid estimates of these
effects because usually only one data set is available. Using data
twice, or taking the average outcomes by treatment status within each
subgroup (using the same data) to estimate the treatment effects, will
yield biased and typically overly-optimistic estimates of the
subgroup-specific treatment effects. Therefore, as described in Section
\ref{validating-estimated-subgroups-via-subgroup-conditional-treatment-effects},
we use resampling-based procedures to alleviate this phenomenon and hope
to estimate these effects reliably. The \pkg{personalized} package
offers two methods for subgroup treatment effect estimation. Both
methods are available via the \code{validate.subgroup()} function.

\hypertarget{repeated-trainingtest-splitting}{%
\subsubsection{Repeated training/test
splitting}\label{repeated-trainingtest-splitting}}

The first method of subgroup-specific treatment effects available in
\code{validate.subgroup()} is prediction-based and requires multiple
replications of data partitioning. For each replication in this
procedure, data are randomly partitioned into training and testing
portions. Then the subgroup identification model is estimated using the
training portion and the subgroup treatment effects are estimated via
empirical averages within subgroups using the testing portion. This
method requires two arguments to be passed to
\code{validate.subgroup()}. The first argument is \code{B}, the number
of replications and the second argument is \code{train.fraction}, the
proportion of samples used for training. Hence \code{1 - train.fraction}
is the portion of samples used for testing.

The main object which needs to be passed to \code{validate.subgroup()}
is a fitted object returned by the \code{fit.subgroup()}. Note that in
order to validate a fitted object from \code{fit.subgroup()}, the model
must be fit with the \code{fit.subgroup()} \code{retcall} set to
\code{TRUE} because the data passed to \code{fit.subgroup()} must be
accessed. The \code{validate.subgroup()} function uses the same
arguments that were passed to the original call of \code{fit.subgroup()}
for fitting during each replication.

The validation process is carried out by fixing the cutpoint value at
the user specified cutpoint from the call to \code{fit.subgroup()}.
However, especially in scenarios with very costly treatments, it may be
of interest to investigate the treatment effects within subgroups
defined by different cutpoints along the range of the benefit score. To
simultaneously run the validation procedure for subgroups defined by
different cutpoints of the benefit score, the user can specify a vector
of benefit score quantiles to \code{validate.subgroup()} via the
\code{benefit.score.quantiles} argument. For example, setting
\code{benefit.score.quantiles = c(0.5, 0.75)} will yield validation
results for subgroups defined by a median cutoff value for the benefit
score and a cutoff value at the 75th quantile of the benefit score, the
latter of which will result in a smaller subgroup assigned to the
treatment that will ideally have a larger treatment effect. The default
value for \code{benefit.score.quantiles} is the vector \code(c(1/6, 2/6,
3/6, 4/6, 5/6)). The results of this can be accessed by setting the
\code{plot.subgroup_validated()} argument \code{type = "conditional"} or
by specifying a vector of indexes via the argument \code{which.quant} of
the \code{print_subgroup_validated()} function.

\begin{CodeChunk}

\begin{CodeInput}
R> # check that the object is an object returned by fit.subgroup()
R> class(subgrp.model.eff)
\end{CodeInput}

\begin{CodeOutput}
[1] "subgroup_fitted"
\end{CodeOutput}

\begin{CodeInput}
R> validation.eff <- validate.subgroup(subgrp.model.eff, 
R+     B = 25,  # specify the number of replications
R+     method = "training_test_replication",
R+     benefit.score.quantiles = c(0.5, 0.75, 0.9),
R+     train.fraction = 0.75)
R> 
R> validation.eff
\end{CodeInput}

\begin{CodeOutput}
family:  gaussian 
loss:    sq_loss_lasso 
method:  weighting 

validation method:  training_test_replication 
cutpoint:           0 
replications:       25 

benefit score: f(x), 
Trt recom = Trt*I(f(x)>c)+Ctrl*I(f(x)<=c) where c is 'cutpoint'

Average Test Set Outcomes:
                               Recommended Ctrl
Received Ctrl -11.2162 (SE = 4.9659, n = 44.64)
Received Trt   -16.4652 (SE = 2.8617, n = 64.4)
                                Recommended Trt
Received Ctrl -16.0059 (SE = 3.1811, n = 58.08)
Received Trt   -9.3996 (SE = 2.0931, n = 82.88)

Treatment effects conditional on subgroups:
Est of E[Y|T=Ctrl,Recom=Ctrl]-E[Y|T=/=Ctrl,Recom=Ctrl] 
                       5.249 (SE = 6.5083, n = 109.04) 
    Est of E[Y|T=Trt,Recom=Trt]-E[Y|T=/=Trt,Recom=Trt] 
                      6.6063 (SE = 4.0641, n = 140.96) 

Est of 
E[Y|Trt received = Trt recom] - E[Y|Trt received =/= Trt recom]:                     
5.4049 (SE = 2.7976) 
\end{CodeOutput}
\end{CodeChunk}

Note that when a larger quantile cutoff is used fewer patients are
recommended the treatment, however the treatment effect among those
recommended the treatment is much larger.

\begin{CodeChunk}

\begin{CodeInput}
R> print(validation.eff, which.quant = c(2, 3))
\end{CodeInput}

\begin{CodeOutput}
family:  gaussian 
loss:    sq_loss_lasso 
method:  weighting 

validation method:  training_test_replication 
cutpoint:           Quant_75 
replications:       25 

benefit score: f(x), 
Trt recom = Trt*I(f(x)>c)+Ctrl*I(f(x)<=c) where c is 'cutpoint'

Average Test Set Outcomes:
                                Recommended Ctrl
Received Ctrl   -13.6323 (SE = 3.057, n = 76.52)
Received Trt  -14.2462 (SE = 1.4602, n = 110.48)
                               Recommended Trt
Received Ctrl -15.5259 (SE = 4.2005, n = 26.2)
Received Trt   -7.2177 (SE = 4.0275, n = 36.8)

Treatment effects conditional on subgroups:
Est of E[Y|T=Ctrl,Recom=Ctrl]-E[Y|T=/=Ctrl,Recom=Ctrl] 
                         0.6139 (SE = 3.3128, n = 187) 
    Est of E[Y|T=Trt,Recom=Trt]-E[Y|T=/=Trt,Recom=Trt] 
                           8.3082 (SE = 6.048, n = 63) 

Est of E[Y|Trt received = Trt recom] - E[Y|Trt received =/= Trt recom]:                    
2.5018 (SE = 2.484) 

<===============================================>

family:  gaussian 
loss:    sq_loss_lasso 
method:  weighting 

validation method:  training_test_replication 
cutpoint:           Quant_90 
replications:       25 

benefit score: f(x), 
Trt recom = Trt*I(f(x)>c)+Ctrl*I(f(x)<=c) where c is 'cutpoint'

Average Test Set Outcomes:
                                Recommended Ctrl
Received Ctrl  -13.8475 (SE = 2.8093, n = 91.88)
Received Trt  -13.0648 (SE = 1.3536, n = 133.12)
                                Recommended Trt
Received Ctrl -16.2354 (SE = 6.8779, n = 10.84)
Received Trt   -7.3182 (SE = 8.6173, n = 14.16)

Treatment effects conditional on subgroups:
Est of E[Y|T=Ctrl,Recom=Ctrl]-E[Y|T=/=Ctrl,Recom=Ctrl] 
                        -0.7826 (SE = 3.4515, n = 225) 
    Est of E[Y|T=Trt,Recom=Trt]-E[Y|T=/=Trt,Recom=Trt] 
                         8.9172 (SE = 11.4587, n = 25) 

Est of E[Y|Trt received = Trt recom] - E[Y|Trt received =/= Trt recom]:                     
0.1748 (SE = 3.3348) 
\end{CodeOutput}
\end{CodeChunk}

\hypertarget{bootstrap-bias-correction-1}{%
\subsubsection{Bootstrap bias
correction}\label{bootstrap-bias-correction-1}}

The second method for estimation of subgroup-conditional treatment
effects described in Section
\ref{validating-estimated-subgroups-via-subgroup-conditional-treatment-effects}
and available in \code{validate.subgroup()} is the bootstrap bias
correction method. The bootstrap bias correction method can be accessed
via the \code{validate.subgroup()} function as follows:

\begin{CodeChunk}

\begin{CodeInput}
R> validation.boot <- validate.subgroup(subgrp.model.eff, 
R+     B = 100,  # specify the number of replications
R+     method = "boot_bias_correction")
R> 
R> validation.boot
\end{CodeInput}

\begin{CodeOutput}
family:  gaussian 
loss:    sq_loss_lasso 
method:  weighting 

validation method:  boot_bias_correction 
cutpoint:           0 
replications:       100 

benefit score: f(x), 
Trt recom = Trt*I(f(x)>c)+Ctrl*I(f(x)<=c) where c is 'cutpoint'

Average Bootstrap Bias-Corrected Outcomes:
                                Recommended Ctrl
Received Ctrl  -11.4446 (SE = 1.875, n = 177.19)
Received Trt  -17.7519 (SE = 1.9165, n = 253.83)
                                 Recommended Trt
Received Ctrl -15.6872 (SE = 1.9325, n = 231.04)
Received Trt   -8.7282 (SE = 1.1628, n = 337.94)

Treatment effects conditional on subgroups:
Est of E[Y|T=Ctrl,Recom=Ctrl]-E[Y|T=/=Ctrl,Recom=Ctrl] 
                      6.3073 (SE = 2.3743, n = 431.02) 
    Est of E[Y|T=Trt,Recom=Trt]-E[Y|T=/=Trt,Recom=Trt] 
                       6.959 (SE = 2.1228, n = 568.98) 

Est of 
E[Y|Trt received = Trt recom] - E[Y|Trt received =/= Trt recom]:                     
6.5025 (SE = 1.7104) 
\end{CodeOutput}
\end{CodeChunk}

\hypertarget{evaluating-performance-of-subgroup-specific-treatment-effect-estimation}{%
\subsubsection{Evaluating performance of subgroup-specific treatment
effect
estimation}\label{evaluating-performance-of-subgroup-specific-treatment-effect-estimation}}

We now generate an independent dataset from the same data-generating
mechanism of the simulation in order to evaluate how well the
subgroup-specific treatment effects are estimated by
\code{validate.subgroup()}.

\begin{CodeChunk}

\begin{CodeInput}
R> x.test <- matrix(rnorm(10 * n.obs * n.vars, sd = 3), 10 * n.obs, n.vars)
R> 
R> 
R> # simulate non-randomized treatment
R> xbetat.test   <- 0.5 + 0.25 * x.test[,21] - 0.25 * x.test[,41]
R> trt.prob.test <- exp(xbetat.test) / (1 + exp(xbetat.test))
R> trt.test    <- rbinom(10 * n.obs, 1, prob = trt.prob.test)
R> 
R> # simulate response
R> delta.test <- (0.5 + x.test[,2] - 0.5 * x.test[,3] - 
R+     x.test[,11] + x.test[,1] * x.test[,12] )
R> xbeta.test <- x.test[,1] + x.test[,11] - 2 * x.test[,12] ** 2 + 
R+     x.test[,13] + 0.5 * x.test[,15] ** 2
R> 
R> xbeta.test <- xbeta.test + delta.test * (2 * trt.test - 1)
R> 
R> y.test <- xbeta.test + rnorm(10 * n.obs, sd = 2)
\end{CodeInput}
\end{CodeChunk}

We then use the \code{predict()} function for objects returned by
\code{fit.subgroup()} to obtain the estimated benefit scores for the
test data:

\begin{CodeChunk}

\begin{CodeInput}
R> bene.score.test <- predict(subgrp.model.eff, newx = x.test)
\end{CodeInput}
\end{CodeChunk}

Finally we evaluate the subgroup-specific treatment effects on the test
data based on the estimated subgroups and compare these values with
confidence intervals from the bootstrap bias correction method:

\begin{CodeChunk}

\begin{CodeInput}
R> ## Effect of control among those recommended control 
R> mean(y.test[bene.score.test <= 0 & trt.test == 0]) -
R+        mean(y.test[bene.score.test <= 0 & trt.test == 1])
\end{CodeInput}

\begin{CodeOutput}
[1] 7.19437
\end{CodeOutput}

\begin{CodeInput}
R> quantile(validation.boot$boot.results[[1]][,1], c(0.025, 0.975), na.rm = TRUE)
\end{CodeInput}

\begin{CodeOutput}
     2.5%     97.5% 
 1.824823 10.732421 
\end{CodeOutput}

\begin{CodeInput}
R> ## Trt effect among those recommended the treatment
R> 
R> mean(y.test[bene.score.test > 0 & trt.test == 1]) -
R+        mean(y.test[bene.score.test > 0 & trt.test == 0])
\end{CodeInput}

\begin{CodeOutput}
[1] 5.957166
\end{CodeOutput}

\begin{CodeInput}
R> quantile(validation.boot$boot.results[[1]][,2], c(0.025,  0.975), na.rm = TRUE)
\end{CodeInput}

\begin{CodeOutput}
     2.5%     97.5% 
 3.266023 11.345697 
\end{CodeOutput}
\end{CodeChunk}

We can see that the true values are contained within the bootstrap
confidence intervals.

\hypertarget{the-plot.subgroupfitted-plot.subgroupvalidated-and-plotcompare-functions}{%
	\subsection[The plot.subgroupfitted(), plot.subgroupvalidated(), and plotCompare() functions]{The \texorpdfstring{\code{plot.subgroup\_fitted()}}{}, \texorpdfstring{\code{plot.subgroup\_validated()}}{}, and \texorpdfstring{\code{plotCompare()}}{} functions}\label{the-plot.subgroupfitted-plot.subgroupvalidated-and-plotcompare-functions}}

The outcomes (or average outcomes) of patients within different
subgroups can be plotted using the \code{plot()} function. In
particular, this function plots patient outcomes by treatment statuses
within each subgroup of patients. Boxplots of the outcomes can be
plotted in addition to densities and interaction plot of the average
outcomes within each of these groups. They can all be generated like the
following with resulting plots in Figures \ref{fig:plot_ex_model_1},
\ref{fig:plot_ex_model_2}, and \ref{fig:plot_ex_model_3}:

\begin{CodeChunk}

\begin{CodeInput}
R> plot(subgrp.model)
\end{CodeInput}
\begin{figure}

{\centering \includegraphics{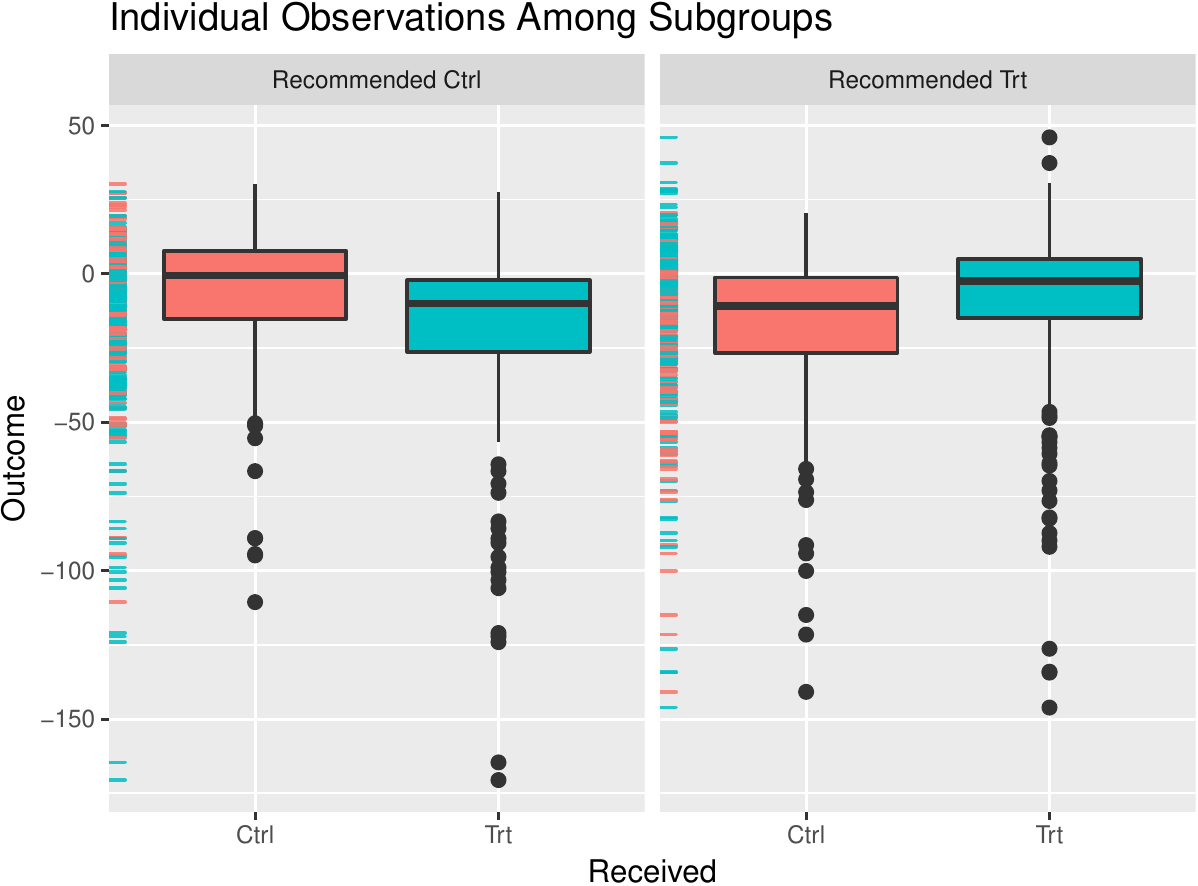} 

}

\caption[Individual outcomes within both the subgroup of patients whose benefit scores are positive and the subgroup of those whose benefit scores are negative]{Individual outcomes within both the subgroup of patients whose benefit scores are positive and the subgroup of those whose benefit scores are negative.}\label{fig:plot_ex_model_1}
\end{figure}
\end{CodeChunk}

\begin{CodeChunk}

\begin{CodeInput}
R> plot(subgrp.model, type = "density")
\end{CodeInput}
\begin{figure}

{\centering \includegraphics{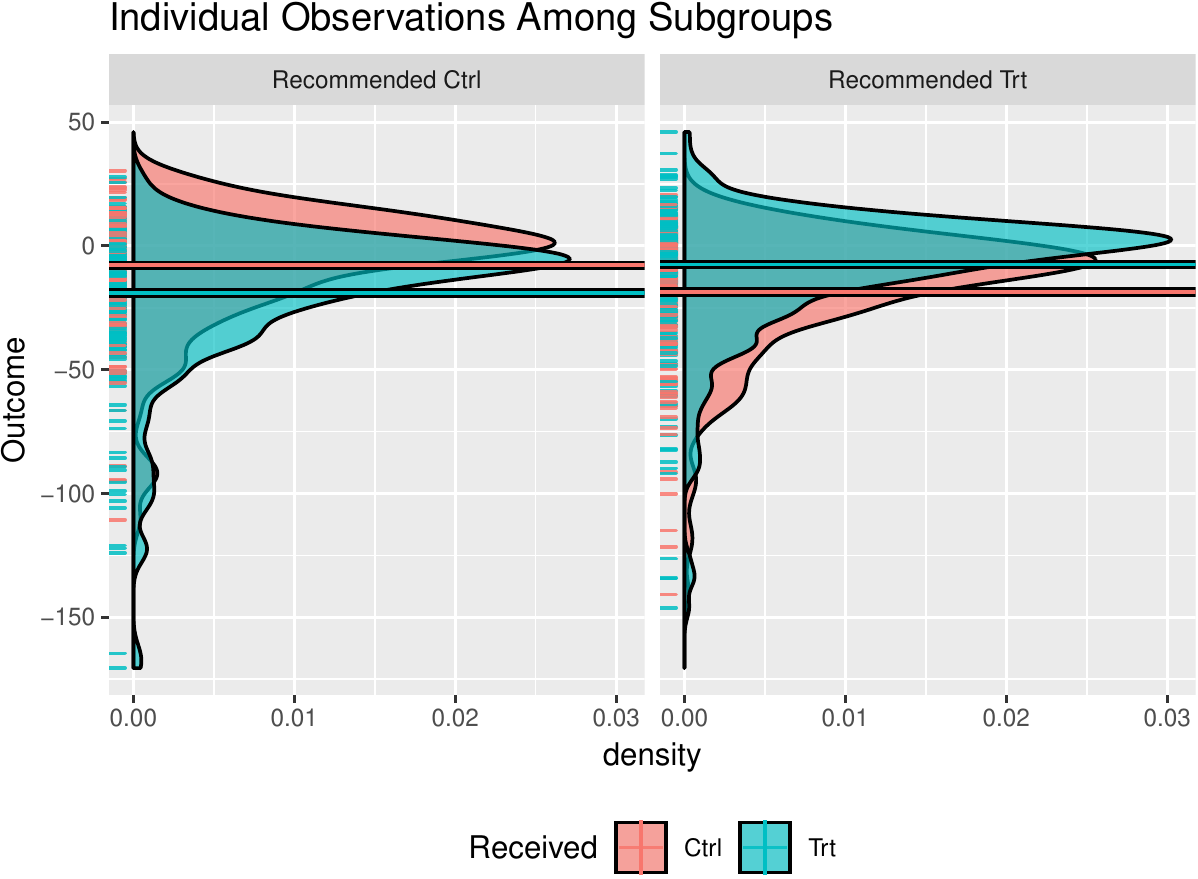} 

}

\caption[Density plots of individual outcome observations among the different subgroups]{Density plots of individual outcome observations among the different subgroups.}\label{fig:plot_ex_model_2}
\end{figure}
\end{CodeChunk}

\begin{CodeChunk}

\begin{CodeInput}
R> plot(subgrp.model, type = "interaction")
\end{CodeInput}
\begin{figure}

{\centering \includegraphics{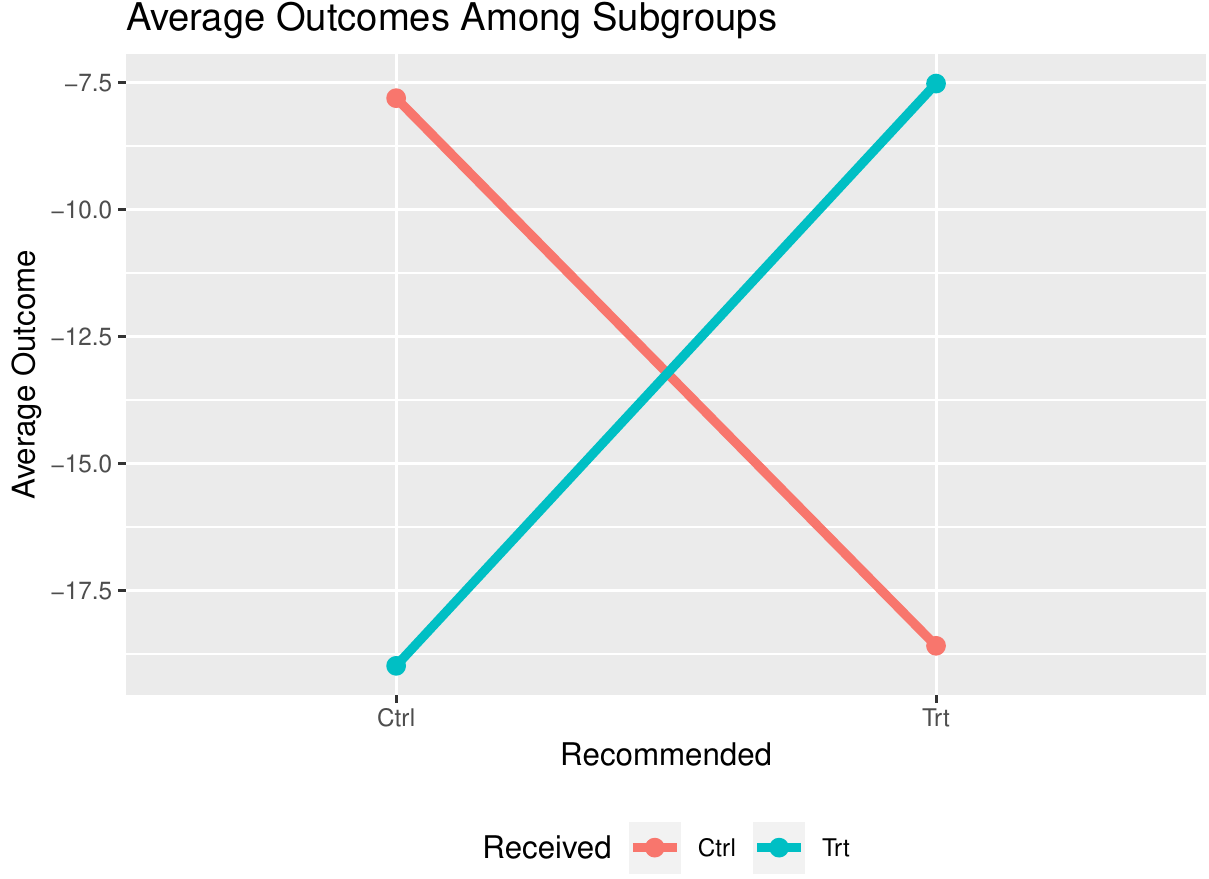} 

}

\caption[Interaction plot of the average outcome values within each subgroup by treatment status]{Interaction plot of the average outcome values within each subgroup by treatment status.}\label{fig:plot_ex_model_3}
\end{figure}
\end{CodeChunk}

For \code{subgroup_fitted} objects, Specifying the \code{plot()}
argument \code{type = "conditional"} displays smoothed means of the
outcomes conditional on each treatment group as a function of the
benefit score. Thus, a meaningful subgroup will be revealed if the
conditional means of the treated and untreated groups are not parallel
in the benefit score. The conditional plot generated from the below code
is in Figure \ref{fig:plot_ex_model_4}.

\begin{CodeChunk}

\begin{CodeInput}
R> plot(subgrp.model, type = "conditional")
\end{CodeInput}
\begin{figure}

{\centering \includegraphics{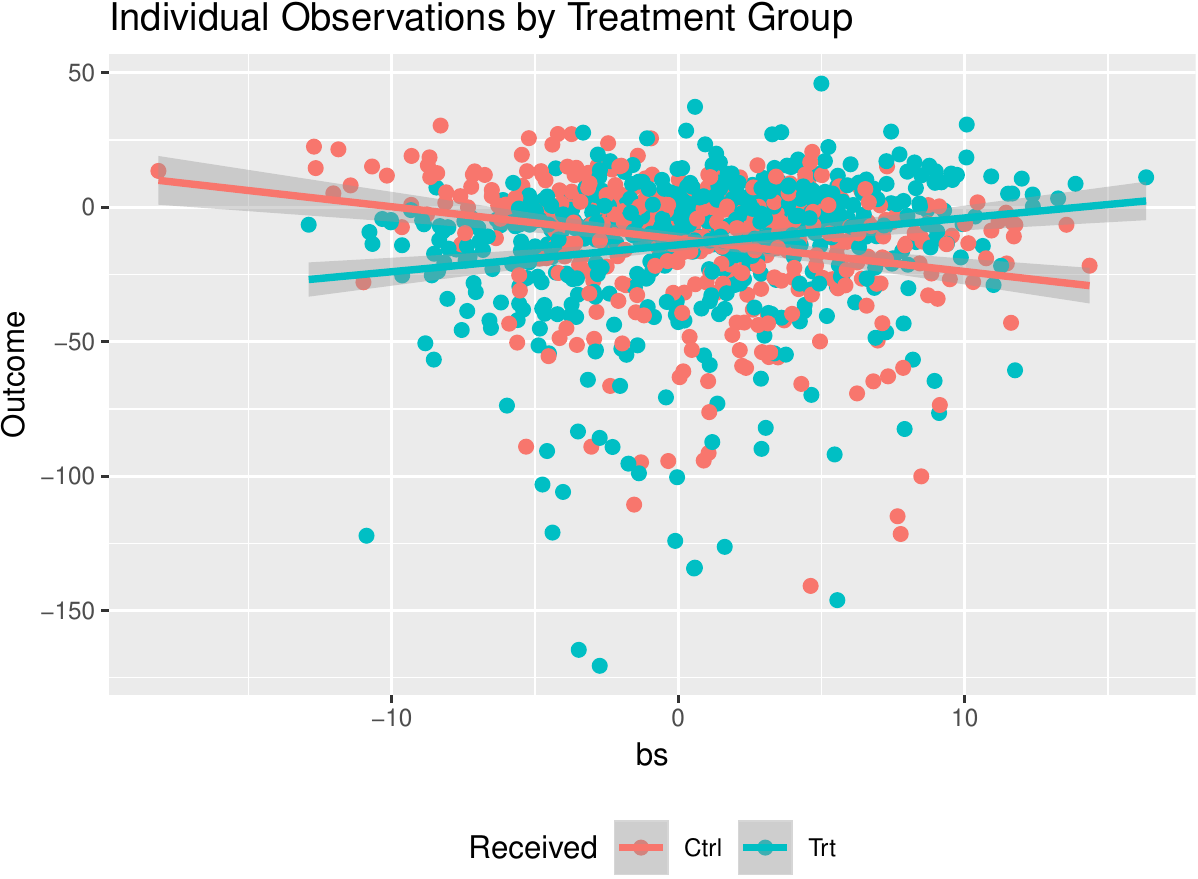} 

}

\caption[Individual observations of outcomes versus estimated benefit score by treatment status with smoothed mean lines by treatment arm]{Individual observations of outcomes versus estimated benefit score by treatment status with smoothed mean lines by treatment arm.}\label{fig:plot_ex_model_4}
\end{figure}
\end{CodeChunk}

Multiple models can be visually compared using the \code{plotCompare()}
function, which offers the same plotting options as the
\code{plot.subgroup_fitted()} function.

\hypertarget{efficiency-augmentation}{%
\subsection{Efficiency augmentation}\label{efficiency-augmentation}}

We now run the repeated training and testing splitting procedure on the
model for continuous outcomes that did not utilize efficiency
augmentation so we can compare with the efficiency-augmented model:

\begin{CodeChunk}

\begin{CodeInput}
R> validation <- validate.subgroup(subgrp.model, 
R+     B = 100,  # specify the number of replications
R+     method = "training_test_replication",
R+     train.fraction = 0.75)
R> 
R> validation
\end{CodeInput}

\begin{CodeOutput}
family:  gaussian 
loss:    sq_loss_lasso 
method:  a_learning 

validation method:  training_test_replication 
cutpoint:           0 
replications:       100 

benefit score: f(x), 
Trt recom = Trt*I(f(x)>c)+Ctrl*I(f(x)<=c) where c is 'cutpoint'

Average Test Set Outcomes:
                              Recommended Ctrl
Received Ctrl  -10.83 (SE = 5.5725, n = 40.21)
Received Trt  -16.314 (SE = 4.9946, n = 59.98)
                                Recommended Trt
Received Ctrl -15.6022 (SE = 3.3228, n = 62.22)
Received Trt   -10.4702 (SE = 3.131, n = 87.59)

Treatment effects conditional on subgroups:
Est of E[Y|T=Ctrl,Recom=Ctrl]-E[Y|T=/=Ctrl,Recom=Ctrl] 
                      5.5198 (SE = 8.5388, n = 100.19) 
    Est of E[Y|T=Trt,Recom=Trt]-E[Y|T=/=Trt,Recom=Trt] 
                        5.132 (SE = 5.019, n = 149.81) 

Est of 
E[Y|Trt received = Trt recom] - E[Y|Trt received =/= Trt recom]:                     
3.9346 (SE = 3.7614) 
\end{CodeOutput}
\end{CodeChunk}

The results across the iterations for either the bootstrap of the
training and testing partitioning procedure can be plotted using the
\code{plot()} function similarly to how the \code{plot()} function can
be used for fitted objects from \code{fit.subgroup()}. Similarly,
boxplots, density plots, and interaction plots are all available through
the \code{type} argument. Example code is below with resulting plot
shown in Figure \ref{fig:plot_ex_model_1a}. For the sake of space, we do
not show the density plot.

\begin{CodeChunk}

\begin{CodeInput}
R> plot(validation)
\end{CodeInput}
\begin{figure}

{\centering \includegraphics{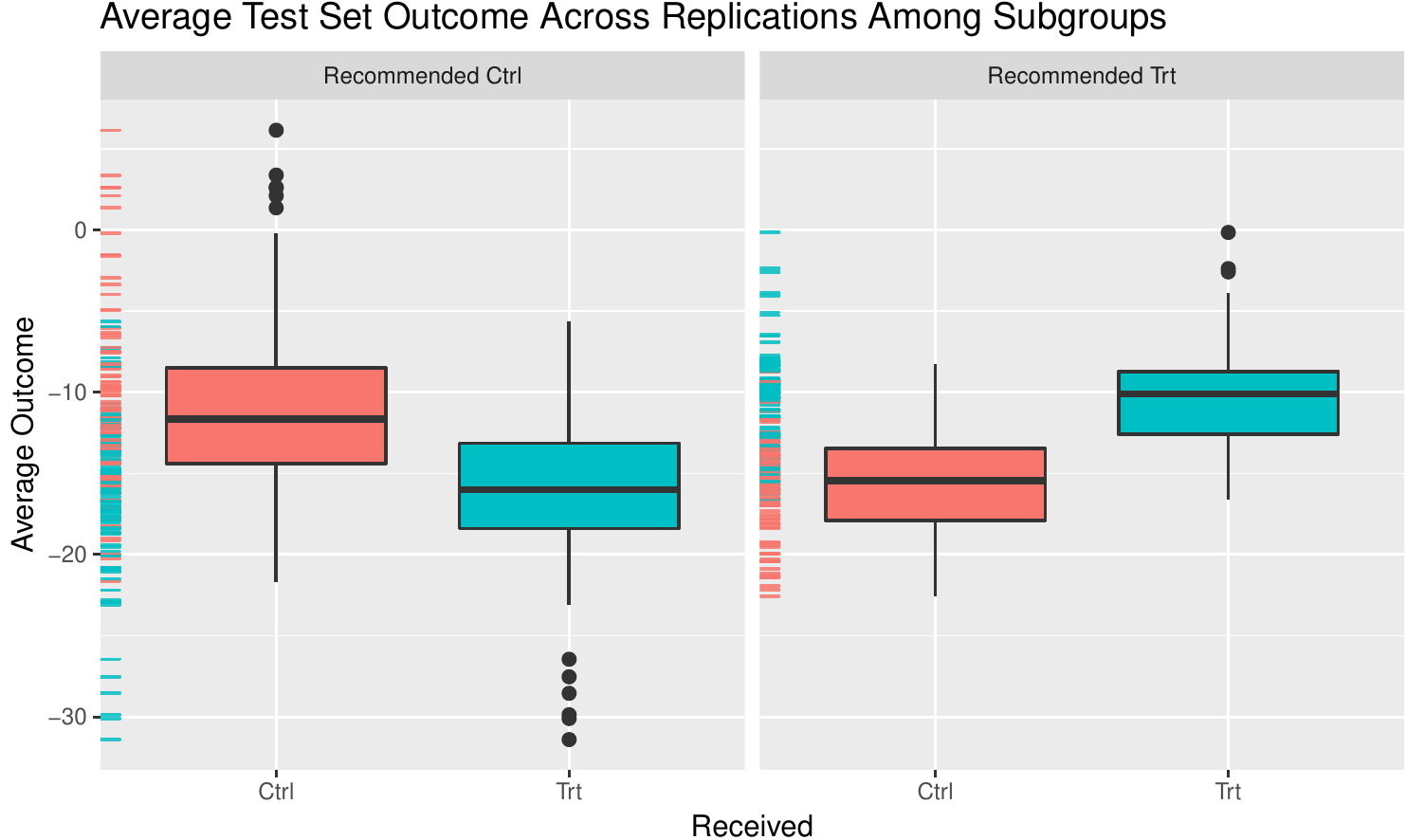} 

}

\caption[Values of average test set outcomes stratified by subgroups and treatment statuses across the training and testing replications]{Values of average test set outcomes stratified by subgroups and treatment statuses across the training and testing replications.}\label{fig:plot_ex_model_1a}
\end{figure}
\end{CodeChunk}

\begin{CodeChunk}

\begin{CodeInput}
R> plot(validation, type = "density")
\end{CodeInput}
\end{CodeChunk}

Specifying the argument \code{type = "conditional"} plots the validation
results conditional on different cutoff values for the benefit score as
specified to \code{validate.subgroup()} via the
\code{benefit.score.quantiles} argument. The resulting conditional plot
generated by the below code is shown in Figure
\ref{fig:plot_ex_model_1c}.

\begin{CodeChunk}

\begin{CodeInput}
R> plot(validation, type = "conditional")
\end{CodeInput}
\begin{figure}

{\centering \includegraphics{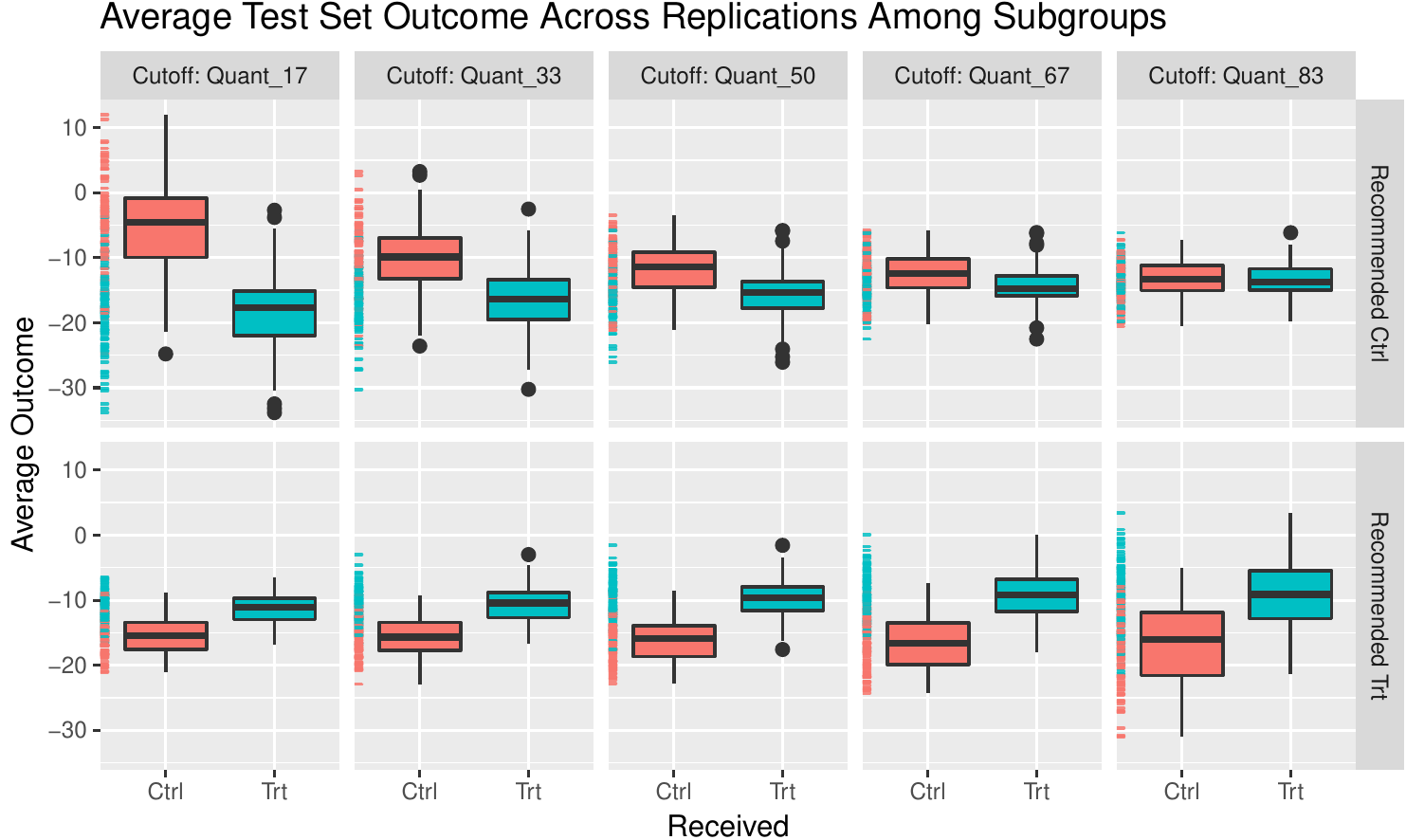} 

}

\caption[Values of average test set outcomes across the training and testing replications stratified by treatment statuses and subgroups as defined by different quantiles of the benefit score]{Values of average test set outcomes across the training and testing replications stratified by treatment statuses and subgroups as defined by different quantiles of the benefit score.}\label{fig:plot_ex_model_1c}
\end{figure}
\end{CodeChunk}

Multiple validated models can be visually compared using the
\code{plotCompare()} function, which offers the same plotting options as
the \code{plot.subgroup_validated()} function. Here we compare the model
fitted using \code{`sq_loss_lasso'} to the one fitted using
\code{`sq_loss_lasso'} and efficiency augmentation. The resulting plot
is shown in Figure \ref{fig:plot_compare_validation_ex}.

\begin{CodeChunk}

\begin{CodeInput}
R> plotCompare(validation, validation.eff)
\end{CodeInput}
\begin{figure}

{\centering \includegraphics[width=0.9\linewidth]{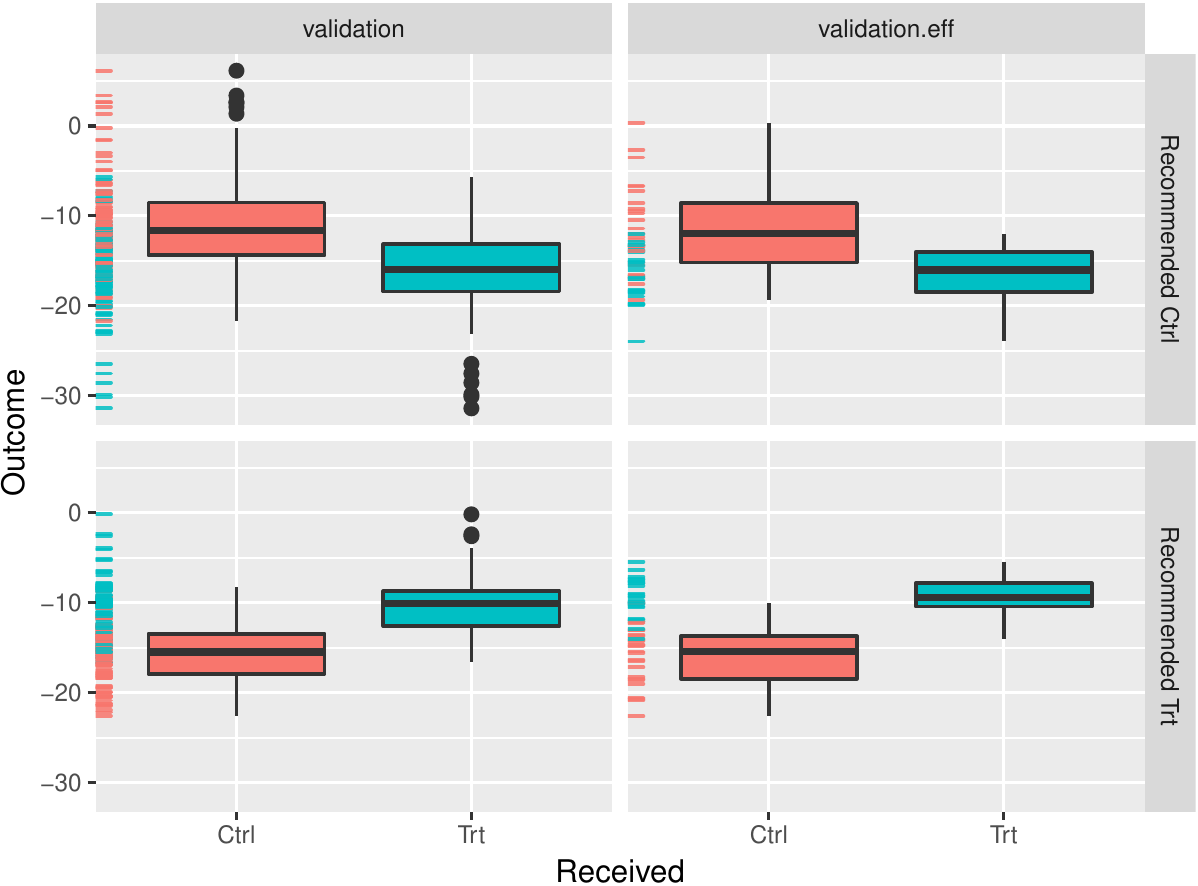} 

}

\caption[Comparison plot of the training and testing validation results for two different models]{Comparison plot of the training and testing validation results for two different models.}\label{fig:plot_compare_validation_ex}
\end{figure}
\end{CodeChunk}

From this comparison plot we can see that the efficiency-augmented model
provides estimated subgroups that result in better overall outcomes when
the recommended treatment is indeed the treatment received.

\hypertarget{example-with-multi-category-treatments}{%
\subsection{Example with multi-category
treatments}\label{example-with-multi-category-treatments}}

To demonstrate, first we simulate data with three treatments. The
treatment assignments will be based on covariates and hence mimic an
observational setting with no unmeasured confounders.

\begin{CodeChunk}

\begin{CodeInput}
R> set.seed(123)
R> n.obs  <- 1000
R> n.vars <- 100
R> x <- matrix(rnorm(n.obs * n.vars, sd = 3), n.obs, n.vars)
R> 
R> # simulated non-randomized treatment with multiple levels
R> # based off of a multinomial logistic model
R> xbetat_1 <- 0.1 + 0.5 * x[,21] - 0.25 * x[,25]
R> xbetat_2 <- 0.1 - 0.5 * x[,11] + 0.25 * x[,15]
R> trt.1.prob <- exp(xbetat_1) / (1 + exp(xbetat_1) + exp(xbetat_2))
R> trt.2.prob <- exp(xbetat_2) / (1 + exp(xbetat_1) + exp(xbetat_2))
R> trt.3.prob <- 1 - (trt.1.prob + trt.2.prob)
R> 
R> prob.mat <- cbind(trt.1.prob, trt.2.prob, trt.3.prob)
R> trt.mat <- apply(prob.mat, 1, function(rr) rmultinom(1, 1, prob = rr))
R> trt.num <- apply(trt.mat, 2, function(rr) which(rr == 1))
R> trt <- as.factor(paste0("Trt_", trt.num))
R> 
R> # simulate response
R> 
R> # effect of treatment 1 relative to treatment 3
R> delta1 <- 2 * (0.5 + x[,2] - 2 * x[,3]  )
R> # effect of treatment 2 relative to treatment 3
R> delta2 <- (0.5 + x[,6] - 2 * x[,5] )
R> 
R> # main covariate effects with nonlinearities
R> xbeta <- x[,1] + x[,11] - 2 * x[,12]^2 + x[,13] + 
R+     0.5 * x[,15] ^ 2 + 2 * x[,2] - 3 * x[,5]
R> 
R> # create entire functional form of E(Y|T,X)
R> xbeta <- xbeta + 
R+     delta1 * ((trt.num == 1) - (trt.num == 3) ) + 
R+     delta2 * ((trt.num == 2) - (trt.num == 3) )
R> 
R> 
R> # simulate continuous outcomes E(Y|T,X)
R> y <- xbeta + rnorm(n.obs, sd = 2)
\end{CodeInput}
\end{CodeChunk}

We will use the \code{factor} version of the treatment status vector in
our analysis, however, the integer values vector, i.e. \code{trt.num},
could be used as well.

\begin{CodeChunk}

\begin{CodeInput}
R> trt[1:5]
\end{CodeInput}

\begin{CodeOutput}
[1] Trt_3 Trt_1 Trt_3 Trt_2 Trt_3
Levels: Trt_1 Trt_2 Trt_3
\end{CodeOutput}

\begin{CodeInput}
R> table(trt)
\end{CodeInput}

\begin{CodeOutput}
trt
Trt_1 Trt_2 Trt_3 
  368   359   273 
\end{CodeOutput}
\end{CodeChunk}

Then we construct a propensity score function that takes covariate
information and the treatment statuses as input and generate a matrix of
probabilities as output. Each row \(i\) of the output matrix represents
an observation and each column \(j\) is the probability that the \(i\)th
patient received the \(j\)th treatment. The treatment levels are ordered
alphabetically (or numerically if the treatment assignment vector is a
vector of integers). Our propensity score model in this example will be
a multinomial logistic regression model with a lasso penalty for the
probability of treatment assignments conditional on covariate
information:

\begin{CodeChunk}

\begin{CodeInput}
R> propensity.multinom.lasso <- function(x, trt)
R+ {
R+     if (!is.factor(trt)) trt <- as.factor(trt)
R+     gfit <- cv.glmnet(y = trt, x = x, family = "multinomial")
R+ 
R+     # predict returns a matrix of probabilities:
R+     # one column for each treatment level
R+     propens <- drop(predict(gfit, newx = x, 
R+         type = "response", s = "lambda.min"))
R+ 
R+     # return the matrix probability of treatment assignments
R+     probs <- propens[,match(levels(trt), colnames(propens))]
R+ 
R+     probs
R+ }
\end{CodeInput}
\end{CodeChunk}

An important assumption for the propensity score is that
\(0 < Pr(T_i = t | \mathbf X) < 1\) for all \(\mathbf X\) and \(t\).
This assumption, often called the positivity assumption, is impossible
to verify. However, in practice validity of the assumption can be
assessed via a visualization of the empirical overlap of our estimated
propensity scores to determine if there is any evidence of positivity
violations. The \code{check.overlap()} function also allows us to
visualize the overlap of our propensity scores for multi-category
treatment applications. The following code results in the plot shown
Figure \ref{fig:check_overlap_multitreat}.

\begin{CodeChunk}

\begin{CodeInput}
R> check.overlap(x = x, trt = trt, propensity.multinom.lasso)
\end{CodeInput}
\begin{figure}

{\centering \includegraphics{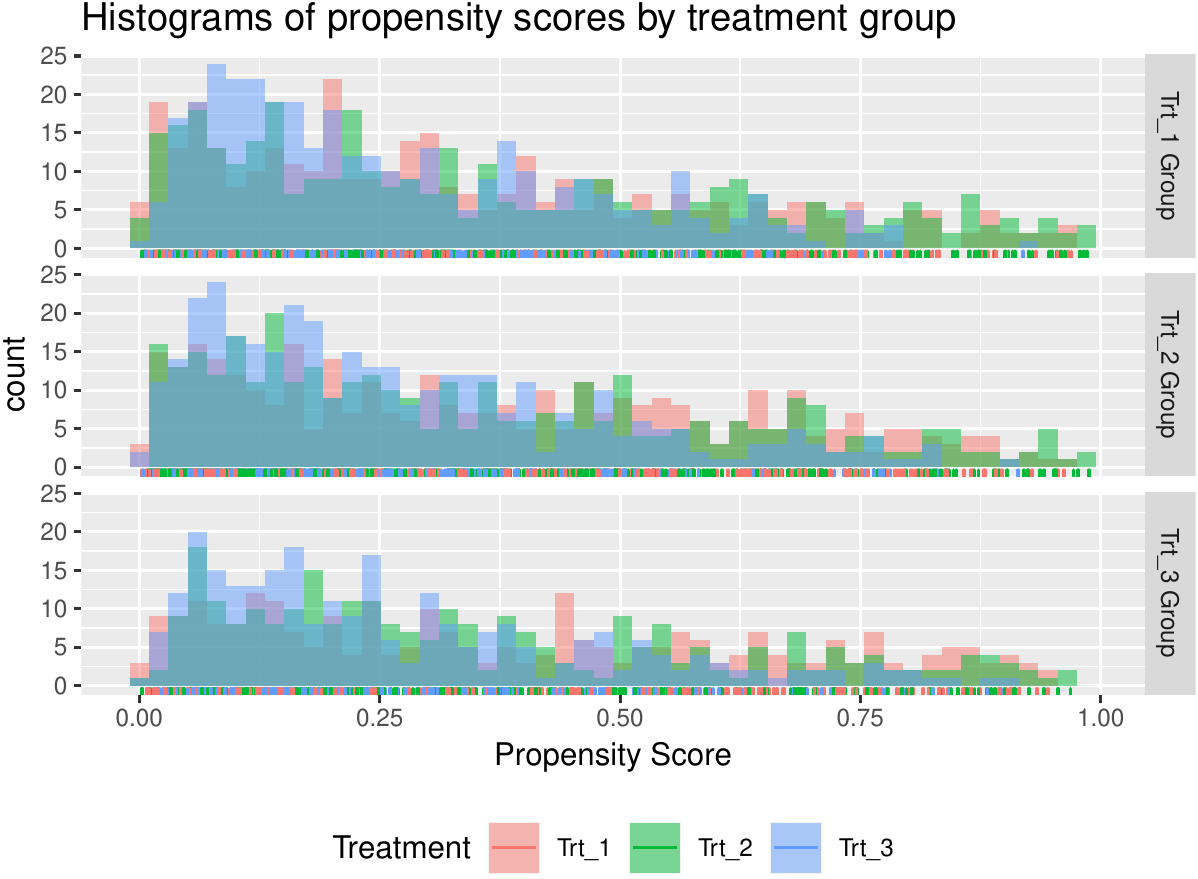} 

}

\caption[Propensity score overlap plot for multi-category treatment data]{Propensity score overlap plot for multi-category treatment data.}\label{fig:check_overlap_multitreat}
\end{figure}
\end{CodeChunk}

Each plot above is for a different treatment group, e.g.~the plot in the
first row of plots is the subset of patients who received treatment 1.
There seems to be no obvious evidence against the positivity assumption.

As the outcome is continuous and there is a large number of covariates
available for our construction of a benefit score, we will use the
squared error loss and a lasso penalty. The model can be fit in the same
manner as for the binary treatment setting, however only linear models
and the weighting method are available. Here we can also specify the
reference treatment (the treatment that the non-reference treatments are
compared with by each benefit score).

\begin{CodeChunk}

\begin{CodeInput}
R> set.seed(123)
R> subgrp.multi <- fit.subgroup(x = x, y = y,
R+     trt = trt, propensity.func = propensity.multinom.lasso,
R+     reference.trt = "Trt_3",
R+     loss   = "sq_loss_lasso")
R> 
R> summary(subgrp.multi)
\end{CodeInput}

\begin{CodeOutput}
family:    gaussian 
loss:      sq_loss_lasso 
method:    weighting 
cutpoint:  0 
propensity 
function:  propensity.func 

benefit score: f_Trt_1(x): Trt_1 vs Trt_3,  f_Trt_2(x): Trt_2 vs Trt_3 
               f_Trt_3(x): 0 
maxval = max(f_Trt_1(x), f_Trt_2(x)) 
which.max(maxval) = The trt level which maximizes maxval
Trt recom = which.max(maxval)*I(maxval > c) + Trt_3*I(maxval <= c) where c is 'cutpoint'

Average Outcomes:
                Recommended Trt_1 Recommended Trt_2 Recommended Trt_3
Received Trt_1  -1.9711 (n = 170) -9.1388 (n = 131) -30.9531 (n = 67)
Received Trt_2 -16.9325 (n = 181)  2.2732 (n = 111) -31.8461 (n = 67)
Received Trt_3 -22.4541 (n = 127) -11.1916 (n = 86)  -2.7072 (n = 60)

Treatment effects conditional on subgroups:
Est of E[Y|T=Trt_1,Recom=Trt_1]-E[Y|T=/=Trt_1,Recom=Trt_1] 
                                         17.6189 (n = 478) 
Est of E[Y|T=Trt_2,Recom=Trt_2]-E[Y|T=/=Trt_2,Recom=Trt_2] 
                                         12.4103 (n = 328) 
Est of E[Y|T=Trt_3,Recom=Trt_3]-E[Y|T=/=Trt_3,Recom=Trt_3] 
                                         28.6932 (n = 194) 

NOTE: The above average outcomes are biased estimates of
      the expected outcomes conditional on subgroups. 
      Use 'validate.subgroup()' to obtain unbiased estimates.

---------------------------------------------------

Benefit score 1 quantiles (f(X) for Trt_1 vs Trt_3): 
     0%     25%     50%     75%    100% 
-18.451  -2.903   2.216   6.885  22.101 

Benefit score 2 quantiles (f(X) for Trt_2 vs Trt_3): 
      0%      25%      50%      75%     100% 
-23.8125  -4.8801  -0.2818   4.7656  26.2459 

---------------------------------------------------

Summary of individual treatment effects: 
E[Y|T=trt, X] - E[Y|T=Trt_3, X]
where 'trt' is Trt_1 and Trt_2

 Trt_1-vs-Trt_3    Trt_2-vs-Trt_3    
 Min.   :-36.902   Min.   :-47.6249  
 1st Qu.: -5.807   1st Qu.: -9.7602  
 Median :  4.432   Median : -0.5636  
 Mean   :  4.415   Mean   : -0.4341  
 3rd Qu.: 13.771   3rd Qu.:  9.5311  
 Max.   : 44.202   Max.   : 52.4918  

---------------------------------------------------

13 out of 200 interactions selected in total by the lasso (cross validation criterion).

The first estimate is the treatment main effect, which is always selected. 
Any other variables selected represent treatment-covariate interactions.


7 out of 100 variables selected for delta 1 by the lasso (cross validation criterion).

                                      Trt_1     V2      V3    V10    V32
Estimates for delta (Trt_1 vs Trt_3) 2.0007 1.0344 -2.1732 0.1916 0.1272
                                         V61     V62     V79
Estimates for delta (Trt_1 vs Trt_3) -0.0834 -0.1028 -0.3649

6 out of 100 variables selected for delta 2 by the lasso (cross validation criterion).

                                       Trt_2      V5    V11    V63     V80
Estimates for delta (Trt_2 vs Trt_3) -0.4728 -2.2508 0.0558 0.3312 -0.1906
                                         V92     V98
Estimates for delta (Trt_2 vs Trt_3) -0.0112 -0.7417
\end{CodeOutput}
\end{CodeChunk}

The \code{summary()} function now displays selected variables for each
of the two benefit scores and shows the quantiles of each benefit score.
We can also plot the empirical observations within the different
subgroups using the \code{plot()} function, however now it is slightly
more complicated. It appears that the average outcome is higher for
those who received the level of the treatment they were recommended than
those who received a different treatment than they were recommended.
Also note that \code{plot.subgroup_fitted()} returns a \code{ggplot}
object \citep{ggplot2} and can thus be modified by the user. The below
example yields Figure \ref{fig:plot_multi_trt_model}.

\begin{CodeChunk}

\begin{CodeInput}
R> pl <- plot(subgrp.multi)
R> pl + theme(axis.text.x = element_text(angle = 90, hjust = 1))
\end{CodeInput}
\begin{figure}

{\centering \includegraphics{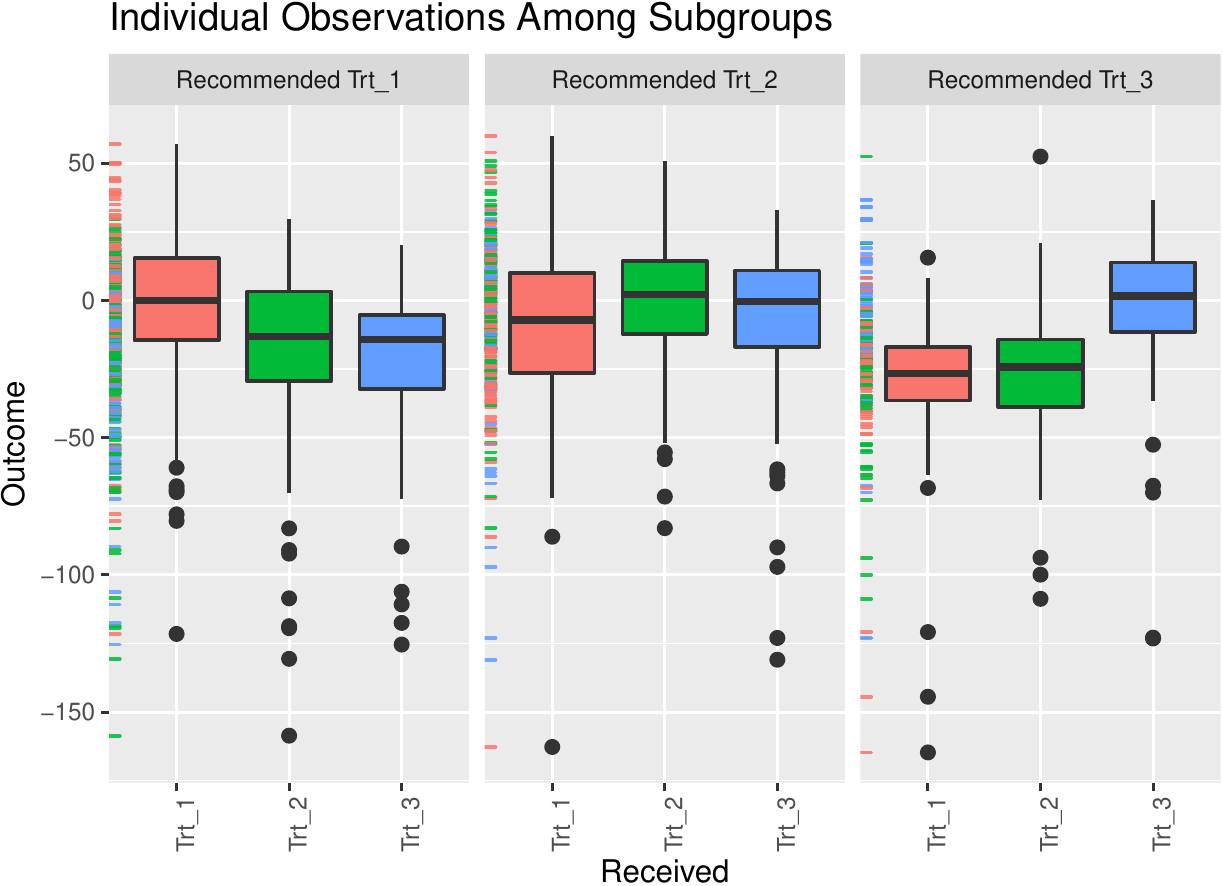} 

}

\caption[Individual outcome observations by treatment group and subgroup]{Individual outcome observations by treatment group and subgroup.}\label{fig:plot_multi_trt_model}
\end{figure}
\end{CodeChunk}

To obtain valid estimates of the subgroup-specific treatment effects, we
perform the repeated training and testing resample procedure using the
\code{validate.subgroup()} function:

\begin{CodeChunk}

\begin{CodeInput}
R> set.seed(123)
R> validation.multi <- validate.subgroup(subgrp.multi, 
R+     B = 100,  # specify the number of replications
R+     method = "training_test_replication",
R+     train.fraction = 0.5)
R> 
R> print(validation.multi, digits = 2, sample.pct = TRUE)
\end{CodeInput}

\begin{CodeOutput}
family:  gaussian 
loss:    sq_loss_lasso 
method:  weighting 

validation method:  training_test_replication 
cutpoint:           0 
replications:       100 

benefit score: f_Trt_1(x): Trt_1 vs Trt_3,  f_Trt_2(x): Trt_2 vs Trt_3 
               f_Trt_3(x): 0 
maxval = max(f_Trt_1(x), f_Trt_2(x)) 
which.max(maxval) = The trt level which maximizes maxval
Trt recom = which.max(maxval)*I(maxval > c) + Trt_3*I(maxval <= c) where c is 'cutpoint'

Average Test Set Outcomes:
                        Recommended Trt_1         Recommended Trt_2
Received Trt_1  -4.02 (SE = 3.29, 19.31%) -8.43 (SE = 3.83, 10.96%)
Received Trt_2 -16.98 (SE = 4.06, 20.25%)   2.49 (SE = 7.77, 9.02%)
Received Trt_3  -20.6 (SE = 3.16, 14.15%) -10.45 (SE = 5.47, 7.53%)
                       Recommended Trt_3
Received Trt_1 -28.34 (SE = 6.91, 6.53%)
Received Trt_2 -25.98 (SE = 5.47, 6.69%)
Received Trt_3  -4.89 (SE = 5.05, 5.57%)

Treatment effects conditional on subgroups:
Est of E[Y|T=Trt_1,Recom=Trt_1]-E[Y|T=/=Trt_1,Recom=Trt_1] 
                                  14.72 (SE = 5.14, 53.7%) 
Est of E[Y|T=Trt_2,Recom=Trt_2]-E[Y|T=/=Trt_2,Recom=Trt_2] 
                                 11.93 (SE = 8.39, 27.52%) 
Est of E[Y|T=Trt_3,Recom=Trt_3]-E[Y|T=/=Trt_3,Recom=Trt_3] 
                                 22.24 (SE = 7.02, 18.78%) 

Est of 
E[Y|Trt received = Trt recom] - E[Y|Trt received =/= Trt recom]:                  
13.76 (SE = 3.04) 
\end{CodeOutput}
\end{CodeChunk}

Setting the \code{sample.pct} argument above to \code{TRUE} prints out
the average percent of all patients which are in each subgroup (as
opposed to the average sample sizes). We can see that about 58\% of
patients were recommended treatment 1 and among those recommended
treatment 1, we expect them to have larger outcomes if they actually
receive treatment 1 as opposed to the other treatments. The estimated
effects are positive within all three subgroups (meaning those
recommended each of the different treatments have a positive benefit
from receiving the treatment they are recommended as opposed to
receiving any another treatments).

We can visualize the subgroup-specific treatment effects using
\code{plot()} as usual with results shown in Figure
\ref{fig:plotcomparemultivalidated}:

\begin{CodeChunk}

\begin{CodeInput}
R> plv <- plot(validation.multi)
R> plv + theme(axis.text.x = element_text(angle = 90, hjust = 1))
\end{CodeInput}
\begin{figure}

{\centering \includegraphics{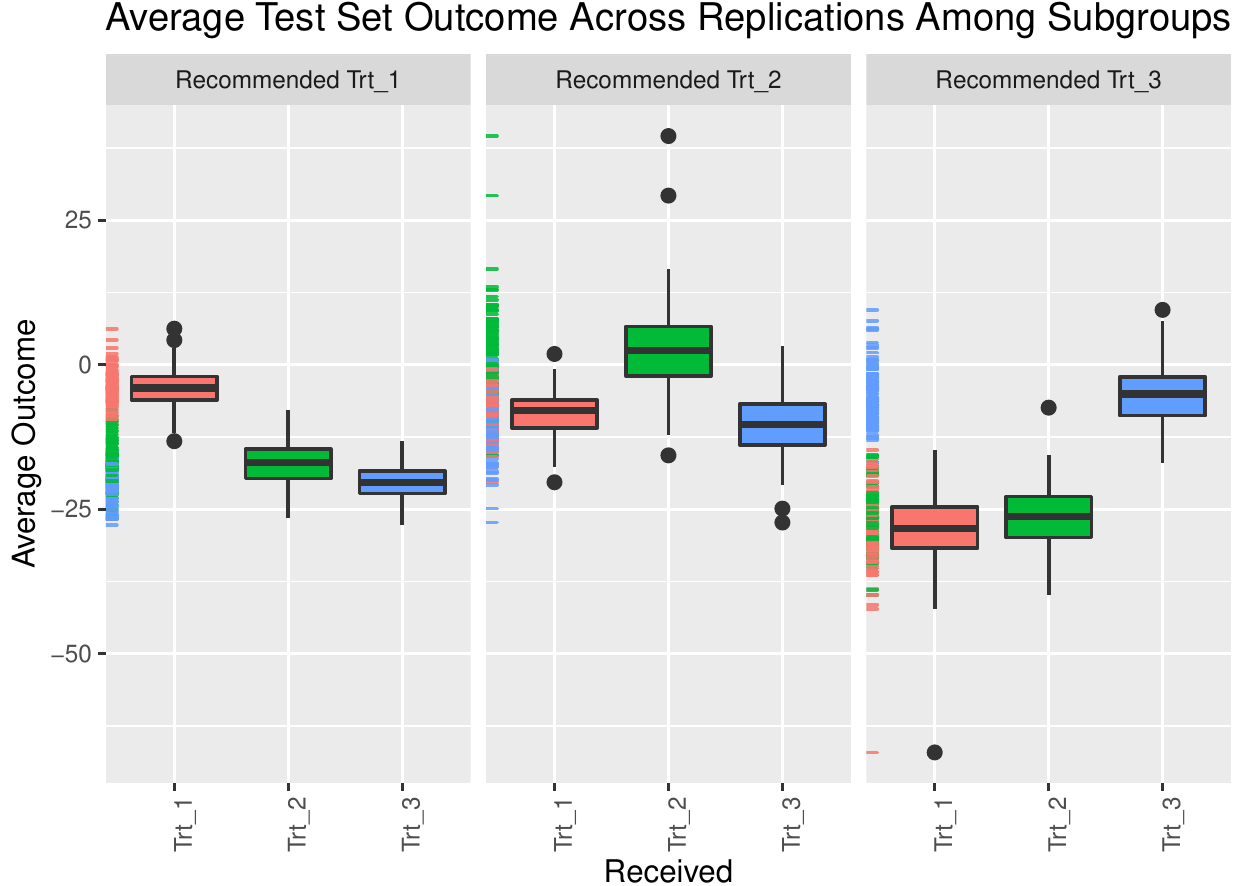} 

}

\caption[Validation results for multi-category treatment data]{Validation results for multi-category treatment data.}\label{fig:plotcomparemultivalidated}
\end{figure}
\end{CodeChunk}

\hypertarget{numerical-comparisons}{%
\section{Numerical comparisons}\label{numerical-comparisons}}

In this section we evaluate the finite sample performance of many
different available methods in the \pkg{personalized} package and
comparator methods available in other packages via a set of numerical
studies. These comparator methods, outside the scope of the
\pkg{personalized} package, are the residual-weighted learning (RWL)
method of \citet{zhou2017residual} as implemented in the
\pkg{DynTxRegime} package, the \(\ell 1\)-PLS approach of
\citet{qian2011performance} (Outcome-Lasso), an outcome-modeling
approach based on Bayesian Additive Regression Trees (Outcome-BART), and
the the model-based trees and forests approach of
\citet{seibold2016model, seibold2017individual} implemented in the
\pkg{model4you} package \citep{model4you}. Comparison with more packages
and methods would be ideal, however the majority of available packages
either do not provide functions for prediction for new patients or are
too computationally demanding. The methods from the \pkg{personalized}
package utilized in this comparison are the A-learning method with the
square loss and a lasso penalty (Sq-A); the weighting method with the
square loss and a lasso penalty (Sq-W); the weighting method the flipped
outcome weighted learning logistic loss and a lasso penalty, i.e.
\(\mathbf M(y, v) = |y|\log(1+\exp\{-\text{sign}(y)v\})\), (FOWL-L-W);
and the weighting method the flipped outcome weighted learning hinge
loss, i.e. \(\mathbf M(y, v) = |y|\text{max}(0, 1-\text{sign}(y)v)\),
(FOWL-H-W). We additionally compare with loss-augmented versions of all
of these aforementioned methods; the corresponding names have ``-Aug''
appended to the end, e.g. ``Sq-W-Aug''.

Covariates were generated as \(\mathbf X= (X_{1}, \dots, X_p)^\top\),
where \(X_2, X_4, X_6, \dots, X_{40}\) are binary random variables with
probability of success 0.25, and the remaining \(p - 20\) elements of
\(\mathbf X\) are generated from a multivariate normal random variable
with variance-covariance matrix 1 on the diagonal and \(\rho^{|i-j|}\)
for the element in the \(i\)th row and \(j\)th column with
\(\rho = 0.75\). Treatment statuses are generated from an observational
study type setting with
\(P(T = 1 | \mathbf X= \mathbf x) = \text{expit} (\beta_{T0} + \boldsymbol \beta_{T}^\top\mathbf x)\),
where \(\text{expit} (x)=1/(1+e^{-x})\) and the first 10 elements in
\(\boldsymbol \beta_{T}\) are generated from a uniform random variable
on \([-0.5, -0.25] \cup [0.25, 0.5]\) and the rest are 0. The intercept
\(\beta_{T0}\) is set such that on average 1/3 of observations would
receive \(T = 1\). The responses are generated from the following two
models:

\begin{description}
\item[Model: 1] $ Y = \boldsymbol \gamma^\top\mathbf X+ T\boldsymbol \beta^\top\mathbf X+ \epsilon$ \hfill
\item[Model: 2] $ Y = \exp(0.5\boldsymbol \gamma^\top\mathbf X) - \exp(0.5\{\nu_1X_{1}X_{2} + \nu_2X_{1}X_{3} + \nu_3X_{2}X_{3} + \nu_4X_{3}X_{4} + \nu_5X_{5}X_{6}\}) + T\boldsymbol \beta^\top\mathbf X+ \epsilon$,\hfill
\end{description}

where the first 10 elements in \(\boldsymbol \gamma\) are generated from
a uniform random variable on \([-c, -0.5c] \cup [0.5c, c]\) and the rest
are 0, \(\nu_i\) for \(i=1,\dots, 5\) are generated from a uniform
random variable on \([-c, -0.5c] \cup [0.5c, c]\), and 10 randomly
chosen elements in \(\boldsymbol \beta\) are generated from a uniform
random variable on \([-1, -0.5] \cup [0.5, 1]\). For the large main
covariate effects setting, \(c = 4/3\) and for the moderate main
covariate effects setting, \(c = 2/3\).

For all methods that apply variable selection, the lasso is used and
10-fold cross validation based on mean-squared error is used for
selection of the tuning parameter. For all methods that require the use
of a propensity score, the propensity score is created by fitting a
binary logistic regression model with a lasso penalty. For all modeling
options in the \pkg{personalized} package that do not use the hinge
loss, the lasso is used for variable selection. For all methods in the
\pkg{personalized} package that use outcome augmentation, a linear model
\texttt{Y $\mathtt{\sim}$ x + x:trt} with the lasso is used to create
the augmentation part. The treatment-covariate interactions are included
in this function so that the main effects can be correctly specified.
However, as the goal in subgroup identification is to estimate
treatment-covariate interactions, the augmentation function we use
averages over the predictions for \texttt{trt =1} and \texttt{trt = -1}.
We note that, under Model 2, this augmentation function is misspecified.

For the BART approach, we use the \pkg{BayesTree} package
\citep{BayesTree} with all covariates and the treatment indicator
included and estimate the benefit score \(\Delta(\mathbf x)\) for each
patient by evaluating the difference in predictions for \texttt{trt = 1}
versus \texttt{trt = -1}. We use the default settings in the
\pkg{BayesTree} package as the default settings are well-known to
perform admirably \citep{chipman2010bart}. Similarly, with the
\(\ell 1\)-PLS approach, we fit a linear model with a lasso penalty for
the outcome, including covariate main effects and treatment-covariate
interactions. The benefit score is estimated in the same way as the BART
approach. The augmentation function needed in creating the residuals for
the RWL method is constructed in the same way as that used for the
outcome augmentation of the \pkg{personalized} package.

The methods are evaluated by investigating the rank correlation between
the true benefit score \(\Delta(\mathbf x)\) with its estimate
\(\widehat{f}(\mathbf x)\) (or a monotone transformation of an estimate
of \(\Delta(\mathbf x)\)) on independent test sets of size 10000.
Methods are also evaluated by their area under the receiver operating
characteristic curves (AUC) with respect to the true underlying
subgroups. The results are displayed in Figures \ref{fig:plotsimres2}
and \ref{fig:plotsimres4}, where ``ME size: large'' indicates the main
effects are large, i.e. \(c = 4/3\), and ``ME size: small'' indicates
\(c = 2/3\). The vertical dashed line separates methods from the
\pkg{personalized} package and other methods. We did not include results
for the outcome weighted learning losses, only the flipped versions of
the outcome weighted learning losses as the flipped versions were
uniformly better. Similarly, the tree-based version of the approach in
\pkg{model4you} is not included, since the forest version of the method
of \pkg{model4you} is uniformly better. However, the results are
available in the Supplementary Material. Under Model 1, the
\(\ell 1\)-PLS method is correctly specified thus can serve as the gold
standard in terms of performance. However, under Model 2, the main
effects are non-linear and the \(\ell 1\)-PLS model is incorrectly
specified. Under Model 2, when the main effect size is small, the
interaction effects dominate the main effects in size and outcome
modeling approaches such as the \(\ell 1\)-PLS method are more robust to
model misspecification than for large main effect sizes under Model 2.

We can see that augmentation with correct specification (Model 1) in the
\pkg{personalized} package can boost performances although not
necessarily so with incorrect specification (Model 2). Understandably,
scenarios with larger main effects are associated with worse performance
for all methods. Of the two tree-based ensemble approaches,
model4you-Forest tends to perform the best under most scenarios. Under
Model 2 and large main effect sizes, the flipped outcome weighted
learning approach with the hinge loss and no loss augmentation works
very well and has the lowest variance. The flipped outcome weighted
learning approach with loss augmentation with the logistic loss is never
the best in any setting, however it is close to the best in all
scenarios and is thus a reasonable choice in data scenarios where not
much is known about the problem \textit{a priori}.

\begin{CodeChunk}
\begin{figure}

{\centering \includegraphics[width=0.95\linewidth]{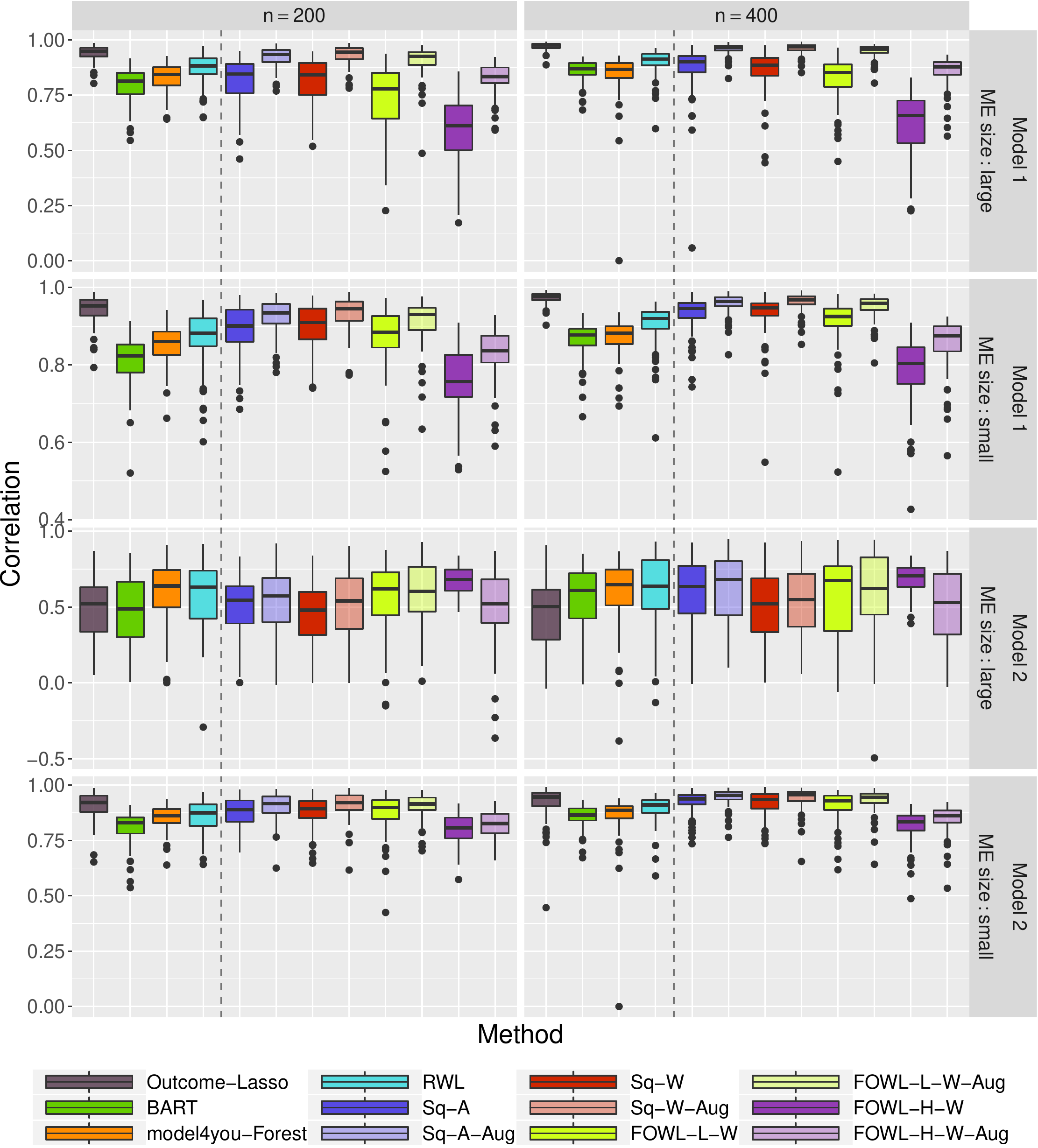} 

}

\caption[Correlations of the estimated benefit scores with the true benefit scores when the number of variables is 50]{Correlations of the estimated benefit scores with the true benefit scores when the number of variables is 50. Results displayed are from 100 independent runs of the simulation.}\label{fig:plotsimres2}
\end{figure}
\end{CodeChunk}

\begin{CodeChunk}
\begin{figure}

{\centering \includegraphics[width=0.95\linewidth]{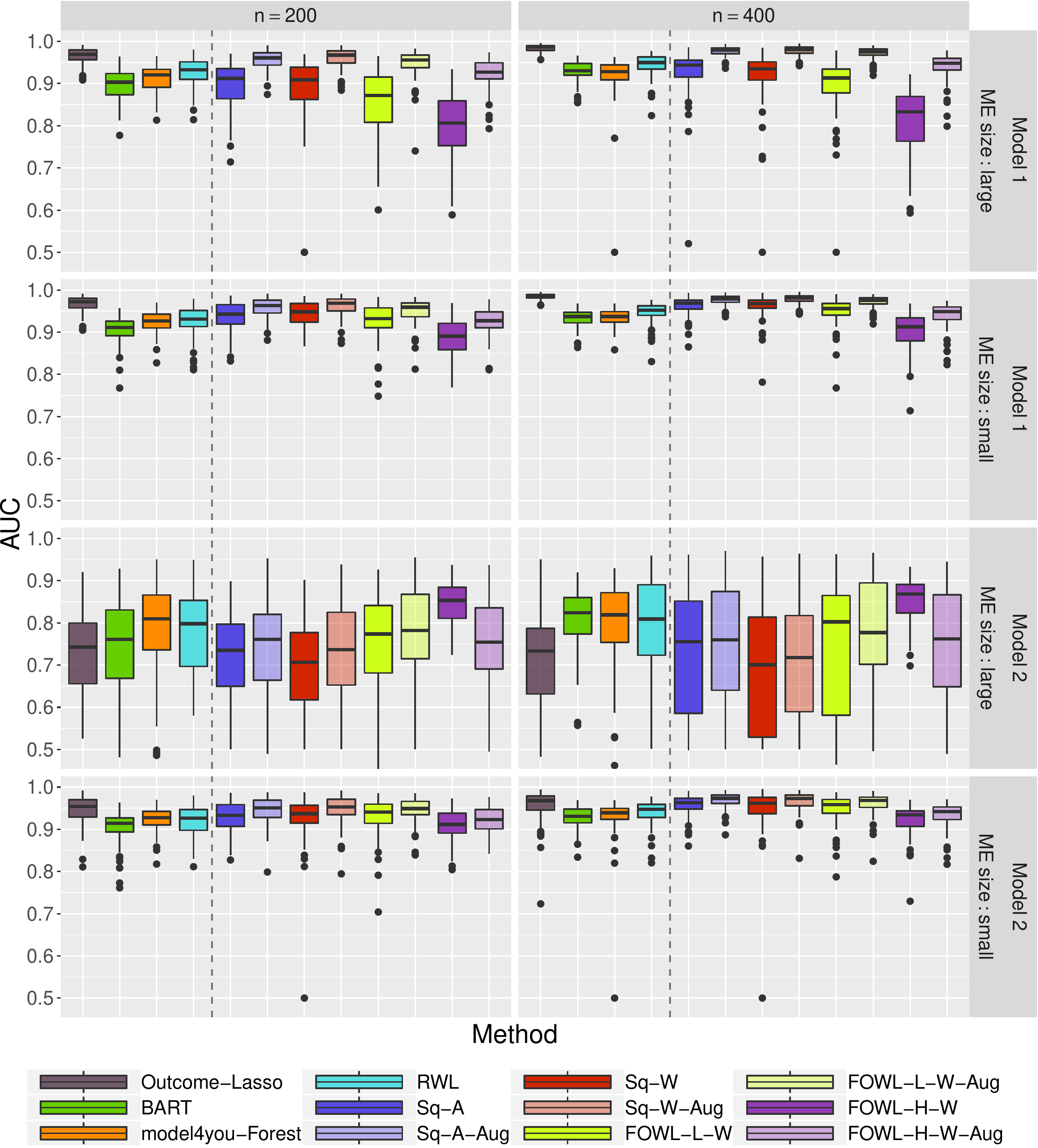} 

}

\caption[AUCs of the estimated benefit groups with respect to the true subgroups when the number of variables is 50]{AUCs of the estimated benefit groups with respect to the true subgroups when the number of variables is 50. Results displayed are from 100 independent runs of the simulation.}\label{fig:plotsimres4}
\end{figure}
\end{CodeChunk}

\hypertarget{analysis-of-national-supported-work-study}{%
\section{Analysis of National Supported Work
study}\label{analysis-of-national-supported-work-study}}

In this section we conduct a subgroup identification analysis for a
study of the effectiveness of a training program designed to help
under-served and under-employed workers gain the requisite skills for
employment. The data came from the National Supported Work Study
\citep{lalonde1986evaluating}. The outcome of interest is whether or not
the earnings of individuals are greater in 1978 than in 1975 before the
training program.

\begin{CodeChunk}

\begin{CodeInput}
R> data(LaLonde)
R> 
R> # whether an individual had a higher salary
R> # in 1978 than 1975
R> y <- LaLonde$outcome
R> 
R> # treatment assignment (employment training vs not)
R> trt <- LaLonde$treat
R> 
R> x.varnames <- c("age", "educ", "black", "hisp", "white", 
R+                 "marr", "nodegr", "log.re75", "u75")
R> 
R> # covariates
R> data.x <- LaLonde[, x.varnames]
R> 
R> # construct design matrix (with no intercept)
R> x <- model.matrix(~ -1 + ., data = data.x)
\end{CodeInput}
\end{CodeChunk}

The data come from a randomized controlled trial where patients were
randomly assigned to either the supported work program or the control
group. Even when the true propensity function is known, it is often more
efficient to estimate it from the data. Hence we use the average number
of those who were in the supported work program as the estimated
propensity score.

\begin{CodeChunk}

\begin{CodeInput}
R> const.propens <- function(x, trt)
R+ {
R+     mean.trt <- mean(trt == "Trt")
R+     rep(mean.trt, length(trt))
R+ }
\end{CodeInput}
\end{CodeChunk}

Here we fit a logistic regression-based estimator using the weighting
method with the lasso penalty. We specify the \code{cv.glmnet()}
argument \code{type.measure = "auc"} to specify the usage of area under
the receiver operating characteristic curve (AUC) as the cross
validation metric for the determination of the lasso tuning parameter.
We use the weighting method here only, since the results for the
A-learning method were very similar.

\newpage

\begin{CodeChunk}

\begin{CodeInput}
R> set.seed(1)
R> subgrp_fit_w <- fit.subgroup(x = x, y = y, trt = trt, 
R+     loss = "logistic_loss_lasso",
R+     propensity.func = const.propens,
R+     type.measure = "auc",
R+     nfolds = 10)
R> summary(subgrp_fit_w)
\end{CodeInput}

\begin{CodeOutput}
family:    binomial 
loss:      logistic_loss_lasso 
method:    weighting 
cutpoint:  0 
propensity 
function:  propensity.func 

benefit score: f(x), 
Trt recom = Trt*I(f(x)>c)+Ctrl*I(f(x)<=c) where c is 'cutpoint'

Average Outcomes:
              Recommended Ctrl  Recommended Trt
Received Ctrl  0.7292 (n = 48) 0.5146 (n = 377)
Received Trt   0.5714 (n = 28) 0.6059 (n = 269)

Treatment effects conditional on subgroups:
Est of E[Y|T=Ctrl,Recom=Ctrl]-E[Y|T=/=Ctrl,Recom=Ctrl] 
                                       0.1577 (n = 76) 
    Est of E[Y|T=Trt,Recom=Trt]-E[Y|T=/=Trt,Recom=Trt] 
                                      0.0914 (n = 646) 

NOTE: The above average outcomes are biased estimates of
      the expected outcomes conditional on subgroups. 
      Use 'validate.subgroup()' to obtain unbiased estimates.

---------------------------------------------------

Benefit score quantiles (f(X) for Trt vs Ctrl): 
     0%     25%     50%     75%    100% 
-0.2034  0.1334  0.1334  0.1334  0.3158 

---------------------------------------------------

Summary of individual treatment effects: 
E[Y|T=Trt, X] - E[Y|T=Ctrl, X]

    Min.  1st Qu.   Median     Mean  3rd Qu.     Max. 
-0.10133  0.06658  0.06658  0.06351  0.06658  0.15661 

---------------------------------------------------

2 out of 10 interactions selected in total by the lasso (cross validation criterion).

The first estimate is the treatment main effect, which is always selected. 
Any other variables selected represent treatment-covariate interactions.

            Trt hispYes marrYes
Estimate 0.1334 -0.3367  0.1825
\end{CodeOutput}
\end{CodeChunk}

To evaluate the impact of the estimated subgroups, we randomly split the
data into a training portion (80\%) and a testing portion (the remaining
20\%), fit a benefit score model on the training portion, and evaluate
the treatment effects within the estimated subgroups on the testing
portion with 500 replications. This allows us to determine whether using
our benefit score to make treatment decisions for patients will result
in better outcomes.

\begin{CodeChunk}

\begin{CodeInput}
R> val_subgrp_w <- validate.subgroup(subgrp_fit_w, B = 500,
R>     method = "training",
R>     train.fraction = 0.80)
\end{CodeInput}
\end{CodeChunk}

\begin{CodeChunk}

\begin{CodeInput}
R> print(val_subgrp_w, digits = 4, sample.pct = TRUE)
\end{CodeInput}

\begin{CodeOutput}
family:  binomial 
loss:    logistic_loss_lasso 
method:  weighting 

validation method:  training_test_replication 
cutpoint:           0 
replications:       500 

benefit score: f(x), 
Trt recom = Trt*I(f(x)>c)+Ctrl*I(f(x)<=c) where c is 'cutpoint'

Average Test Set Outcomes:
                           Recommended Ctrl               Recommended Trt
Received Ctrl  0.706 (SE = 0.1478, 5.5283%) 0.5228 (SE = 0.055, 53.3186%)
Received Trt  0.6239 (SE = 0.2045, 3.3062%) 0.602 (SE = 0.0617, 37.8469%)

Treatment effects conditional on subgroups:
Est of E[Y|T=Ctrl,T=Recom]-E[Y|T=/=Ctrl,T=Recom] 
                   0.0826 (SE = 0.2455, 8.8345%) 
  Est of E[Y|T=Trt,T=Recom]-E[Y|T=/=Trt,T=Recom] 
                  0.0793 (SE = 0.0824, 91.1655%) 

Est of 
E[Y|Trt received = Trt recom] - E[Y|Trt received =/= Trt recom]:                     
0.0763 (SE = 0.0761) 
\end{CodeOutput}
\end{CodeChunk}

The results over the 500 training and testing replications are displayed
in Figure \ref{fig:plot}.

\begin{CodeChunk}

\begin{CodeInput}
R> plot(val_subgrp_w)
\end{CodeInput}
\begin{figure}

{\centering \includegraphics{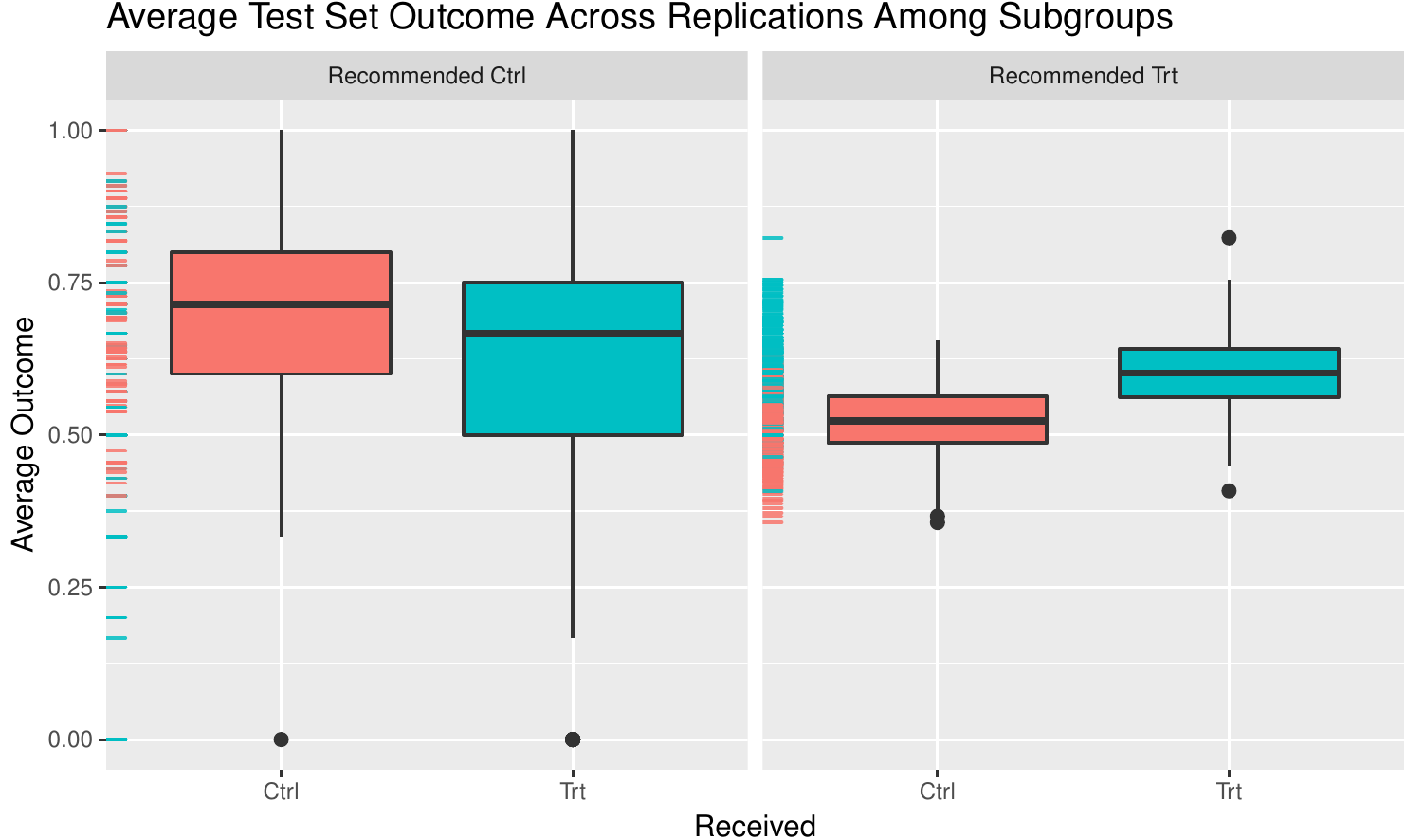} 

}

\caption[Average test set outcomes across training and testing replications stratified by subgroup and treatment status]{Average test set outcomes across training and testing replications stratified by subgroup and treatment status.}\label{fig:plot}
\end{figure}
\end{CodeChunk}

The \code{plotCompare()} function allows us to inspect the treatment
effects within subgroups on the testing datasets across all repetitions
of the validation procedure in comparison with the estimated subgroup
treatment effects using the training data. This comparison, with an
additional comparison with the estimates generated from the bootstrap
bias correction approach, is displayed in Figure
\ref{fig:plot_validate_subgrps}. The estimates of the subgroup treatment
effects based on the training data is likely overly-optimistic. However,
the training/testing procedure allows us to correct a possible
overfitting. We can see that among those who are recommended to receive
the employment training, those who actually received the employment
training were more likely to have a higher salary than those who did not
receive the training. On the other hand, among those who were not
recommended the training, those who did not receive the training may
have a slightly higher salary. However there seem to be much more
variation. Indeed the estimate of the benefit of employment training is
attenuated for the validation-based estimates compared with the biased
empirical estimates. Across the replications approximately 91\% of the
samples were recommended to receive employment training.

\begin{CodeChunk}

\begin{CodeInput}
R> val_subgrp_w_boot <- validate.subgroup(subgrp_fit_w, B = 500,
R>     method = "boot")
\end{CodeInput}
\end{CodeChunk}

\begin{CodeChunk}

\begin{CodeInput}
R> print(val_subgrp_w_boot, digits = 4, sample.pct = TRUE)
\end{CodeInput}

\begin{CodeOutput}
family:  binomial 
loss:    logistic_loss_lasso 
method:  weighting 

validation method:  boot_bias_correction 
cutpoint:           0 
replications:       500 

benefit score: f(x), 
Trt recom = Trt*I(f(x)>c)+Ctrl*I(f(x)<=c) where c is 'cutpoint'

Average Bootstrap Bias-Corrected Outcomes:
                           Recommended Ctrl                Recommended Trt
Received Ctrl 0.7029 (SE = 0.0506, 6.5235%) 0.5216 (SE = 0.0262, 52.2706%)
Received Trt  0.6167 (SE = 0.0785, 3.9729%)    0.5941 (SE = 0.03, 37.233%)

Treatment effects conditional on subgroups:
Est of E[Y|T=Ctrl,T=Recom]-E[Y|T=/=Ctrl,T=Recom] 
                  0.0863 (SE = 0.0827, 10.4964%) 
  Est of E[Y|T=Trt,T=Recom]-E[Y|T=/=Trt,T=Recom] 
                  0.0725 (SE = 0.0356, 89.5036%) 

Est of 
E[Y|Trt received = Trt recom] - E[Y|Trt received =/= Trt recom]:                     
0.0698 (SE = 0.0354) 
\end{CodeOutput}
\end{CodeChunk}

The bootstrap bias correction approach yields very similar estimates of
the subgroup-conditional treatment effects. However, the bootstrap bias
correction approach has smaller standard errors. This aligns with the
findings of \citet{Foster2011}, who noted that cross validation type
approaches lead to excessively high standard errors.

\begin{CodeChunk}

\begin{CodeInput}
R> plotCompare(subgrp_fit_w, val_subgrp_w, val_subgrp_w_boot, type = "int")
\end{CodeInput}
\begin{figure}

{\centering \includegraphics{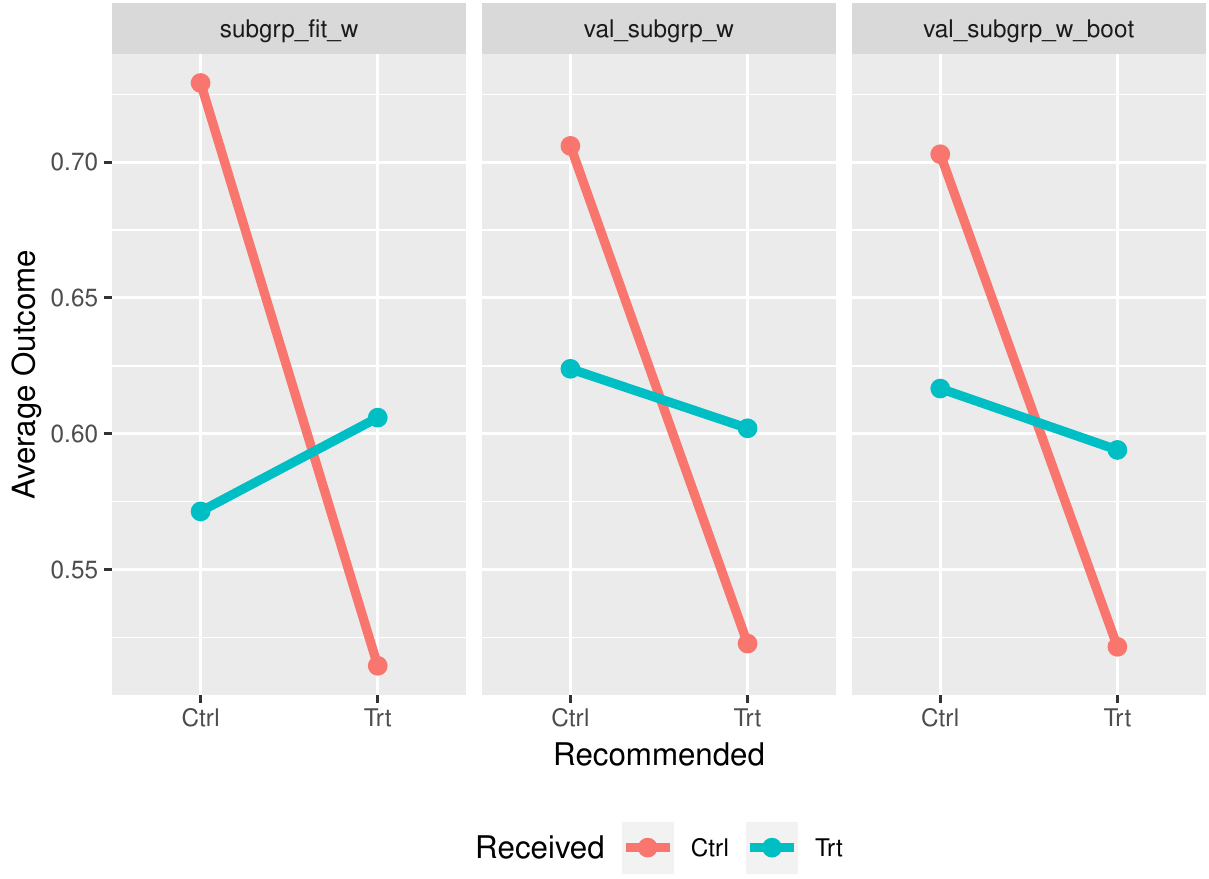} 

}

\caption[Interaction plot showing the difference in the empirical averages of outcomes on the training data compared with the average test set results across the training and testing replications]{Interaction plot showing the difference in the empirical averages of outcomes on the training data compared with the average test set results across the training and testing replications.}\label{fig:plot_validate_subgrps}
\end{figure}
\end{CodeChunk}

\hypertarget{discussion}{%
\section{Discussion}\label{discussion}}

The \pkg{personalized} package provides simple-to-use routines for
subgroup identification and personalized medicine via the general
subgroup identification framework of \citet{chen2017ageneral}. The
methods available under this framework cover a wide variety of models,
outcomes, and loss functions all under a unified code structure. The
underlying code is also designed to incorporate new models and loss
functions that fall under the purview of the framework of
\citet{chen2017ageneral}. The \pkg{personalized} package offers multiple
methods for validating the impact of estimated subgroups and various
ways of visualizing and inspecting the estimated subgroups and the
resulting subgroup treatment effects. We hope to make subgroup
identification and personalized medicine available to more statisticians
and practitioners by making the entire subgroup identification analysis
process as simple, understandable, and general as possible.

\hypertarget{acknowledgments}{%
\section{Acknowledgments}\label{acknowledgments}}

Research reported in this article was partially funded through a
Patient-Centered Outcomes Research Institute (PCORI) Award
(ME-1409-21219). The views in this publication are solely the
responsibility of the authors and do not necessarily represent the views
of the PCORI, its Board of Governors or Methodology Committee.

\bibliography{subgroupingreferences.bib}

\end{document}